\documentclass{emulateapj}

\newcommand{\ffcomm}{}
\newcommand{\refcomm}{}

\newcommand{\chandra}{\emph{Chandra}}
\newcommand{\xmm}{XMM-\emph{Newton}}

\slugcomment{To appear in ApJS, October 2005}

\shorttitle{Bright X-ray flares in COUP}
\shortauthors{Favata et al.}

\begin{document}

\title{Bright X-ray flares in Orion young stars from COUP: evidence
  for star-disk magnetic fields?}

\author{F. Favata}%\altaffilmark{1}}
\affil{Astrophysics Division, Research and Science Support Dept. of
ESA, Postbus 299, 2200 AG Noordwijk, The Netherlands}
\email{Fabio.Favata@rssd.esa.int}

\and

%\author{E. Flaccomio\altaffilmark{2}, F. Reale\altaffilmark{2}, G.
%  Micela\altaffilmark{2} and S. Sciortino\altaffilmark{2}}
\author{E. Flaccomio}
\affil{INAF-Osservatorio Astronomico di Palermo Giuseppe S. Vaiana,
  Piazza del Parlamento 1, 90134 Palermo, Italy}

\and\author{F. Reale} \affil{Dipartimento di Scienze Fisiche \&
  Astronomiche, Sezione di Astronomia, Piazza del Parlamento 1, 90134
  Palermo Italy}

\and\author{G. Micela and S. Sciortino}
\affil{INAF-Osservatorio Astronomico di Palermo Giuseppe S. Vaiana,
  Piazza del Parlamento 1, 90134 Palermo, Italy}

\and\author{H. Shang}%\altaffilmark{3}} 
\affil{Institute of Astronomy and Astrophysics, Academia Sinica, P.O.
  Box 23-141, Taipei 106, Taiwan}

\and\author{K.\,G. Stassun} \affil{Department of Physics and
  Astronomy, Vanderbilt University, Nashville, TN 37235, USA}

\author{E. D. Feigelson}%\altaffilmark{3}} 
\affil{Department of
  Astronomy and Astrophysics, 525 Davey Laboratory, Pennsylvania State
  University, University Park, PA 16802, U.S.A}

\begin{abstract}
  
  We have analyzed a number of intense X-ray flares observed in the
  \emph{Chandra} Orion Ultradeep Project (COUP), a 13 days observation
  of the Orion Nebula Cluster (ONC), concentrating on the events with
  the highest statistics (in terms of photon flux and event duration).
  Analysis of the flare decay allows to determine the physical
  parameters of the flaring structure, in particular its size and
  (using the peak temperature and emission measure of the event) the
  peak density, pressure and minimum confining magnetic field. A total
  of 32 events, \refcomm{representing the most powerful $\simeq 1$\%
    of COUP flares}, have sufficient statistics and are sufficiently
  well resolved to grant a detailed analysis. A broad range of decay
  times are present in the sample of flares, with $\tau_{\rm lc}$ (the
  $1/e$ decay time) ranging from 10 to 400 ks.  Peak flare
  temperatures are often very high, with half of the flares in the
  sample showing temperatures in excess of 100 MK. Significant
  sustained heating is present in the majority of the flares.  The
  magnetic structures which are found, from the analysis of the
  flare's decay, to confine the plasma are in a number of cases very
  long, with semi-lengths up to $\simeq 10^{12}$ cm, implying the
  presence of magnetic fields of hundreds of G (necessary to confine
  the hot flaring plasma) extending to comparable distance from the
  stellar photosphere.  These very large sizes for the flaring
  structures ($L \gg R_*$) are not found in more evolved stars, where,
  almost invariably, the same type of analysis results in structures
  with $L \le R_*$. As the majority of young stars in the ONC are
  surrounded by disks, we speculate that the large magnetic structures
  which confine the flaring plasma are actually the same type of
  structures which channel the plasma in the magnetospheric accretion
  paradigm, connecting the star's photosphere with the accretion disk.


%\keywords{Stars: structure -- Planets: exoplanets -- macros: \LaTeX \ }

\end{abstract}


\section{Introduction}
\label{sec:intro}

Beginning in their protostellar (Class I) phase, young stellar objects
(YSOs) have long been known as copious sources of X-ray emission
(\citealp{fm99}; \citealp{fm2003}). The presence of frequent flares,
observed both in accreting and non-accreting sources, shows that the
emitting plasma is magnetically confined, and in most cases also
magnetically heated. In more evolved (Class III, non-accreting) YSOs
the X-ray emission mechanism is likely to be very similar to the one
operating in similarly active older stars, involving a scaled up
version of the solar corona, with the emitting plasma being entirely
confined in magnetic loop structures and heated by similar mechanisms
as in less active stars (most likely magnetic energy dissipation
caused by the shear at the loop's footpoints, e.g.\ 
\citealp{pgn+2004}). In accreting sources (Class I and Class II), on
the other hand, recent evidence points to the X-ray emission mechanism
being, at least in part, different: statistically, in the ONC
(\citealp{fms2003}; \citealp{pre+2005}) accreting YSOs are observed to
be less X-ray luminous than non-accreting ones, although the actual
mechanism causing this difference is not understood.  Two accreting
YSOs have been observed to date in X-rays at high spectral resolution:
TW~Hya, (\citealp{khs+2002}; \citealp{ss2004}), and BP~Tau
\citep{srn+2005}.  In both cases the O\,{\sc vii} He-like triplet
shows very low $f/i$ ratios, which can be caused by very high
densities and/or by a very high UV ambient flux. Such low $f/i$ is not
observed in any ``normal'' (ZAMS, MS or more evolved) coronal source.
As e.g.\ discussed by \cite{dra2004} both effects are most likely
associated with the accretion stream (obviously not present in
non-accreting stars).  While the X-ray emission from TW Hya is very
soft ($T\simeq 3$ MK), and thus could be explained as (largely) driven
by the accretion shock, both cool and hot plasma (up to 30 MK) is
present in BP Tau; the hot plasma cannot originate in the accretion
shock (not enough gravitational energy is available), and thus in
these cases accretion-driven X-ray emission likely coexists with a
more ``canonical'' corona.

An obvious question to ask, also in the light of the recent results
concerning TW Hya and BP Tau, is therefore whether the different types
of YSOs have different types of X-ray emitting structures (and thus of
confining magnetic fields) in terms of size, location, structuring and
density. One of the few diagnostic tools which can provide an insight
in this question is the analysis of intense X-ray flares. In this
context, a flare is defined as a sudden impulsive rise in the plasma
temperature, immediately followed by a rise in the X-ray luminosity of
the source (typically by a factor of at least a few times the
quiescent luminosity), and later followed by a slower, roughly
exponential decay of both the X-ray luminosity and temperature. These
types of events are often observed both in the Sun (see e.g.\ the
review by \citealp{pf2002}) and in most active stars, and their study
has been the subject of a copious literature (see e.g.\ the relevant
chapter reviewing the topic in \citealp{fm2003}).

The \chandra\ Orion Ultradeep Project (COUP) is an unique long (13
days span with 9.7 days of effective exposure) X-ray observation of
the Orion Nebula Cluster (ONC). The details of the COUP observation
are described by \cite{gf+2005}. With its long, nearly uninterrupted
observation, the COUP sample offers an unique opportunity to study
flares in YSOs, and thus to study the type of magnetic structures
present in young stars, and in particular in stars which are either
actively accreting or which are surrounded by inactive disks. The
magnetic field structure in YSOs, is expected to be different from the
one in older, isolated stars. Magnetic field lines (and thus possibly
magnetic loops) connecting the stellar photosphere and the disk have
been postulated in the context of the magnetospheric \citep{mhc94}
model of accretion, in which the accreting plasma originates from the
disk and is channeled onto the stellar photosphere by the magnetic
field.  Magnetically funneled accretion both explains the observed
broad optical emission line profiles \citep{mhc94} and the ejection of
high-velocity bipolar outflows \citep{sns+2000}.

%As shown e.g.\ by \cite{mhc94} the channeling
%magnetic field allows the material to accrete onto the star at
%free-fall speed, thus naturally explaining the large line widths
%observed e.g.\ in the Balmer lines in Class~II YSOs. \textbf{Discuss
%  F. Shu's work here.}

However, no direct observational evidence for such extended magnetic
structures has been available up to now. Strong magnetic fields have
been measured in the photosphere of YSOs using Zeeman splitting, and
fields strengths up to 6 kG have been determined (see the review of
\citealp{jv2005}; also \citealp{jvs+2004}). Such measurements cannot
however determine the field's structure, and whether it is dominated
by a large-scale, dipole structure, or by higher-order multipoles
imposing a smaller scale to the field. In particular, the measurement
of photospheric fields does not allow determination of how the field
extends into the circumstellar region, and in particular into the
disk. Magnetic flux tubes are required by the magnetospheric model of
accretion to channel accreting material from the inner edge of the
disk to the photosphere.  Analysis of the flares' decay can help to
determine if the same flux tubes are also the seat of more energetic
processes, i.e.\ if the plasma is heated to temperatures resulting in
X-ray emission and in the attendant flaring activity.

A number of intense X-ray flares on PMS stars have been previously
observed and studied. Prior to the \xmm\ and \chandra\ observations
some 6 flares on PMS stars had been analyzed in detail (see review by
\citealp{fm2003}), although a large number of fainter events (for
which therefore physical parameters could not be derived) had also
been observed -- see e.g.\ \cite{snh2000} for a survey of flares in
the Taurus-Auriga-Perseus complex. \cite{fmr2001} analyzed four of the
bright events on YSOs known at the time, finding, in all cases, that
the flaring loop structures were of modest size ($L \le R_*$), similar
to what observed in active main sequence and more evolved stars.

A number of flaring events on YSOs have been observed with \chandra\ 
and \xmm\ (see e.g.\ \citealp{int+2003} for a \chandra\ study of
flares in the $\rho$ Oph region), but very few have been analyzed in
detail.  \cite{gmf+2004} analyzed a flare observed with \chandra\ on
the YSO LkH$\alpha$312 in M78, also finding modest-sized loop
structures ($L = 0.2$--$0.5 ~R_*$).

Therefore, the analysis of the flares observed up to now on YSOs has
resulted in magnetic structures which are in all respects similar to
the ones found in older stars, supporting a view in which the
coronae of YSOs are similar to the ones of their more evolved
counterparts, and supplying no evidence for the long magnetic
structures expected to connect the star with its disk. As it will be
shown in this paper, the analysis of the large flares occurring in the
COUP sample provides a different picture: in addition to flares
occurring in star-sized or smaller structures, similar to what found
up to now in both PMS and older stars, a number of events are found
taking place in very long magnetic structures, with lengths of several
stellar radii. As discussed in Sect.~\ref{sec:discussion}, loops of
this size around fast rotating stars are unlikely to be stable if
anchored on the star alone, and therefore such structures are likely
to be extending from the star to the disk, providing observational
evidence for the type of structures postulated by magnetospheric
accretion models.

While COUP is the best sample available to date for the study of the
size of coronal structures on YSOs, its limitations should also be
stressed. The main one is the relatively limited photon statistics of
\chandra\ observations of YSOs in the ONC: only the most intense and
longer lasting flares will have sufficient statistics to allow for the
time-resolved spectral analysis needed by the approach used here.
Therefore, the physical parameters derived in the present paper are
not necessarily representative of the ``average'' flaring structure in
the coronae of YSOs (in the ONC or elsewhere); rather they must be
interpreted as the structures associated with the most intense flares
present in these coronae.  Structures associated with less intense
flares (most likely smaller magnetic loops than the ones associated
with very intense flares) are certainly present, and given the large
large number of shorter and less intense flares visible in the light
curve of these stars (which cannot be analyzed in detail due to the
low statistics) these are common. Indeed, even in our biased sample a
number of compact flaring events is present, and, like what observed
in older stars, most likely confined in loops anchored on the stellar
photosphere, with no interaction with the disk.

The present paper is structured as follows: the selection of our
sample is described in Sect.~\ref{sec:sample}, the analysis procedures
adopted in Sect.~\ref{sec:anal}, the results from the analysis are
presented in Sect.~\ref{sec:results}, which includes detailed modeling
of a sample flare (on COUP 1343, Sect.~\ref{sec:coup1343}). An example
of a complex flaring event (COUP 450, which cannot be analyzed with
the method adopted here) is discussed in Sect.~\ref{sec:coup450}, and
the implications from the present work are discussed in
Sect.~\ref{sec:discussion}. The conclusions are summarized in
Sect.~\ref{sec:concl}, while Appendix~\ref{sec:individ} discusses in
detail a number of notable individual events.

\section{Sample selection}
\label{sec:sample}

We have studied all flares in the COUP sample (described in detail
elsewhere, \citealp{gf+2005}) which have sufficient statistics for the
analysis performed here. From a set of simulations we initially
determined the minimum number of photons in a spectrum necessary to
derive reliable spectral parameters given the typical X-ray spectra of
COUP stars.  A minimum of 750 photons was found to represent a
reasonable compromise between the need for statistics and being able
to analyze as large as possible a number of events.

To subdivide the light curve in different interval, a Maximum
Likelihood (ML) algorithm was used, the same as also used and
discussed by \cite{whf+2005}. Briefly, the rationale behind the ML
approach is to recursively subdivide the light curve in a number of
segments, with the property that the source's count rate is compatible
with being constant during each segment. The minimum number of photons
which are comprised in a given ML block is a configurable parameter.
While for many applications allowing blocks with as little as 1 photon
is useful (e.g.\ to allow to track fast variability), for our analysis
a minimum of 750 photons per ML blocks has been imposed (to allow for
meaningful spectral analysis). In all the following whenever the term
``ML block'' is used, it implies ML blocks with a minimum of 750
photons each.

The ML block approach identifies (as discussed by \citealp{whf+2005})
a 'characteristic' level, essentially a quiescent level during the
observation, as well as blocks which are significantly above this
level and during which the X-ray luminosity is constant.  Note that
our constraint of $\ge 750$ photons per block results, in a number of
cases, in significant ``under-binning'' of the light curve, i.e.\ the
light curve is clearly variable within a given block.  This is for
example visible in the case of COUP~649 (see Sect.~\ref{sec:src649}).

\begin{figure}
%\epsscale{0.75}
\plotone{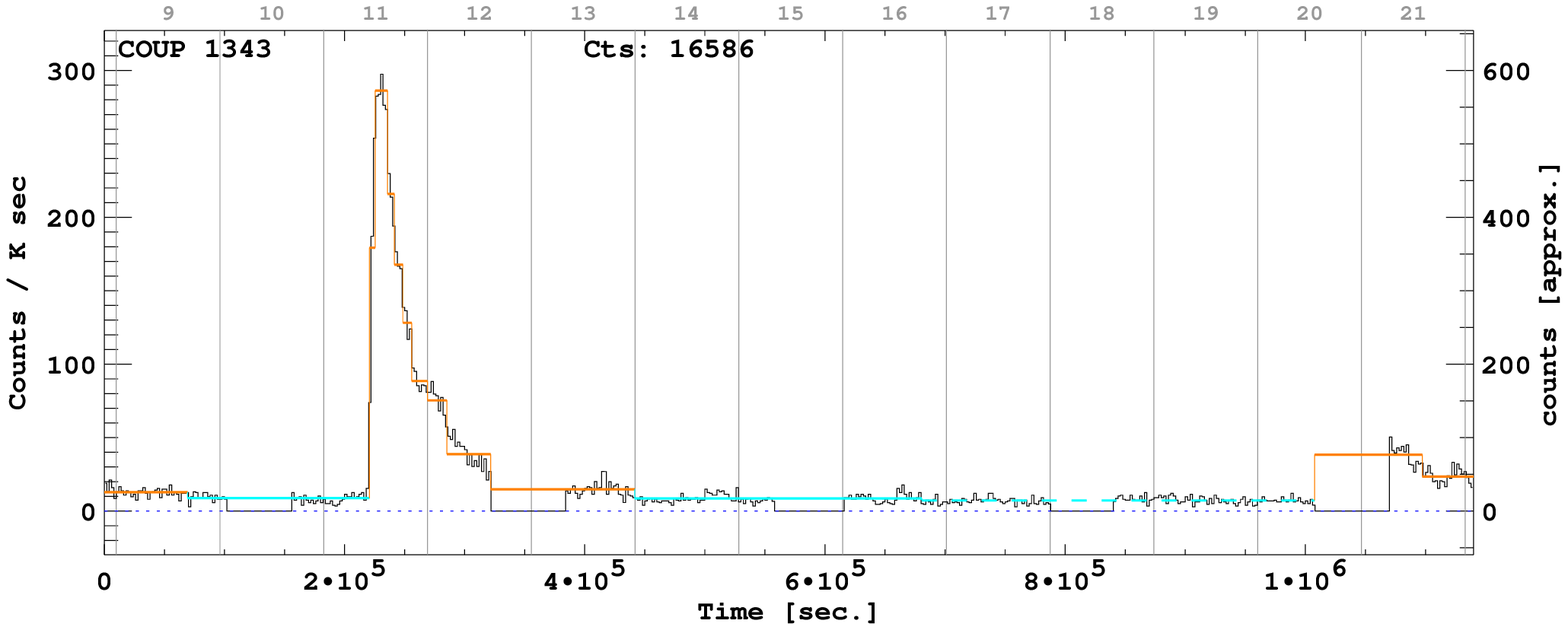}
\plotone{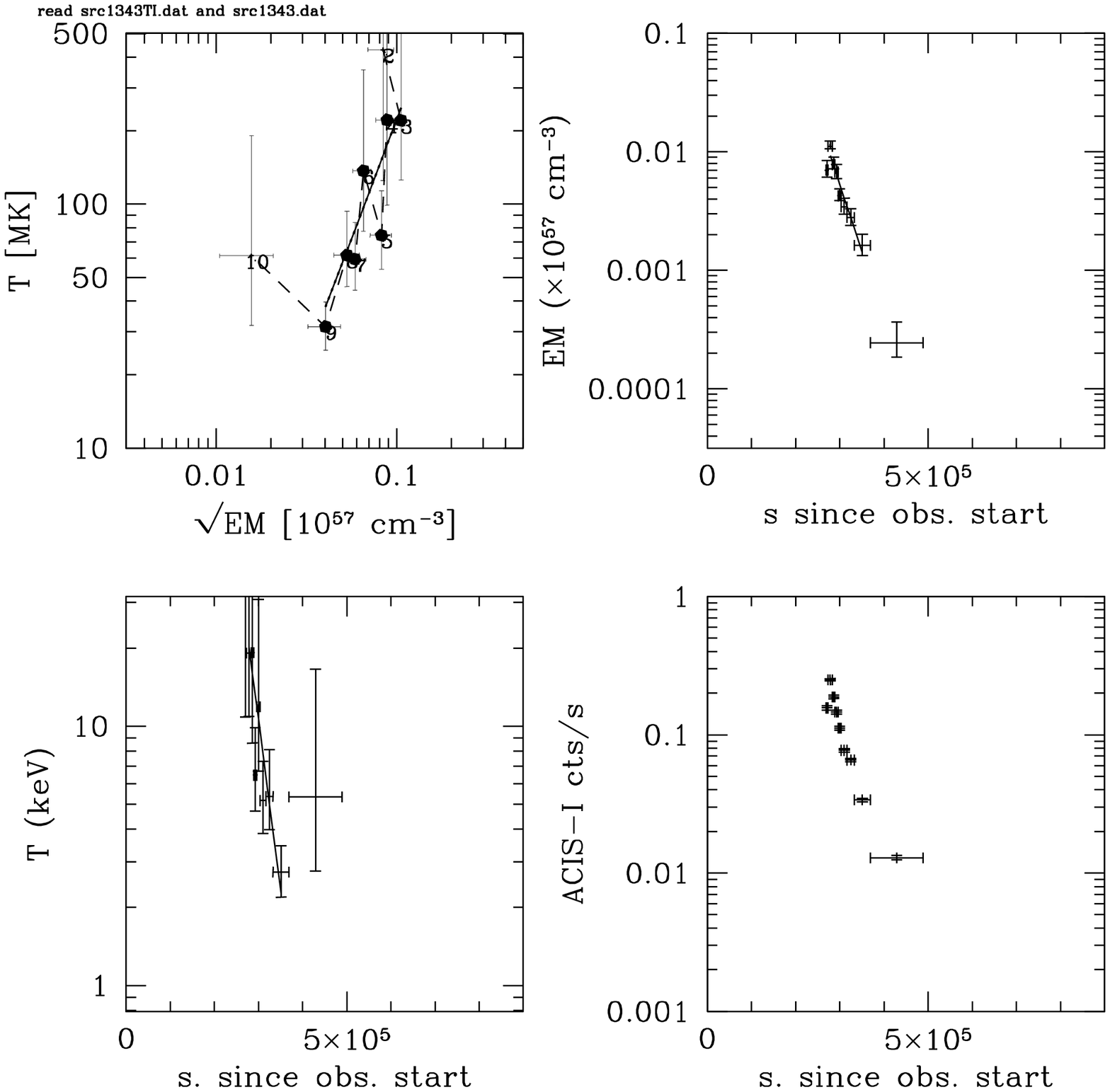}
\caption{Top panel: the COUP light curve of source 1343; the ML blocks
  at the characteristic level are plotted in light blue and the ML
  blocks with a higher level are plotted in orange. The source's ACIS
  count rate is given in the left axis, while the right axis gives the
  approximate number of counts in each 2 ks bin. Time is given in
  seconds from the beginning of the observation in the bottom axis,
  and in calendar days (January 2003) in the top axis. Bottom panel:
  evolution of the flare. The ML blocked light curve with only the
  flaring event shown is plotted in the bottom right quadrant, while
  the temperature and emission measure evolution are plotted in the
  bottom left and top right quadrants respectively. The top left
  quadrant shows the evolution of the flare in the the $\log T$--$\log
  \sqrt{E\!M}$ plane, with the points connected by a dashed line to
  allow to follow the event's evolution.  See the text
  (Sect.~\ref{sec:anacoup}) for a complete description of the
  diagrams.\label{fig:src1343}}
\end{figure}

Candidate flares for analysis have been initially hand-selected by
inspection of the atlas of light curves produced by the ML blocking
for the complete COUP sample, and choosing all sources which had at
least three contiguous ML blocks above the characteristic values.
These are candidate flaring events with sufficient information to
grant a more detailed analysis. For example the light curve of
COUP~1343 is plotted in Fig.~\ref{fig:src1343}, where (as in all other
light curves) the ML blocks at the characteristic level are plotted in
light blue and the ML blocks with a higher level are plotted in
orange.  The latter ML blocks form the candidate flaring events. For
COUP 1343, in addition to the flare being analyzed (at the beginning
of the observation) there are additional ML blocks above the
characteristic level, which have been ignored as they are not part of
the main flare. This is also common in a number of other sources.

Following the spectral analysis described below, the evolution of each
candidate flaring event in the $\log T$--$\log \sqrt{E\!M}$ plane (a
proxy for the $\log T$--$\log \sqrt{n_e}$ plane, as described in
Sect.~\ref{sec:anal}) has been examined.  Events evolving according to
the pattern discussed by \cite{rbp+97} show, after the flare peak, a
decay of both the temperature and the emission measure, more or less
regular depending on the event characteristics. The event shown in
Fig.~\ref{fig:src1343} shows this 'canonical' behavior: as visible in
the the $\log T$ vs.\ $\log \sqrt{E\!M}$ diagram (where the event
evolves temporally in the clockwise direction), the temperature peaks
while the emission measure is still rising (block 2), and has already
started to decay when the emission measure peaks (block 3). Then
(blocks 4--9) both the temperature and the emission measure decay
regularly. Two smaller flares are also visible in the light curve, one
on day 13 and the other on day 21. As not enough statistics is
available in each case, they have not been further considered.

We discarded for the purpose of the present analysis all events for
which the evolution did not proceed in a sufficiently regular way
(e.g.\ events for which the temperature did not decay). These can
either be flares with different underlying mechanisms, which cannot
therefore be analyzed within the framework used here (such as e.g.\ 
the two-ribbon flares seen in the Sun) or different types of
variability (e.g.\ induced by rotational modulation, or by the
emergence of new active regions, etc.). Only events which could be
analyzed in an useful way (i.e.\ producing well constrained physical
parameters within the assumed framework) have been retained in the
final sample. One example of a particularly complex and intense flare,
which cannot be analyzed with the approach used here, is shown and
discussed in Sect.~\ref{sec:coup450}.

Clearly, as mentioned in Sect.~\ref{sec:intro}, such procedure does
not produce an unbiased sample. The most obvious bias is toward events
which have sufficient X-ray flux and duration to yield enough
time-resolved spectra to allow the analysis. The resulting physical
parameters are therefore unlikely to be representative of the
'typical' flaring structures present in YSOs in the ONC, but will
rather represent the more extreme cases (and thus the largest magnetic
structures present). \refcomm{Quantifying the biases introduced by our
  selection procedure is not obvious. An indication can however come
  from comparison with the study of \cite{whf+2005}, who have
  performed an analysis of flares in young solar analogs in COUP using
  an automatic procedure. The young solar analog sample consists of 28
  out of 1616 COUP sources. A total of 41 distinct flares were
  identified on 26 of them, for an average of $\simeq 1.5$ flares per
  stars during the COUP observation. If the statistics on solar
  analogs were valid for the complete COUP sample (an assumption which
  needs to be verified) this would imply some 2400 flares in the
  observation. The 32 events analyzed here represent only 1\% of the
  COUP population of flares expected extrapolating the statistics for
  solar-mass stars. We are therefore studying (by necessity, imposed
  by photon statistics) the tip of an much larger iceberg.}

\section{Analysis method}
\label{sec:anal}

The analysis of the decay of flares is a classic tool to derive the
size of the flaring structure, and thus by inference other quantities
such as the plasma density and the confining magnetic field (as e.g.\ 
discussed for the Sun by \citealp{cvp70}). Stellar flares have been
observed since the beginning of stellar X-ray astronomy, and they have
been analyzed in a variety of ways, most of which proceeded by analogy
to the characteristics of (spatially resolved) solar flares. In the
case of the Sun, flares occur in localized regions in the corona,
often in single magnetic loop structures, which can remain mostly
unchanged during the flare evolution.

While the heating mechanism causing the flare still remains not
understood in its details, it is generally assumed to be unlocked by
the release of energy stored in the magnetic field. The impulsive
energy release generates either a strong thermal front or particle
beams, which, channeled by the magnetic field, hit the photosphere and
cause plasma evaporation; the plasma fills the coronal loop(s), and
produces the observed increase in X-ray luminosity. The decay of the
temperature and the X-ray luminosity is caused jointly by the
radiative losses and by heat conduction back to the photosphere.

Both cooling mechanisms depend on the geometry of the flaring
structure, and thus the decay time scale of flares is related to the
size of the coronal structure (or structures). Stellar flares have
been used to infer coronal structure sizes since the early stages of
coronal X-ray astronomy (e.g.\ \citealp{pts90}).  However, up and
until the late '90s, it was largely assumed that the heating event was
impulsive, and concentrated at the beginning of the flare, coincident
with the observed impulsive rise in temperature, so that long decay
times always led to infer large coronal structures, in the form of
very long loops. In the '90s studies of solar flares showed that the
heating mechanisms often extends well into the decay phase, so that
sustained heating and thermodynamic cooling processes are competing,
in some cases resulting in long decay times also for flares confined
in compact structures.  Thus, analyses of stellar flares based on the
assumption of impulsive heating would likely result in longer coronal
structures than actually present on the star (see for a detailed
review of past results \citealp{fm2003}).

\cite{sss+93} showed that, in solar flares, the slope of the flare
decay in the temperature-density plane is a sensitive diagnostic of
the presence of sustained heating.  \cite{rbp+97} later extended the
approach showing that the slope $\zeta$ of the flare decay in a $\log
T$ vs.\ $\log n_e$ diagram gives a quantitative measure of the
time-scale of sustained heating, which can be used to correct the
observed decay time scale and to derive the intrinsic (thermodynamic)
decay time, thus allowing us to obtain the actual size of the flaring
structure. \cite{rbp+97} developed a methodology for the analysis of
stellar flares, based on grids of hydrodynamic models of flaring
loops, which was verified on solar flares (for which the size of
individual flaring structures can be obtained from imaging
observations), and which has been applied to a significant number of
stellar events on a variety of stars, including YSOs. In the vast
majority of cases, the analysis of stellar events has shown that
sustained heating is present, so that flaring structures in stellar
coronae are generally smaller than previously thought.

\subsection{Uniform cooling loop modeling}
\label{sec:models}

The rise time in both solar and stellar flares is almost invariably
observed to be significantly shorter than the decay time, and
therefore it has been often assumed, in deriving the flare's physical
parameters, that the heating is impulsive, concentrated at the
beginning of the event, and that the decay takes place in an
``undisturbed'' fashion from a loop in near-equilibrium at the flare's
peak, according to the thermodynamic cooling time of the plasma. The
two cooling processes which determine the decay time of the flare are
thermal conduction downwards to the chromosphere and radiation, each
with its characteristic $1/e$ decay time,
\begin{equation}
\tau_{\rm cond} \simeq {3nkT \over \kappa T^{7/2}/L^2} ~~~~~~~
\tau_{\rm rad} \simeq {3nkT \over n^2 P(T)}
\end{equation}
where $n$ is the plasma density, $\kappa$ is the thermal conductivity
and $P(T)$ is the plasma emissivity per unit emission measure. Both
decay times depend on the loop's length, $\tau_{\rm cond}$ explicitly
through the $L^{2}$ term, $\tau_{\rm rad}$ implicitly through the
density's dependence. The effective cooling time of the loop is the
combination of the two,
\begin{equation}
{1\over\tau_{\rm th}} \simeq {1\over\tau_{\rm cond}} + {1\over\tau_{\rm rad}}
\end{equation}

More generally it has been shown by \cite{srj+91} that the decay time
of a flaring loop starting from equilibrium and decaying freely is
linearly related to its half length, through
\begin{equation}
L = {{\tau_{\rm th} \sqrt{T_{\rm pk}}}\over {3.7 \times 10^{-4}} }
\label{eq:serio}
\end{equation}
where $L$ is in units of $10^9$ cm, $\tau_{\rm th}$ in sec and $T_{\rm
  pk}$ is the peak temperature of the plasma in the flaring loop in
units of $10^7$ K. A fast decay thus implies a short loop, while a
slow decay implies a long loop. In all the following, unless otherwise
specified, the term ``length'' will be used to indicate the loop's
half length, from one of the footpoints to the loop apex.

In the 80's and 90's a number of approaches to the analysis of stellar
flare decays have been used (as discussed in detail by
\citealp{rea2002}; see also \citealp{fm2003} for a review of previous
literature). While differing in their details, they were mostly
equivalent to the use of Eq.~\ref{eq:serio} above, and their key
assumption was that the flaring loop always decays, after a short
heating episode, in an undisturbed fashion.

If heating does not switch off abruptly but rather decays slowly, the
plasma is subject to prolonged heating which extends into the decay
phase of the flare. As a result, the actual flare decay time
determined from the flare's light curve, $\tau_{\rm lc}$, will be
longer than the intrinsic decay time $\tau_{\rm th}$, and simple
application of Eq.~\ref{eq:serio} will result in an over-estimate of
the size of the flaring loop.  If sustained heating is present the
observed flare decay time must be ``corrected'' to obtain a reliable
estimate of the flaring region's size.

In the case of solar flares \cite{sss+93} showed that the slope
$\zeta$ of the flare decay in a $\log T$ vs.\ $\log n_e$ diagram
provides a diagnostic of the presence of sustained heating, with a
shallower slope (slower temperature decay) indicating an event with
strongly sustained heating. This approach was extended to the analysis
of stellar flares by \cite{rbp+97}, who showed that $\zeta$ provides a
quantitative diagnostic of the ratio between the intrinsic and
observed decay times, i.e.\ 
\begin{equation}
{\tau_{\rm lc} \over \tau_{\rm th}} = F(\zeta)
\end{equation}
 
The approach of \cite{rbp+97} was based on numerical simulations of
flaring loops, under the assumption of an exponentially decaying
heating function (with a time scale $\tau_{\rm H}$), and it was
validated by comparing its results with the measured size of a number
of (spatially resolved) solar flares.

The actual functional form and parameters of $F(\zeta)$ depends on the
bandpass and spectral response of the instrument used to observe the
X-ray emission from the flare, and therefore needs to be separately
determined for each instrument. Note that for stellar observations no
density determination is normally available; in this case, the
quantity $\sqrt{E\!M}$ is used as a proxy to the density, under the
assumption that the geometry of the flaring loop does not vary during
the decay.

For ACIS observations the relationship
between $\zeta$ and the ratio between the observed and intrinsic flare
decay time has been calibrated for the present analysis, resulting in
\begin{equation}
{\tau_{\rm lc} \over \tau_{\rm th}} = F(\zeta) = {{0.63} \over {\zeta - 0.32}} + 1.41
\label{eq:fzeta}
\end{equation}
which is valid for $0.32 < \zeta \lesssim 1.5$. The limits of
applicability correspond on one side ($\zeta \simeq 1.5$) to a freely
decaying loop, with no heating ($\tau_{\rm H} = 0$), to the other
($\zeta = 0.32$) to a sequence of quasi-static states for the loop
($\tau_{\rm H} \gg \tau_{\rm th}$), in which the heating time scale is
so long as to mask the loop's intrinsic decay. Note that while it is
not possible to have loops with $\zeta > 1.5$, flares with $\zeta <
0.32$ are possible if the heating is not simply exponentially decaying
but if it increases again after an initial decay.

\refcomm{The temperature in Eq.~\ref{eq:serio} is the peak temperature
in the flaring loop, which will contain plasma with a distribution of
temperatures. The relationship between the maximum temperature
present in the loop and the temperature determined by performing a
simple 1-temperature fit to the integrated X-ray spectrum emitted by
the loop has also been calibrated here. In the case of
observations performed with ACIS is }
\begin{equation}
T_{\rm pk} = 0.068 \times T_{\rm obs}^{1.20}
\label{eq:tpk}
\end{equation}
where both temperatures are in K.

Putting together the above elements, the length of the flaring
structure is therefore given by
\begin{equation}
L = {{\tau_{\rm lc} \sqrt{T_{\rm pk}}}\over {3.7 \times 10^{-4}~ F(\zeta)} }
\label{eq:size}
\end{equation}

Once the length of the flaring loop has been derived, it is possible
to infer additional physical parameters of the flaring regions, with
some additional assumptions. Given (from the analysis of the peak
spectrum) a peak emission measure for the flare, a density can be
derived if a volume is available. The above analysis only gives the
loop length; if an assumption is made about the ratio between the
loop's radius and its length ($\beta = r/L$) one can derive a
volume,
\begin{equation}
V = 2 \pi \beta^2 L^3
\end{equation}
and the resulting plasma density at flare peak
\begin{equation}
  \label{eq:dens}
  n \simeq \sqrt{EM \over V} \simeq \sqrt{EM \over 2 \pi \beta^2 L^3}
\end{equation}

The minimum magnetic field necessary to confine the flaring plasma can
be simply estimated as
\begin{equation}
B \simeq \sqrt{8 \pi p} = \sqrt{8 \pi knT}
\label{eq:b}
\end{equation}
where $p$ is the plasma pressure derived from the density and
the temperature.

For the solar corona (the only one accessible to imaging observations)
typically $\beta \simeq 0.1$, and this value has been used in most
previous analyses of stellar flares. We have also assumed, in all of
our estimates, $\beta = 0.1$, unless otherwise specified. The
consequences of
differences in $\beta$, depending on the loops characteristics, are
discussed in Sects.~\ref{sec:uncert} and~\ref{sec:results}.

\subsection{Uncertainties and error analysis}
\label{sec:uncert}

%\textbf{F. Reale, pls. check/integrate}

Both statistical and systematic uncertainties are of course present in
the length estimate for the flaring loop obtained through
Eq.~\ref{eq:size}. The main statistical uncertainty comes from the
uncertainty in the determination of $\zeta$ and therefore of
$F(\zeta)$.  Uncertainties in the determination of the peak
temperature and of the decay time are usually significantly smaller
than the uncertainty deriving from the $F(\zeta)$, which, given the
hyperbolic form of the function, becomes potentially very significant
at the lower end of the range of validity of $\zeta$. In our analysis,
the error bar reported for the length estimate is the one obtained by
propagation of the statistical errors on $\zeta$, $T_{\rm pk}$ and
$\tau_{\rm lc}$, and it is dominated by the error on $\zeta$.

One possible additional source of uncertainty would be the lack of
accurate knowledge about the star's gravity. Gravity may in principle
bring systematic effects to the diagnostics, because it makes the
plasma drain faster during the decay phase, flattening the flare's
decay in the $\log T$--$\log \sqrt{E\!M}$ plane, mimicking the
presence of sustained heating \citep{rm98}.  In a number of cases, no
estimates of either the radius or the mass of the parent star are
known, so that no determination of the surface gravity is possible.
However, given the low values of surface gravity in YSOs (with their
large radii, typically $g \simeq 0.1\,g_\odot$), the reduction of the
gravity at distances comparable to the stellar radius from the
surface, and the very high temperatures observed at flare peaks, the
pressure scale height of the flaring plasma is likely to be always
larger than the length of the flaring structures, so that gravity is
unlikely to be an additional source of uncertainty.

More subtle is the issue of systematics, i.e.\ of uncertainties
deriving from assumptions in the models which are not reflected by
reality. The modeling of the flaring plasma structure is based on
relatively simple physics, robust in its assumptions. The modeling is
simplified by the one-dimensional nature of the problem, with the
magnetic field confining the plasma except along the field lines, so
that the precise shape of the flaring structure is not
important. Another assumption is that the sustained heating is an
exponentially decaying function; of course this is a simple
parameterization which has no underlying physical reasoning behind
it. However, as long as the temperature indeed decreases, and the
slope in the $\log T$--$\log \sqrt{E\!M}$ plane is steeper than the
one of the quasi-static sequence locus ($\zeta > 0.32$) the heating
must be a decreasing function of time. Its detailed functional form in
this case has a limited impact on the loop length, so that the
resulting values are robust against the details of the heating
function. \refcomm{Also, as shown by \cite{rbp2002} strong,
long-lasting flares must be strongly confined by the magnetic and
would decay much faster than observed if the plasma were to break
confinement. }

The derived parameters (density, magnetic fields) are subject to an
additional uncertainty relative to the loop's aspect ratio $\beta$.
Our simple assumption about $\beta$ clearly cannot hold for a loop
with constant cylindrical cross section when this has a length of
several stellar radii (as is the case in some of the events analyzed
here). In this case the footpoints would cover the whole stars or be
larger than it, a clearly unrealistic configuration. Thus, such loops
must, at the footpoints, have smaller aspect ratios. At the same time,
though, a loop of such length is likely to follow a dipole-like
magnetic field configuration at large distances from the star (with
the higher-degree multipole terms becoming less important with
distance). This implies that it will likely have an expanded
cross-section at the apex (where most of the flaring plasma will be)
than at the footpoints.  These two effects compensate each other
(although to an unknown degree), and we have thus decided to still
maintain $\beta = 0.1$ in the analysis, keeping in mind that for the
longer loops the ``effective $\beta$'' may be somewhat smaller.

%\textbf{discuss additional uncertainties on derived parameters}

\subsection{Application of the analysis to the COUP data}
\label{sec:anacoup}

To determine the evolution of COUP flares in the $\log T$ vs.\ $\log
\sqrt{E\!M}$ diagram, the temperatures and emission measure of the
flaring component have been separately determined for each ML block.
All spectral fits were performed within the XSPEC package, using the
\textsc{mekal} spectral emissivity model for coronal equilibrium
plasma, \refcomm{for consistency with the rest of the COUP analysis as
described in \cite{gf+2005}}. An example of the analysis is shown in
Fig.~\ref{fig:src1343}: 
the top panel shows the raw ACIS light curve (at a resolution of 2
ks), with superimposed the light curve rebinned in ML blocks. The
segments in light blue are the ones representative of the
``characteristic'' level for the source's emission, while the ones in
orange are the ones significantly above the characteristic level, and
which form the basis for the flare analysis. The count rate for the
light curve is given in cts/ks on the left vertical axis, and in
(approximate) counts per 2 ks bin on the right vertical axis. The time
scale is given in ks from the observation start on the bottom axis and
in calendar days (January 2003) in the top axis. Also indicated are
the radius and mass of the star (if known, in solar units) and the
total number of photons in the light curve.

In a number of cases the flaring source showed significant pile-up in
the ACIS detector during the flare. In this case the pile-up free
light-curve was examined (obtained by extracting the photons from a
ring around the peak of the PSF, as described in Sect.~6 of
\citealp{gf+2005}).  However, in most cases the statistics of the
pile-up free light curve are so much lower that insufficient photons
were left for a detailed analysis of the flare. Only in the case of
COUP~891 (Fig.~\ref{fig:src891}) could the flare be analyzed, in spite
of the presence of pile-up. The flares in COUP~107, 245,290, 394, 430
and 881, which would have had sufficient statistics and sampling, were
lost from our sample because of pile-up.

The source's ``characteristic spectrum'' has been determined by
performing an absorbed multi-temperature fit to the integrated
spectrum of all ML blocks compatible with the characteristic level of
the source. The purpose of this fit was only to provide a noise-free,
phenomenological description of the source's characteristic emission,
\refcomm{which is effectively a form of background to be subtracted
from the flaring emission. Our fit provides a model of the background
to be subtracted, which is more robust and less sensitive to
statistical fluctuations than the subtraction of the characteristics
spectrum itself, which in a number of cases has rather low
statistics.} Therefore enough temperature components have been
included until the fit has yielded a satisfactory $\chi^2$, and
individual metal abundances have been left free to vary,
\refcomm{independent of whether their best-fit value was physically
meaningful or not.} 

For each ML block with emission above the characteristic level, a
spectral fit to the emission in excess of the characteristic level has
been performed. In addition to the characteristic component model with
frozen fit parameters, we include a single additional absorbed thermal
component, with global metal abundance set to $Z = 0.3\,Z_\odot$, thus
determining the emission measure and temperature of the flaring
component. \refcomm{The absorbing column density $N({\rm H})$ was left
  free to vary. Previous works on CCD resolution X-ray spectroscopy of
  very active stars, including PMS stars, shows that they typically
  have low coronal abundance, around $Z = 0.3\,Z_\odot$, which we have
  therefore chosen here. While coronal abundance variations have been
  observed during intense stellar flares, at the temperatures of
  interest here the plasma is fully dominated by continuum emission,
  and therefore the exact value of the metallicity -- which often
  cannot be well constrained -- has little influence on the best-fit
  temperature and emission measures.} In all cases this simple
one-temperature fit resulted in a satisfactory description of the
spectrum, although a number of events have spectral peculiarities
(regarding variations of abundance and absorbing column density) which
will be discussed in detail in a future paper. These do not affect any
of our conclusions.

The bottom four-quadrant panel in Fig.~\ref{fig:src1343} shows the
results of our analysis in the case of COUP 1343. The bottom right
quadrant shows the ML blocked light curve with the time axis expanded
for clarity (only the flaring event is shown, and none of the
characteristic level blocks), while the upper right and lower left
quadrants show, respectively, the time evolution of the emission
measure and of the temperature of the flaring component, as determined
from the analysis of each of the ML flaring blocks, in a semi-log
scale. The best fit exponential to the decay is also plotted for both
quantities \refcomm{as a continuous line}.  The top left quadrant
shows the evolution of the flare in the the $\log T$--$\log
\sqrt{E\!M}$ plane, with the points connected by a dashed line to
allow to follow the event's evolution. \refcomm{Points are numbered
  accordingly to the sequential number of the ML block}. The blocks
which have been used to determine the flare's evolution are marked by
a dot, and the best-fitting decay is plotted as continuous line. Its
slope $\zeta$ is the one used in Eq.~\ref{eq:fzeta}.  For all such
plots we have chosen to adopt the same scale in both axes, to allow an
immediate comparison of different events.

\begin{figure}
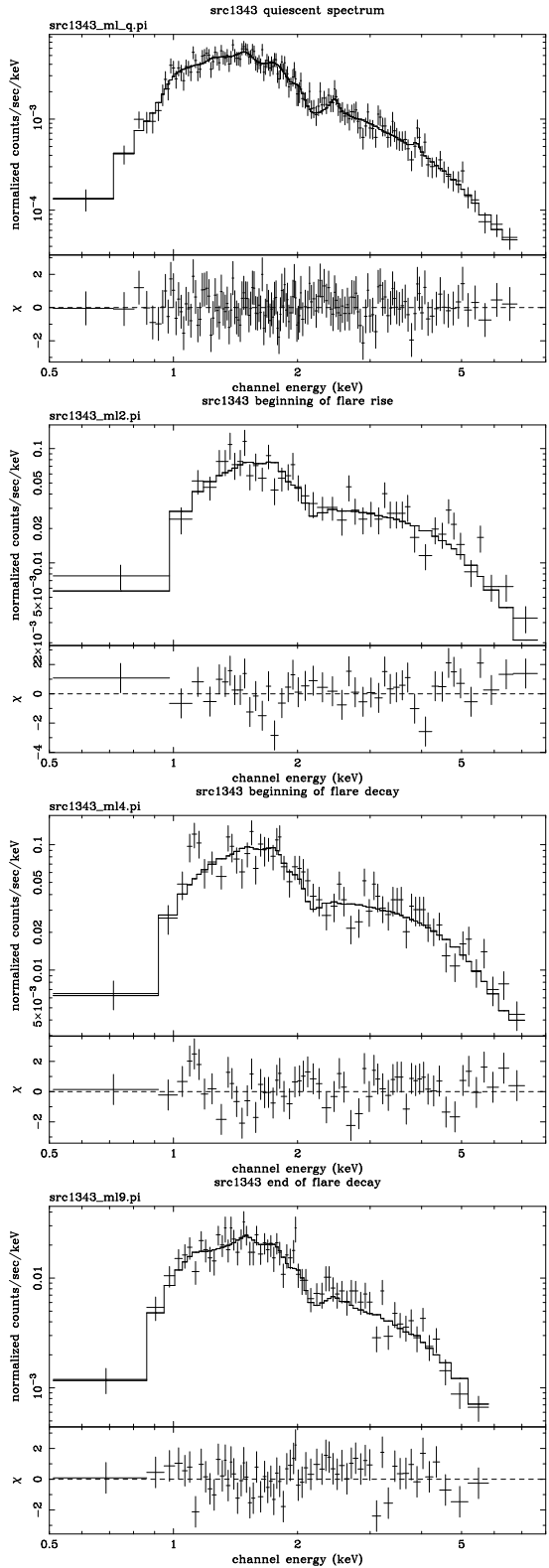

%\epsscale{0.5}
%\includegraphics[angle=90]{f1.eps}
\includegraphics[angle=270, width=7.2cm]{f2a.ps}
\includegraphics[angle=270, width=7.2cm]{f2b.ps}
\includegraphics[angle=270, width=7.2cm]{f2c.ps}
\includegraphics[angle=270, width=7.2cm]{f2d.ps}
\caption{\ffcomm{Top left panel: the characteristics
    spectrum of COUP 1343, together with its best multi-temperature
    fit; top right: spectrum collected during segment 2 (at the
    beginning of the flare rise); bottom left: spectrum collected
    during segment 4 (at the beginning of the flare decay); bottom
    right: spectrum collected during segment 9 (at the end of the
    flare decay). The flare spectra are shown together with the
    best-fit single-temperature model.}
\label{fig:spec1343}}
\end{figure}

\refcomm{In Fig.~\ref{fig:spec1343} we plot four representative ACIS
spectra for COUP~1343, illustrating the type and quality of the
spectra used in the present paper. The top left panel of
Fig.~\ref{fig:spec1343} shows the characteristic spectrum of
COUP~1343, with the 3-temperature fit used to describe it. This
spectrum has been integrated for $\simeq 540$ ks, and has 3821 source
file counts. The other panels show (clockwise) the spectrum at the
beginning of the flare's rise (block 2, when the plasma is
hottest), at the beginning of the flare's decay (block 4) and at
the end of the decay (block 9, when the plasma has already cooled
to a temperature very similar to the quiescent value). The flare
spectra contain 738 photons over 4.7 ks, 1058 photons over 5.6 ks and
1214 photons over 36 ks, respectively.}

For a large number of sources (14 out of 32), the peak temperature
derived by the above procedure converges to a value higher than 10
keV, although invariably with large error bars. Detailed analysis of
the fits shows that the fit procedure cannot determine with good
accuracy the temperature of thermal components hotter than $\simeq
100$ MK on ACIS spectra; when hotter plasma is present, statistically
equivalent fits can be obtained by arbitrarily fixing the temperature
for this hot component to any value between 100 and 600 MK. We have
thus decided to adopt a conservative approach, and for all cases for
which the maximum best-fit temperature $T_{\rm obs} > 100$ MK the
value 100 MK has been used in Eq.~\ref{eq:tpk} (yielding $T_{\rm pk} =
270$ MK). This will result in smaller loop sizes derived from
Eq.~\ref{eq:size} than it would be the case if the actual maximum peak
temperature were used. However the dependence of $L$ on $T_{\rm pk}$
is relatively weak (going as $\sqrt{T_{\rm pk}}$); thus, even if the
peak temperature were to be underestimated, in a few cases perhaps by
up to a factor of two, this would result in an additional uncertainty
in the size of the flaring loop of $\simeq 40\%$.

\section{Results}
\label{sec:results}

The physical parameters derived for each flare are listed in
Table~\ref{tab:res}. For each source we report the maximum observed
temperature $T_{\rm obs}$, the peak X-ray luminosity $L_{\rm X}$, the
peak temperature at the loop apex $T_{\rm pk}$ determined from
Eq.~\ref{eq:tpk}, the slope $\zeta$ of the flare decay in the $\log T$
vs.\ $\log \sqrt{E\!M}$ diagram, the ratio of the observed to the
intrinsic decay times $F(\zeta)$ (determined from Eq.~\ref{eq:fzeta}),
the observed decay time scale (in ks) $\tau_{\rm lc}$ determined by
fitting the flare decay binned in ML blocks. The length of the flaring
loop $L$ (from Eq.~\ref{eq:size}) is given in units of $10^{10}$ cm,
as well as, for stars which have a radius estimate published, in units
of the stellar radius. The density $n_e$ computed from the peak
emission measure and from the volume of the flaring loop (Eq.\ 
\ref{eq:dens}) is given in units of $10^{10}$ cm$^{-3}$, and the
minimum magnetic field necessary to confine the flaring plasma at the
loop apex $B$ (Eq.\ \ref{eq:b}) is given in Gauss.  Also given (when
known) are the equivalent width of the Ca\,{\sc ii} IR triplet, a
diagnostic of active accretion processes, and whether the source has
\refcomm{$\Delta(K-L)$ and $\Delta(I-K)$ excesses.} Note that the peak
X-ray luminosity is computed after subtraction of the characteristic
emission from the flaring one.

\begin{deluxetable*}{rrrrlrrlrlrrrr}
\tabletypesize{\scriptsize}
%\rotate
\tablecaption{Flare parameters from the ONC COUP sample\label{tab:res}}
\tablewidth{0pt}
\tablehead{
\colhead{Src.} & \colhead{$T_{\rm obs}$} & \colhead{$L_{\rm X}$} &
\colhead{$T_{\rm pk}$} & \colhead{$\zeta$} & \colhead{$F(\zeta)$} &   
\colhead{$\tau_{\rm lc}$} & \colhead{$L$} & \colhead{$L/R_*$} &
\colhead{$n_e$} & \colhead{$B$} & \colhead{Ca\,{\sc ii}} & $\Delta(K-L)$&
$\Delta(I-K)$\\
 & \colhead{MK} & \colhead{$10^{30}$} & \colhead{MK} &
% & \colhead{MK} & \colhead{$10^{30}$ erg s$^{-1}$} & \colhead{MK} &
 &  &   
\colhead{ks} & \colhead{$10^{10}$ cm} & & \colhead{$10^{10}$
  } & \colhead{G} &  
%  cm$^{-3}$} & \colhead{G} &  
\colhead{\mbox{\AA}} & mag & mag
}
\startdata
%Src.& Tobs& & Lx  & Tpk & \zeta       & F(z)& taulc& L_{10}        &L/R*  &ne10 & B   &Ca ii  & L   & D(I-K)
7    & 100 & 41 &270 & $0.40$\tablenotemark{a}          & 9.5 & 9.5  & 4.5 (--)      & 0.1  & 110 & 1030& \nodata& \nodata & $ -0.02$\\
28   & 80  &100 &208 & $0.47 \pm 0.10$ & 5.7 & 81.1 & 55 (17--93)   & 1.9  & 4.7 & 180 & 1.6    & \nodata & 0.30 \\
43   & 58  & 13 &142 & $2.06 \pm 0.22$ & 1.8 & 61.8 & 112 (75--134) & 5.5  & 0.5 & 51  & 1.4    & \nodata & 0.50\\
90   & 100 &240 &270 & $0.35\pm0.22$   & 20.2& 33.0 & 7.2 (0--41)   & 0.4  & 146 & 1170& 1.6    & \nodata & 0.08\\
141  & 70  & 19 &177 & $0.46$\tablenotemark{a}          & 5.9 & 22.6 & 13.7          &  0.6 & 16  & 310 & $-17.8$& \nodata & 0.43 \\
223  & 68  & 58 &170 & $1.29$\tablenotemark{a}          & 2.1 & 28.3 & 48.3          & 2.5  & 4.0 & 154 & 1.7    & \nodata & 0.48 \\
262  & 100 & 41 &270 & $0.79 \pm0.40$  & 2.8 & 122.3& 198 (43--315) & 18.0 & 0.4 & 62  & 2.3    & n & 0.48 \\
332  & 48  & 19 &113 & $4.75$\tablenotemark{a}          & 1.6 & 397.0& 733           & \nodata   & 0.04& 12  & \nodata & y & \nodata \\
342  & 100 & 44 &270 & $2.60 \pm 3.04$   & 1.7 & 52.9 & 141 (0--290)  & \nodata   & 0.7 & 80  & \nodata & n & \nodata \\
%%454  & 84  & 26 &219 & $4.02\pm0.13$   & 1.6 & 127.4& 323 (259--426)& 10.1 & 0.14& 32  & 2.1    & y & \\
454  & 84  & 26 &219 & $4.02\pm1.13$   & 1.6 & 127.4& 323 (259--426)& 10.1 & 0.14& 32  & 2.1    & y & 0.20 \\
597  & 39  & 7 & 87 & $0.40\pm0.28$    & 9.8 & 85.7 & 22 (0--75)    & 1.6  & 5.4 & 128 & 4.5    & y &\nodata \\
649  &  80 & 22 &206 &  $0.60$\tablenotemark{a}         & 3.7 & 60.7 & 64 (--)       & 4.2  & 1.8 & 113 & 0.0    & n & 0.43\\
669  & 79  & 34 &203 &  $1.63\pm0.65$  & 1.9 & 45.6 & 92 (69--111)  & 5.1  & 1.1 & 88  & \nodata & n & 0.36\\
720  & 77  & 22 &197 &  $\le 1.87$   & 3.2 & 110.9& 132 (0--614)& \nodata & 0.6 & 62 & \nodata   & \nodata & \nodata \\
% 720  & 77  & 22 &197 &  $0.67\pm1.20$   & 3.2 & 110.9& 132 (388--614)& \nodata   & 0.6 & 62  & \nodata     & -- & \\
752  & 120 & 190 &339 & $\le 0.46$     & 6.0 & 78.7 & 65 (0--74)    & 5.6  & 5.4 & 250 & 1.1    & \nodata & 0.62 \\
848  & 100 & 40 &270 & $6.53\pm2.14$     & 1.5 & 55.2 & 162 (139--190)& 11.8 & 0.5 & 70  & 0.0    & y & 0.50 \\
891  & 100 & 110 &270 & $1.68 \pm0.27$ & 1.7 & 73.1 & 173 (150--200)& 5.1  & 1.6 & 120 & 1.8    & \nodata & 1.10 \\
915  & 100 & 300 &270 & $0.98\pm0.12$  & 2.4 & 41.5 & 78 (60--105)  & \nodata & 2.4 & 223 & \nodata & \nodata & \nodata\\
960  & 49  & 98 &115 & $0.36\pm0.13$   & 15.6& 19.8 & 3.7 (0--14)   & 0.3  & 260 & 1020& 0.0    & \nodata & $-0.51$\\
971  & 86  & 97 &226 &  $1.48\pm0.23$  & 2.0 & 17.3 & 35 (32--40)   & 1.6  & 7.9 & 250 & 1.8    & \nodata & \nodata\\
976  & 100 & 32 &270 & $2.96$\tablenotemark{a}          & 1.7 & 28.2 & 76 (--)       & 12.0 & 1.5 & 120 & 0.0    & y & \nodata\\
997  & 74  & 28 &190 & $0.62\pm0.25$   & 3.5 & 31.5 & 34 (7--62)    & \nodata   & 5.0 & 180 & \nodata & n & \nodata\\
1040 & 75  & 420 &190 & $\le 0.62$  & 8.4 & 19.4 & 8.7 (0--22)   & \nodata   & 142 & 970 & \nodata & \nodata & \nodata\\
1083 & 100 & 14 &270 & $1.00\pm0.11$   & 2.4 & 124.6& 235 (186--302)& \nodata   & 0.2 & 40  & \nodata & n & \nodata\\
1114 & 100 & 95 &270 & $\le 0.77$   & 2.8 & 43.6 & 68 (0--84)    & \nodata   & 2.8 & 160 & $-1.5$ & \nodata & \nodata \\
1246 & 100 & 72 &270 & $0.90\pm0.18$   & 2.5 &22.5  &40 (32--47)    & 3.8  & 5.6 & 240 & 0.0    & \nodata  & 0.75\\
1343 & 100 & 164 &270 & $1.95\pm0.51$  & 1.8 & 38.9 & 96 (79--113)  & \nodata   & 2.3 & 150 & \nodata & y & \nodata\\
1384 & 55  & 23 &133 & $0.70\pm0.25$   & 3.1 & 50.7 & 51 (20--82)   & 3.0  & 2.4 & 110 & 1.9 & \nodata & 0.56 \\
1410 & 100 & 37 &270 & $\le 0.57$   & 6.1 &152.7 & 110 (0--200) & 55.0 & 1.0 & 95  & 0.0    & \nodata  & 2.30 \\
1443 & 69  & 51 &174 & $0.62\pm0.12$   & 3.5 & 35.4 & 26 (22--55)   & \nodata   & 5.8 & 190  & \nodata & \nodata & \nodata\\
1568 & 100 & 780 &270 & $\le 0.58$  & 15.1& 12.4 & 3.7 (0--15)   & 0.4  & 120 & 3480 & \nodata    & \nodata  & 0.19 \\
1608 & 96  & 16 &258 & $\le 0.70$   & 3.7 & 70.6 & 82 (0--99)   & 6.7  & 1.0 & 96   & $-1.3$ & n & 1.41 \\

\enddata

%% Text for table notes should follow after the \enddata but before
%% the \end{deluxetable}. Make sure there is at least one \tablenotemark
%% in the table for each \tablenotetext.

\tablecomments{See main text (Sect.~\ref{sec:results}) for a
  description of individual columns. The value for $\zeta$ is given
  together with the uncertainty range deriving from the fit to the
  flare decay in the $\log T$--$\log \sqrt{E\!M}$ plane. When $\zeta$
  is derived from only 2 decay points no uncertainty is given in the
  Table. In this case no uncertainty range is given for $L$; this
  value, as well as the derived values for $n_e$ and $B$, are in this
  case to be treated with caution. }

\tablenotetext{a}{$\zeta$ was derived from only two points in the
flare decay and therefore no uncertainty range is given.} 

\end{deluxetable*}
%\end{deluxetable}

The derived quantities ($n_e, B$) are computed assuming $\beta = 0.1$.
For long loops (of several stellar radii) this assumption may be
incorrect, as such large loops would have footpoints covering much of
the stellar surface (even though loop expansion may partially offset
this, see Sect.~\ref{sec:uncert}). However, the dependence of $n_e$
and $B$ on $\beta$ is moderate, with $n_e \propto \beta^{-1}$ and $B
\propto \beta^{-0.5}$. Thus, the values in Table~\ref{tab:res} can be
easily scaled to any value of $\beta$.

The flares analyzed here are all X-ray bright, but not exceptionally
so, when compared to other intense flares already observed in YSOs.
The peak X-ray luminosities reported in Table~\ref{tab:res},
determined by fitting the spectrum of the peak ML block for the flare,
range from $7 \times 10^{30}$ erg s$^{-1}$ up to $8 \times 10^{32}$
erg s$^{-1}$ for the most X-ray bright event. The latter one takes
place on COUP 1568 (discussed in detail in
Appendix~\ref{sec:individ}), at $M = 2.6\,M_\odot$ one of the most
massive stars in our sample. Thanks to the large offset angle, the
source does not suffer from pile-up even if the peak ACIS count rate
approaches 1 cts/s. While certainly very high, the peak luminosity of
$8 \times 10^{32}$ erg s$^{-1}$ is comparable to the values observed
in other PMS flares, e.g.\ the $10^{33}$ erg s$^{-1}$ determined at
peak in a similar band by \cite{tkm+98} for the flare observed with
the ASCA satellite on the weak line T Tau V773 Tau.

More remarkable are the high temperatures present in COUP flares.  Up
to recently, flaring events with temperatures reaching up to 100 MK
where considered exceptional. In the COUP sample, on the other hand,
very high temperatures are common. While the ACIS spectral response
does not allow to properly constrain temperatures in excess of 100 MK,
the best fit values for the hottest ML block in the flare exceed 100
MK in about half of the events studied here. The fact that the
temperature decay shows a clear regular pattern even when the best-fit
values are above 100 MK (as for example is the case for COUP 1568),
gives confidence in the fact that the peak observed temperatures are,
in a number of cases, very high, hundreds of MK, even though their
values cannot be accurately determined using ACIS spectra (see
Sect.~\ref{sec:anacoup}).

For the ONC YSO sources studied here, the nominal sizes of the flaring
structures vary from a fraction of the stellar radius (e.g. COUP 7,
90, 141), with absolute sizes of order $10^{11}$ cm, to very large
structures, up to a few times $10^{12}$ cm, i.e., 10--20 stellar
radii. As discussed below, the largest events with a reliable size
estimate (with small uncertainties) typically have $L \simeq 5\,R_*$.

A number of intense flares on active coronal sources have been
analyzed to date with the approach used here. These include a number
of YSOs, both accreting and non-accreting (YLW 15, HD~283572,
LkH$\alpha$,92, V773 Tau, \citealp{fmr2001}), the nearby ZAMS star AB
Dor \citep{mpr+2000}, a number of dMe flare stars (\citealp{frm+2000};
\citealp{fmr+2000}, \citealp{rgp+2004}), and a number of active binary
systems as well as Algol \citep{fs99}.  In all cases, without
exception, the size of the flaring structure derived from the analysis
is at most comparable to the stellar radius, and in most cases it's
smaller. On the other hand the present analysis of intense flares on
YSOs in the ONC results, in a number of cases, in very large loop
sizes, extending to several stellar radii, showing them to be events
of a different nature from the flares analyzed to date.

The strongest evidence for large loop sizes in the COUP sample comes
from well resolved long-lasting events with little sustained heating.
In these cases, one is largely observing the undisturbed thermodynamic
decay of the flaring structure, which, as discussed in
Sect.~\ref{sec:anal}, is directly related to the size of the flaring
structure. One such example is COUP~1343, for which our analysis gives
$\zeta = 1.95 \pm 0.51$, and $F(\zeta) = 1.8$, resulting in a loop
size of $\simeq 1\times10^{12}$ cm. The presence of a small amount of
residual heating is derived from the tail of the light curve, which
decays more slowly, and indeed the detailed simulation of
Sect.~\ref{sec:coup1343} shows that for the first 50 or so ks the
decay of both the emission measure and the temperature is well
reproduced by a freely decaying loop of $L = 1\times10^{12}$ cm. In
fact, if we were to relax our conservative assumption on the peak
temperature of the loop (with the best-fit temperature clipped to 100
MK) and use the nominal best-fit temperature, the resulting loop size
would be even larger.

Other similar cases of well determined flaring structure sizes include
COUP 669, 891, 971 and 1246. In all these cases $F(\zeta) \la 2.5$,
with a well constrained value of $\zeta$, leading to a reliable
determination of $L$ from Eq.~\ref{eq:size}. Both COUP 971 and 1246
have $L \simeq 4 \times 10^{11}$ cm, while COUP 669 and 1343 have $L
\simeq 1\times10^{12}$ cm. With $L = 1.7\times 10^{12}$ cm, COUP 891
is the largest of the well constrained loop structures. In all these
cases, the uncertainties for $L$ resulting from the uncertainty in the
determination of $F(\zeta)$ are of order 20--30\%. Apart from COUP 971
(which, at $L \simeq 1.6\,R_*$, is likely to be a large ``normal''
coronal flare, with both loop footpoints anchored on the stellar
photosphere), the well determined loops have $L \simeq 4$--$5\,R_*$,
which, as discussed in Sect.~\ref{sec:discussion}, is the typical
corotation radius for a low-mass YSO (also likely to be the disk
truncation radius), supporting the hypothesis that these loops may
indeed be structures linking the star and the accretion disk.

For the large events with a well constrained loop size, the peak
densities are all in the range $1$--$8\times 10^{10}$ cm$^{-3}$, and
the equipartition magnetic fields (determined from Eq.~\ref{eq:b}, at
the top of the loop) range from 40 to 250 G.  In the assumption of a
simple dipole geometry (very likely to dominate at large distance from
the stellar surface) the photospheric field would be
\begin{equation}
B_{\rm ph} \simeq (L - R_*)^3 \times B_{\rm eq}
\label{eq:bdip}
\end{equation}

Assuming an average equipartition field of 100 G and a typical loop
length of $5\,R_*$, the corresponding field strength at the stellar
surface would therefore be of order 6 kG, at the top of the range of
the field strengths determined in YSOs by Zeeman splitting
\citep{jv2005}.

In a few flares of Table~\ref{tab:res}, we obtain $\zeta \gg 1.5$
(e.g.\ COUP 332, 454, 848, 976), outside the range of validity of
Eq.~\ref{eq:fzeta}. Probably, in these events the analyzed part of the
decay is too short to cover a significant part of the decay path, and
the slope is not yet well defined.  Cases which result in long flaring
structures in the presence of strong sustained heating are to be
regarded with caution, as the strong heating dominates the decay,
hiding the intrinsic decay of the cooling plasma and potentially
introducing a larger uncertainty in the results. Also, the hyperbolic
form of Eq.~\ref{eq:fzeta} will tend to include lengths much smaller
than the 'nominal' results of Eq.~\ref{eq:size} within the final error
bar for $L$. One such example is COUP~1410, for which $\zeta = 0.45 \pm
0.12$ and $F(\zeta) = 6.1$. While the nominal loop length is $L =
1.1\times 10^{12}$ cm, the uncertainty range resulting from the
uncertainty in $\zeta$ is $L = 10^{11}$--$2 \times 10^{12}$ cm, with a
factor of ten error bar at the lower end.

%%% Add comments on large flare sizes.

A number of individual flaring events of particular interest are
discussed in detail in Appendix~\ref{sec:individ}, where the
individual light curves and the time evolution of the spectral
parameters are also shown.

\subsection{Detailed simulation of the flare on COUP 1343}
\label{sec:coup1343}

Although the method used here has been widely applied in the
literature, and its reliability tested for example in the case of the
eclipsed flare observed by SAX on Algol (\citealp{sf99};
\citealp{fs99}), as discussed in the above Section no previous
analysis based on this approach has resulted in such large loop sizes
(of order $\simeq 0.1$ AU). It is thus legitimate to ask: 1) whether
such gigantic loops can be still described with standard loop models,
closed magnetic structures where plasma is confined and moves and
transports energy along the magnetic field, and 2) whether the
diagnostics we are using (in particular the slope $\zeta$ in the $\log
T$--$\log \sqrt{E\!M}$ plane) are still valid. To check on the
applicability of the method to such extreme regimes, and to validate
its use on the present sample, we have modeled one of the
well-resolved large flares from the COUP sample in detail. We have
used a time-dependent hydrodynamic model of plasma confined inside a
coronal loop, as already done for other more regular stellar flares
(\citealp{rps+88}; \citealp{rgp+2004}).

We have chosen to model the large event on COUP 1343, in which the
peak count rate is approximately $30\,\times$ larger than the
characteristic rate and which appears, from the analysis performed
using the approach described in Sect.~\ref{sec:anacoup}, to have
evidence for little sustained heating. The event lasts some 130 ks,
and both the rise phase and most of the decay are not interrupted by
gaps in the observations.  Additionally, the decay in the $\log
T$--$\log \sqrt{E\!M}$ plane is well defined. The event and its
analysis with the analysis method used here (as described in
Sect.~\ref{sec:anacoup}) is shown in Fig.~\ref{fig:src1343}.

\begin{figure}
\plotone{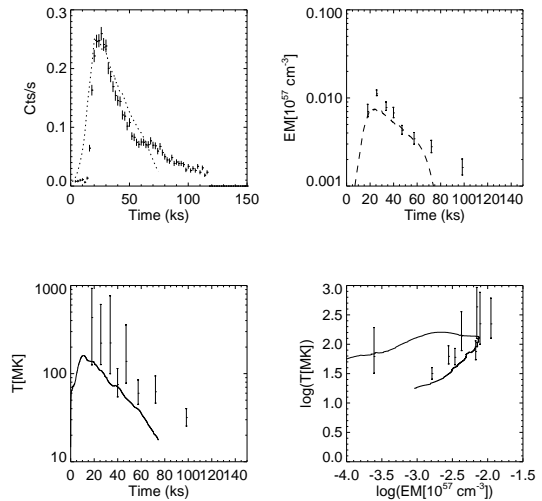}
\caption{The results of the detailed hydrodynamic modeling of the
  large flare on COUP 1343. The top left panel shows the flare's light
  curve (as seen by ACIS), the top right panel the evolution of the
  emission measure, and the bottom left panel the evolution of the
  temperature. The bottom right panel shows the evolution of the flare
  in the $\log T$--$\log \sqrt{E\!M}$ plane. Continuous and dashed
  lines show the evolution of parameters from the hydrodynamic model,
  while the points with the vertical error bars are the values
  determined from the time-resolved spectral analysis of COUP~1343.
  \label{fig:src1343model}}
\end{figure}

To model the event on COUP 1343 in detail, we have assumed the plasma
to be confined in a loop with constant cross-section and half-length
$L = 10^{12}$ cm, symmetric around the loop apex. The flare simulation
is triggered by injecting a heating pulse in the loop which is
initially at a temperature of $\sim 20$ MK (which is the
\refcomm{dominant temperature in the characteristic spectrum as well
as the} temperature observed at the end of flare decay).  We have
chosen to deposit two heat pulses with a Gaussian spatial distribution
of intensity 10 erg cm$^{-3}$ s$^{-1}$ and width $10^{10}$ cm (1/100
of the loop half-length) at a distance of $2 \times 10^{10}$ cm from
the footpoints, i.e.\ very close to the footpoints themselves
(\citealp{rgp+2004}). After 20 ks the heat pulses are switched off
completely, with no residual heating in the decay, as indicated by the
density-temperature diagnostics for this specific flare. We have
computed the evolution of the loop plasma by solving the
time-dependent hydrodynamic equations of mass, momentum and energy
conservation for a compressible plasma confined in the loop
(\citealp{psv+82}; \citealp{bpr+97}), including the relevant physical
effects such as the plasma thermal conduction and radiative losses
(computed assuming solar metal abundances -- at these high
temperatures however the line losses are a minor contributor, so that
the results are robust against differences in the plasma metal
abundance). The gravity component along the loop is computed assuming
a radius $R_* = 3~ R_{\odot}$ and a surface gravity $g_* = 0.1~
g_{\odot}$, typical of low-mass pre-main sequence objects.  The
results of the simulations are shown in Fig.~\ref{fig:src1343model},
together with the observational results for COUP 1343.

The computed evolution of the flare's parameters largely resembles the
evolution computed for other stellar flares (e.g.\ 
\citealp{rgp+2004}), although on larger scales.  The heat pulses make
the loop plasma heat rapidly ($\simeq 1$ hour) to temperatures above
200 MK and the initially denser chromospheric plasma near the loop's
footpoints expands dynamically upwards at speeds above 2000 km/s, to
reach the loop apex on similar timescales (also about 1 hour). After
this first impulsive phase, the temperature does not change much and
the evaporation continues substantially but less dynamically. After
the end of the heat pulse, the plasma begins to cool while still
filling the loop. The density begins to decrease only two hours later.

From the density and temperature distribution of the plasma along the
model loop, we can synthesize the expected plasma X-ray spectrum
filtered through the ACIS spectral response, and the relative spectral
parameters as they would be determined through a spectral fit to the
ACIS data. We have computed the emission in the coronal part of the
loop, i.e.\ above the loop transition region, assuming a hydrogen
column density $N(\rm {H}) = 10^{22}$ cm$^{-2}$ (i.e.\ the value
obtained from the fit to the source's characteristic level spectrum).
The dashed line in Fig.~\ref{fig:src1343model} (top left panel) shows
the ACIS light curve resulting from the simulation, integrated above 1
keV, assuming a loop cross-section radius of $2 \times 10^{10}$ cm,
i.e.\ $\beta = 0.02$.  The smaller $\beta$ than customary ($\beta =
0.1$ being assumed elsewhere) is necessary to provide a good fit to
the flare's decay; the small $\beta$ also results in footpoints still
small relative to the photosphere. The spectra obtained from the loop
modeling have been fit with 1-$T$ model spectra, and the resulting
evolution of the emission measure and of the temperature, as well as
the $\log T$--$\log \sqrt{E\!M}$ diagram are also shown in the figure.
A visual comparison of the model results with the actual light curve
and with the evolution of the measured spectral parameters indicates
that the loop simulation results are in good qualitative agreement
with the data and that, therefore, the data are compatible with the
flare occurring in a single giant flaring structure. The modeling
could be further improved by changing some of the simulation's
parameters; for example the light curve's agreement for $t \ga 50$ ks
could be improved by including a low residual heating active during
the late decay (as done e.g. by \citealp{rgp+2004} in their simulation
of a large flare on Proxima Cen). However, such changes would not
alter the large-scale features of the flaring loop, in particular its
large size, and thus for the purpose of the present work they are not
necessary.

One additional diagnostic provided by detailed simulations such as the
one discussed here is the rise time of the event. This is not
available with the more general analysis employed for the complete
sample of flares discussed in the present paper, which only uses the
decay phase. In the case of COUP 1343 the rise phase is very steep,
with a very peaked light curve. Not all impulsive heating functions
will produce such steep rise; we find that to produce the steep rise
it is necessary to have strongly localized heating deposited very
close to the loop footpoints. More diffuse heating, or heating
localized near the loop apex would produce a smoother light curve.

%Only one of the sources in our sample (source 141) has strong Ca\,{\sc
%  ii} emission, and thus evidence for strong ongoing
%accretion. Interestingly, source 141 

\section{Complex flaring events}
\label{sec:coup450}

While the present paper deals with all the flaring events in the COUP
sample which can be modeled within the framework described by
\cite{rbp+97}, a number of events are present in the sample which do
not show such regular behavior, and which cannot therefore be analyzed
with the same approach. Perhaps the most dramatic one is the flare on
COUP 450, which, coincidentally, was also observed at mm wavelengths.
COUP 450 is associated with a well known radio source (GMR-A,
\citealp{gmr87}) and is identified with star HC~573 from
\cite{hc2000}.

\begin{figure}
\plotone{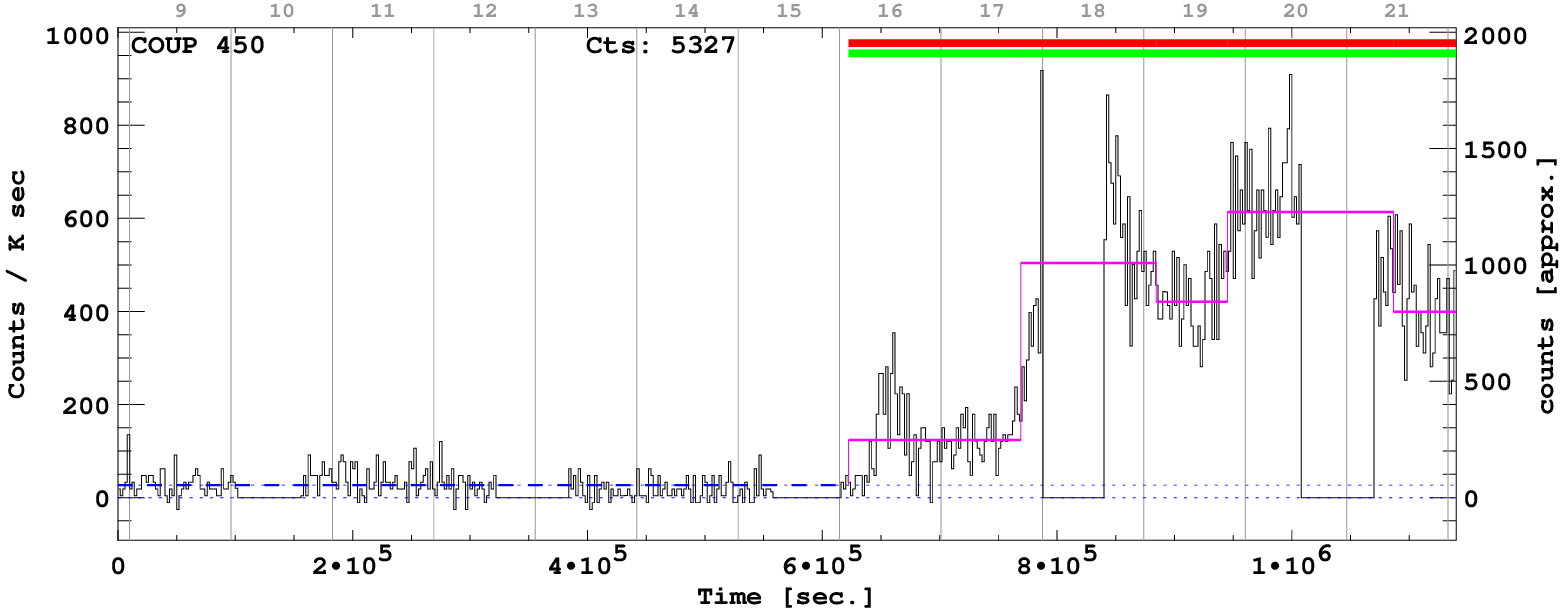}
\plotone{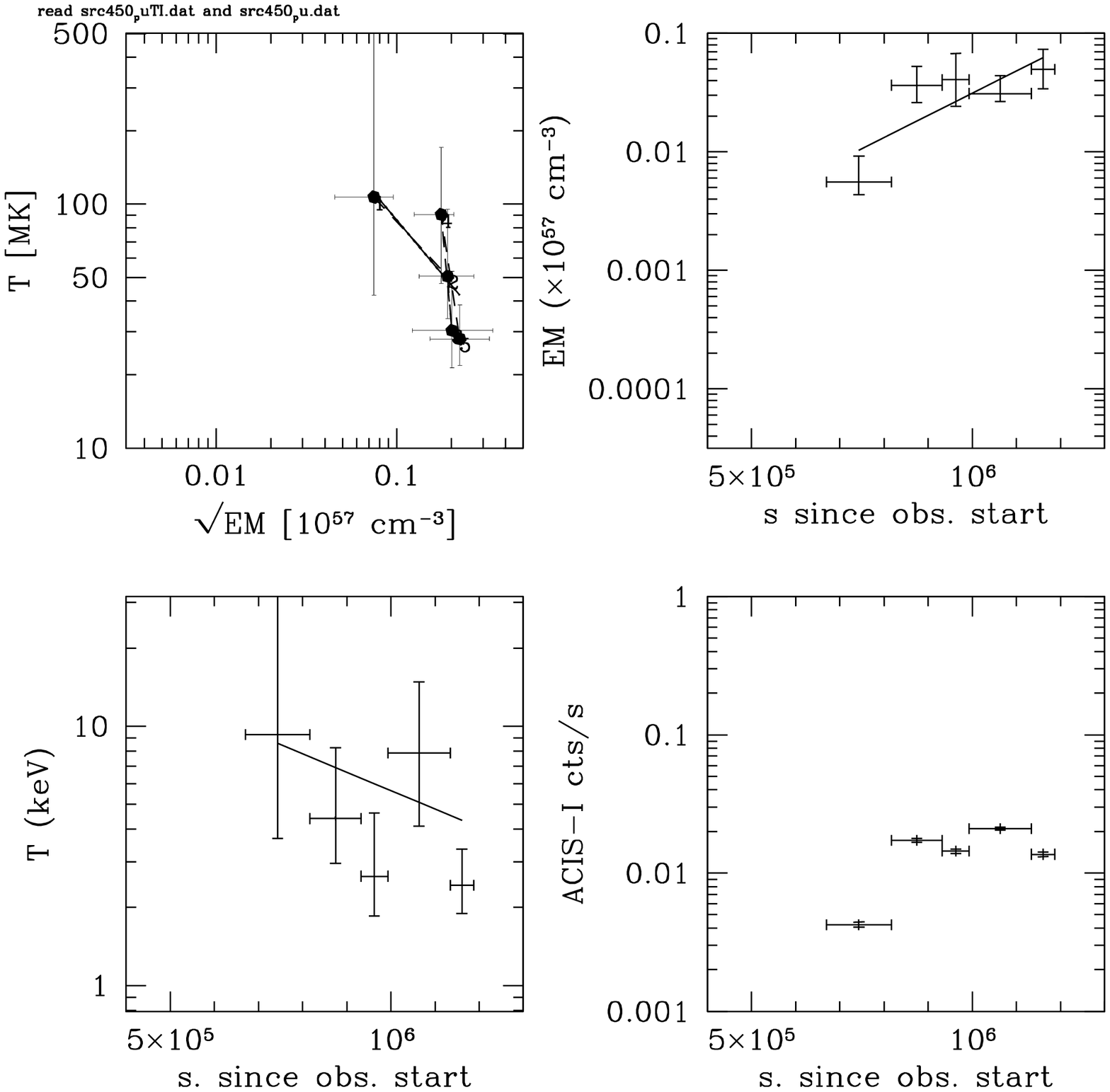}
\caption{Top panel: the COUP light curve of source 450. Bottom panel:
  evolution of the flare. See caption to Fig.~\ref{fig:src1343} for
  details. The orange and green bars indicate the extent of the
  flaring intervals. \label{fig:src450}}
\end{figure}

The X-ray event has been briefly reported by \cite{gfg+2003}, and the
COUP light curve for the event is shown in Fig.~\ref{fig:src450}. As
discussed in detail by \cite{bpl+2003}, the source also flared in the
mm regime, although the mm flare started significantly later (about
two days after the start of the X-ray flare). The mm flare lasted some
20 days, with erratic variability continuing for tens of days
thereafter. The COUP observation only covers the first day or so of
the mm event (which has a much spottier time coverage), and no obvious
correlation between the details of the two light curves is present.
The counterpart to COUP 450 has been studied by \cite{bpl+2003} who
(using IR Keck spectroscopy) identify it as a heavily absorbed K5 T
Tauri star with average photospheric magnetic fields $B \simeq 2-3$ kG
determined from Zeeman-splitted infrared lines.

The COUP 450 event is not exceptional either in terms of its peak
luminosity (with a rise of $\simeq 20\, \times$ in X-ray count rate at
peak) or in terms of its duration (other events with durations of a
few days are present in COUP). It is however one of the most complex
events in terms of its light curve shape, and it lies at the extreme
of complex flares, which lack a clear, well defined decay phase. Such
events cannot therefore be analyzed with the method used here. In this
specific case, the analysis is also hampered by the very strong degree
of pile-up present in the \chandra\ observation, so that while the raw
count rate ensures a 'high-resolution' view of the light curve, most
events are lost to pile-up, and the residual pile-up free events
suffer from very limited statistics. Nevertheless it is possible to
still resolve the light curve in a number of intervals, allowing us to
monitor the evolution of the spectral parameters.

The time evolution of the flare can be seen in the bottom panels of
Fig.~\ref{fig:src450}. The very significant amount of pile-up in the
ACIS observation results in a very small pile-up free count rate. In
the light curve shown in the top panel of Fig.~\ref{fig:src450} the
purple line shows the ML blocks of the pile-up free events: although
the raw count rate approaches 1 cts/sec, the pile-up free ML blocks
have durations of tens of ks, which clearly subsample the event's
evolution. Even so, however, the irregular evolution of the event is
readily visible. No regular decay is present; the emission measure does
not decrease, and the temperature shows an irregular behavior, with a
strong increase in the next to last bin. Whether this is evidence of
strong reheating within the same flaring structure or whether this is
a separate structure undergoing flaring (perhaps induced by the first
event) cannot be inferred from our data, specially because of the
coarse time sampling caused by pile-up.

The lack of a regular evolution for this event does not allow us to
make (using the approach adopted in the present paper) any inference
about the size and characteristics of the flaring region, and thus
about the flaring mechanism and the underlying physics. The clear lack
of a decay does point in this case to a complex heating function, with
a highly time-variable release of energy in the flaring plasma, and --
perhaps -- to the involvement of different structures in the flaring,
rather than to the single, confined plasma structure inferred from the
analysis of the simpler events discussed in Sects.~\ref{sec:anal}
and~\ref{sec:results}.

\section{Discussion}
\label{sec:discussion}

%\textsc{Still missing: data and discussion re: flare luminosity and
%  energetics.} 

%\textsc{To be added: discuss the issue of the loop diameters, which
%  must be thin in comparison w/ the length.}

%\textsc{Emphasize that all sort and lengths of loops are present.}

%%%%%%%%%%

In a significant number of cases the size of the flaring structures
derived here for intense flares in COUP sources is large (several
stellar radii) with absolute sizes up to $10^{12}$ cm.  In a few of
them, the evidence for large loop sizes is strong, because we could
constrain negligible heating to be present during the decay.
Significant heating would have made the diagnosed lengths shorter and
at the same time more uncertain.  Given the high plasma temperatures
at the peak of the flare ($T>100$ MK) the magnetic field necessary to
confine the plasma typically is of a few hundred G.  This scenario is
very different from the one observed in more evolved stars, ZAMS or
MS, in which the same type of analysis of flare decay as performed
here has invariably yielded relatively compact magnetic structures $L
\le R_*$, compatible with scaled-up version of the geometry observed
on the Sun.  Also the limited number of flares previously studied with
the same technique on YSOs \citep{fmr2001} have shown evidence for
compact structures.  This is not in contradiction with the evidence
from the COUP sample, as the YSOs studied by \cite{fmr2001} all are in
the nearest SFRs (thus allowing studies of much less intense flares)
and were the result of shorter observations (and therefore
biased against the long lasting events well represented in COUP).

Such long loops are unlikely to be anchored on the star alone, as the
centrifugal force would rip them open and eject the plasma. The
corotation radius is defined as the distance at which a test gas
particle in keplerian orbit around the central star has the same
angular velocity as the star, 
\begin{equation}
R_{\rm c} = (GM_\ast/\Omega^2_\ast)^{1/3}
\end{equation}

The effective gravitational potential (gravity plus centrifugal) takes
the form of a vanishing gradient at the corotation radius, which is a
saddle point, allowing only for magnetically dominated plasma to be
present above and below the equatorial plane.  The plasma favors
moving inward while inside of the corotation radius, but flowing
outward while outside of the corotation radius.  Unless the magnetic
field is strong enough to provide sufficient tension therefore
enforcing corotation, the field itself would easily be opened up by
plasma flowing downhill in the effective gravitational
potential.

Following \cite{ms87} -- hereafter MS87 -- and assuming the simplest
stellar dipole field, a scaled-up version of the Sun for the stellar
activity (with the magnetic field strength proportional to the
rotational velocity of the star, i.e.\ the $p=1$ case of MS87, so that
$L_{\rm X}\propto B^2$), Table 1 in MS87 gives a list of sizes of the
largest closed loops that can possibly exist for a rotating solar-mass
star under similar condition.  By simply re-scaling with our stellar
parameters, for loops of temperatures on the order of 100 MK or
higher, the largest equatorial extent for closed magnetic loops (in
the notation of MS87, $C$-cusp point, see their Fig.~1), will likely
be in the range of 3--4 $R_\ast$, for a slow rotator (6--8 days, or
$\alpha/\alpha_\odot \sim 4$--$5$ in the notation of MS87, where
$\alpha$ is the rotation rate of the star), and high plasma
temperature ($\zeta_d\propto\alpha$ and $(\zeta_d)_\odot=4$, where
$\zeta_d$ is the ratio of magnetic to thermal energy in the corona).
\refcomm{Similar considerations have recently been applied by
  \cite{jar2004} in the context of 'coronal stripping' in fast
  rotating, supersaturated stars.}

Our flaring sources have typical rotational periods (when they are
known) of a few days, up to a week.  For typical CTTSs periods of 6--8
days, the corresponding corotation radius is about $9\times 10^{11}$
to $1.2\times 10^{12}$ cm, equivalent to $\simeq 5\,R_\ast$, if
$R_\ast = 3\,R_\odot$.  Loop structures larger than this would have
difficulty anchoring on the star alone without very strong magnetic
field tension to counter balance the centrifugal force. What is then
the magnetic geometry which can support coherent fields of hundreds of
G over $10^{12}$ cm? In the case of YSOs one obvious possibility is to
have magnetic flux tubes connecting the stellar photosphere with the
disk, at the co-rotation radius or somewhat outside it. As e.g.\
discussed by \cite{ssg+97}, the twisting of the magnetic field lines
induced by the differential (Keplerian) rotation of the inner disk rim
and the photosphere will twist the flux tubes, resulting in longer
loops and presumably in the stressed field configurations which drive
the flaring.  A good candidate among COUP flaring sources for
star-disk flaring structures is COUP 1083, a moderate mass star with
$P_{\rm rot} = 5.9$ d and with a flare implying $L = 2.4 \times
10^{12}\,{\rm cm} = 34\,R_\odot$, in the presence of moderate
sustained heating, with an uncertainty range $1.9$--$3.0 \times
10^{12}\,{\rm cm}$.  Even assuming a very young YSO, with a large
radius $R = 3\,R_\odot$,the loop length is $L \simeq 10\, R_*$. Given
the above consideration on stability of large magnetic loops in
rotating stars, it is unlikely that this type of loops can be anchored
on the star alone.

%% Reference to Isobe?

Magnetic flux tubes extending from the inner rim of the disk to the
stellar photosphere provide a natural location for the intense flares
found here on a number of sources. An additional open question is what
heating mechanism can heat the plasma in these long loop to such high
temperatures. While the general coronal heating mechanism is still not
fully established, the most modern magnetohydrodynamic simulations
(e.g.\ \citealp{pgn+2004} and \citealp{ssa+2004}) point to the heating
being due to shuffling of the loop's footpoint, a natural result of
convective motions in the photosphere. The photospheric convection is
certainly present in accreting YSOs, and shearing of the disk would
provide an equivalent, natural source of loop footpoint shuffling on
the disk side.

It should be stressed that the COUP flaring sources analyzed in the
present paper present a large variety of flaring structures. While the
long flaring loops which likely connect the star with the accretion
disk are well represented in the sample, so are also the more compact
structures which have very similar characteristics with the ones
observed on more evolved stars, ZAMS or MS. Thus, it appears from the
present analysis that two types of (flaring) coronal structures can
coexist on YSOs: one which appears similar to the one present in
older, more evolved stars (and thus likely to be a scaled-up version
of the solar corona) and one made of very extended magnetic
structures, which is instead peculiar to YSOs, being dependent on the
presence of disks on which to anchor one footpoint of the very long
loops. While unfortunately none of the COUP sources has more than one
flare sufficiently bright to be analyzed with the present technique,
some sources (notably COUP 597 and COUP 891, both discussed in
Appendix~\ref{sec:individ}) show, in addition to the long events
linked with long loops, some short events, too short to be analyzed,
but likely originating in compact loops (given the small $\tau_{\rm
  lc}$), so that it would appear that both types of coronae could
exist, at the same time, in a given YSO.

\cite{whf+2005} have presented an analysis of flare characteristics
and frequency in ``young suns'' in the COUP database.  Only two of our
objects are in common with the sample of \cite{whf+2005}, namely COUP
223 and 262. While no statistical conclusions can be drawn from only
two objects, the flare on COUP 262 results in a long coronal structure
($L \ge 3.6~R_*$, see discussion in Sect.~\ref{sec:coup262}), showing
that these extended magnetic structures were indeed present in the
young Sun.

Very recently, \cite{lmr+2005} have obtained evidence for the presence
of long magnetic structures, in T Tau S: their radio VLBI observations
show that, in addition to a compact source which they identify as the
star's magnetosphere, a fainter streak of radio emission is present,
extending to a distance of about $30 ~R_\odot$ from the star, which
they state ``may result from reconnection flares at the star-disk
interface''. The size of the radio structure ($\ge 10 R_*$, T Tau S
having a radius $R \simeq 3 R_\odot$) is very similar to the loop
lengths we have derived here from the analysis of the large X-ray
flares, providing independent support for the presence of such long,
coherent magnetic structures in YSOs.

\subsection{The role of disks and of accretion}

To investigate the role of disks in the presence of long magnetic
structures one would ideally like to have a clear separation of the
sample in stars with disk and stars without. Unfortunately, the
\ffcomm{available diagnostics do not allow us to do this in an
  unambiguous way. The most complete investigation on the presence of
  disks in the ONC is the work of \cite{hsc+98}, who have compared the
  results of a number of indicators for the presence of disks in the
  ONC, including IR excesses in a number of color bands and Ca\,{\sc
    ii} emission. Emission in Ca\,{\sc ii} should indicate the
  presence of active accretion, and thus will miss ``quiescent''
  disks.  NIR excesses are often used as an indicator of the presence
  of disks, and \cite{hsc+98} have quantitatively modeled the expected
  excesses in a number of bands, using models very similar to the ones
  of \cite{mch97} and converging on similar conclusions. As discussed
  in detail in \cite{hsc+98} the IR excess is expected to be produced
  by the disk material being heated both by the stellar radiation and
  by the friction induced by accretion. In practice, its magnitude
  depends on a number of factors, such as the relative importance of
  accretion vs.\ the size of the disk's inner hole, the relative
  contrast between the disk emission and the photosphere and the
  system inclination. Disks with low accretion rate and with a large
  inner hole, generating only a far-IR excess could be difficult to
  see in the NIR (see also the discussion in \citealp{mch97}).}
  
\ffcomm{\cite{hsc+98} conclude that the excess $\Delta(I-K)$
  (determined by taking into account the star's spectral type and thus
  effective temperature and intrinsic colors and its reddening) is the
  most effective single indicator of the presence of disks;
  measurements in in the $L$ bands would have be even better but were
  not available to \cite{hsc+98} except for few stars. However, while
  $\Delta(I-K)$ clearly is an effective indicator on a statistical
  basis, its value for determining whether individual stars have a
  disk is much less clear.  The distributions shown e.g.\ in Fig.~10
  of \cite{hsc+98} have strong tails to negative values (as far as
  $\Delta(I-K) \simeq -1$) and a significant number of stars with
  Ca\,{\sc ii} emission (thus, accreting systems) do not have
  $\Delta(I-K)$ excesses. Even more significantly, as discussed in
  Sect.~7.3 of \cite{hsc+98}, a number of sources with disks visible
  in the HST images (``proplyds'', from the \citealp{ow96} sample) do
  not have IR excesses, clearly showing that some sources with disks
  escape detection through IR excesses.}

\begin{figure}
\plotone{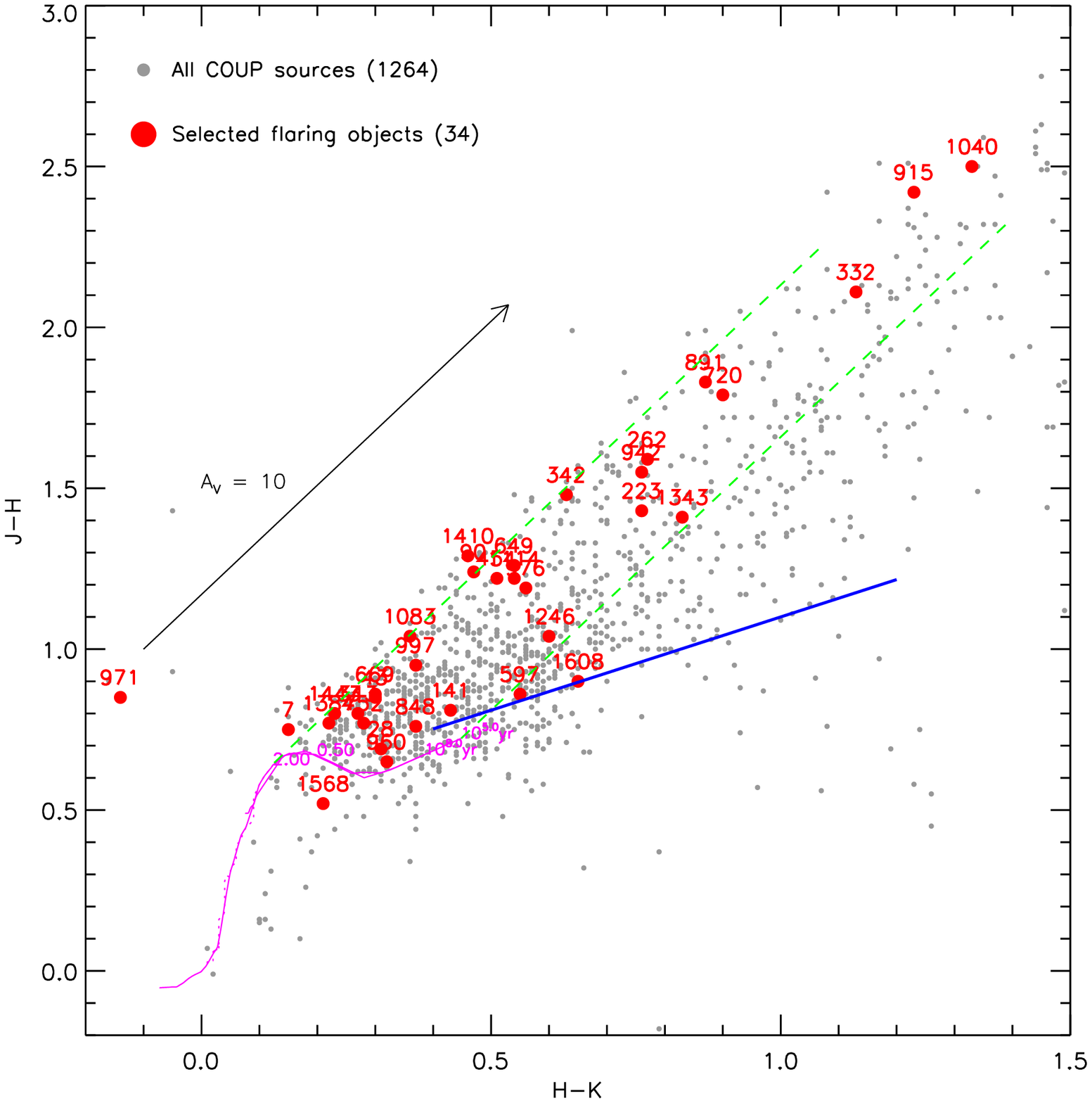} \plotone{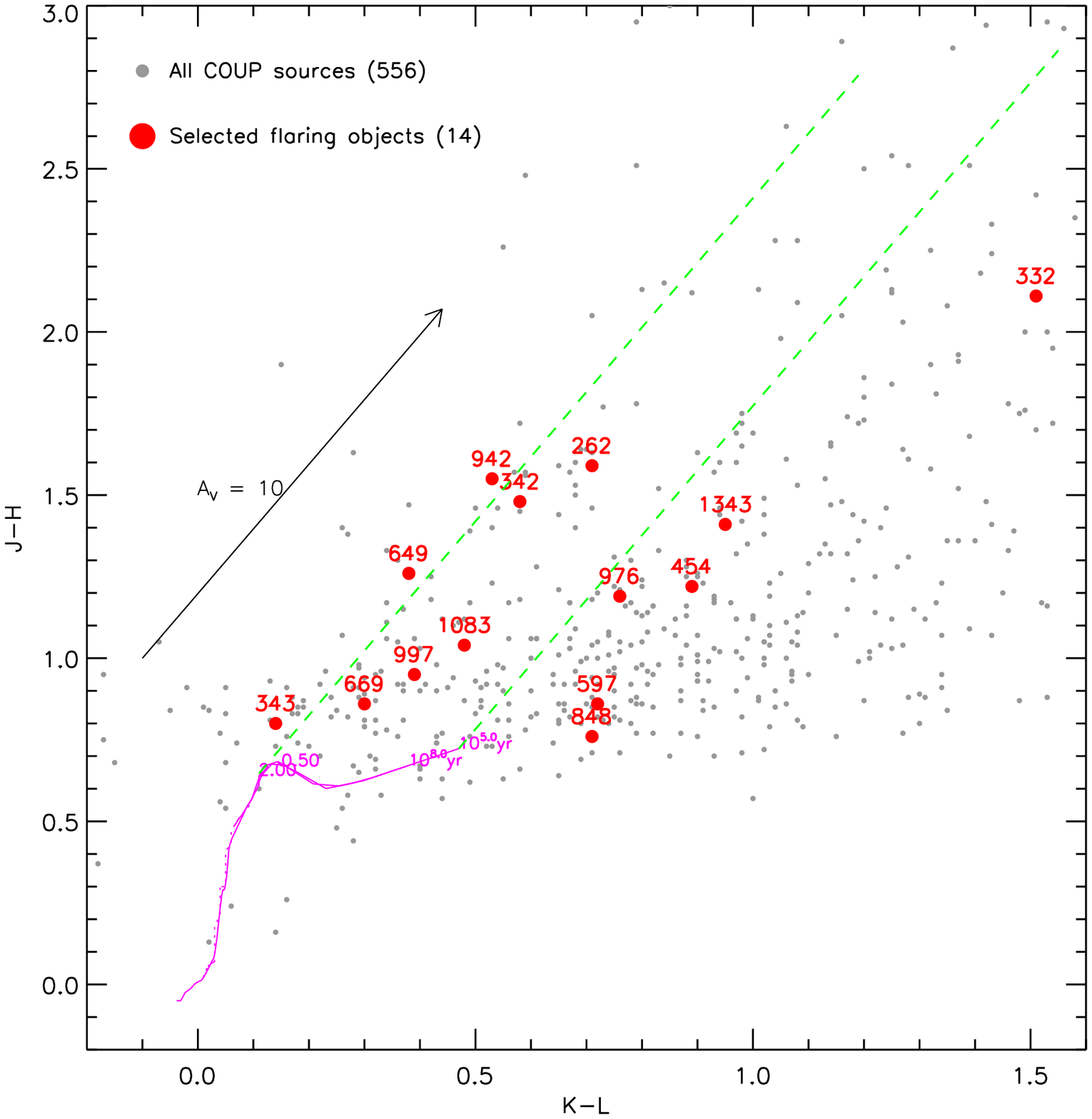}
\caption{Top panel: the $J-H$ vs.\  $H-K$ NIR color-color diagram for
  all COUP sources for which the information is available (grey
  points). Individual sources analyzed in the present paper are marked
  as red points, with a label showing the COUP number. The purple
  lines are two isochrones, for $10^{5}$ and $10^{8}$ yr (nearly
  indistinguishable from each other), while the blue line represents
  the locus of CTTS. The two green dashed lines delimit the region of
  the plane which can be explained as reddened diskless stars; stars
  to the right of this region are showing a significant NIR excess
  indicative of the presence of a disk. Bottom panel: same plot for
  $J-H$ vs.\ $K-L$ colors. This diagram is a more sensitive probe of
  the presence of cooler disks, although many of them will peak in the
  far IR and thus will also not be visible in the $L$ band.
  \label{fig:ircc}}
\end{figure}

\ffcomm{In the present paper, we have tried to analyze whether a
  relationship exists between the detection of long flaring structures
  and the presence of disks from the available indicators, bearing
  however in mind the above caveat. In Table~\ref{tab:res} we report,
  for all the COUP sources analyzed in our sample, the available data
  from the compilation of \cite{gf+2005} -- the original source for
  most of the IR data is \cite{hsc+98} -- regarding Ca\,{\sc ii}
  emission and excess in $I-K$ colors. We also report whether the
  sources have an excess in the NIR $JHKL$ bands from inspection of
  Fig.~\ref{fig:ircc}, although (quoting \citealp{hsc+98} literally)
  ``calculating disk fractions simply by counting the relative number
  of stars outside of and inside of reddening vectors in observed
  color-color diagrams is clearly naive''. }

Fig.~\ref{fig:ircc} plots the IR color-color diagrams for all COUP
sources for which $JHKL$ NIR photometry is available (from
\citealp{gf+2005}), with the subsample studied in the present paper
singled out. The slanted blue line in Fig.~\ref{fig:ircc} is the
observational mean locus of CTTSs in Taurus derived by \cite{mch97}.
All sources lying to the right of the reddened photospheric locus in
Fig.~\ref{fig:ircc} (bounded by the green dashed lines) have NIR
colors compatible (bearing in mind the above caveat) with the presence
of disks similar to the ones present in Taurus YSOs. For objects whose
NIR colors place them in the reddened photospheric region the
situation is however much less clear \ffcomm{as they could still have
  disks, perhaps with low accretion and large interior holes,
  generating negligible IR 
  excesses.  Indeed, as discussed by \cite{mch97} -- see their Fig.~5
  -- the most important parameter driving the NIR excess is the
  accretion rate, and for a given color excess there is a minimum
  accretion rate, thus disk heating rate, needed to generate a NIR
  excess, independent of central hole size and of the disk
  inclination.  Also, large holes ($R_{\rm hole} \gg R_*$) will
  produce no NIR excess in the $JHKL$ bands, with any excess shifted
  redward of $3\,\mu$m.}

From Fig.~\ref{fig:ircc} it is evident that none of the COUP sources
with strong flares has a significant $H-K$ excess (with the exception
of COUP 1608 and, marginally, COUP 597), while 6 sources (out of
13 for which $L$ photometry is available) show a $K-L$ band excess.
Of these 6 (COUPs 332, 454, 597, 848, 976, 1343), 5 have no $H-K$
band excess, while COUP 597 is the only one possibly displaying
excesses in both $H-K$ and $K-L$. In the framework of the \cite{mch97}
models, COUP 597 is the only one whose $JHKL$ colors would be
compatible with the presence of an accreting disk with a moderate-size
central hole, while the $JHKL$ colors of COUPs 332, 454, 848, 976
and 1343 (with $K-L$ excess but no $H-K$ excess) would be compatible
with the presence of a central disk with a larger size central hole
(few stellar radii) and low accretion rate. The remaining sources in
our flaring sample have no measurable excess in these bands,
\ffcomm{which would be compatible either with their being diskless
  sources or their having disks characterized by a very low accretion
  rate and with large inner holes.}

\refcomm{An additional complexity comes from the fact that star-disk
magnetic coupling requires a gaseous disk at the corotation radius,
while NIR excesses are only sensitive to the dust content of the
disk. But, given sufficient heating, the dust might be sublimated
relatively near the star leaving behind the needed gas disk without an
NIR excess. As discussed by \cite{mch+2003}, for sufficiently large
accretion rates, the inner rim lies beyond the corotation radius, so
that pure gaseous disks must extend inside the dust rim.}

\ffcomm{The situation is however clearly more complex, as evident from
  the fact that some sources which fall, in the color-color diagrams
  of Fig.~\ref{fig:ircc}, in the region of normal reddenened
  photospheres with no disks have significant $\Delta(I-K)$ excesses.
  Unfortunately the overlap between the two samples is not very large
  (13 out of 32 sources have $JHKL$ colors, 19 out of 32 have
  $\Delta(I-K)$ determinations, and only 6 have both), but all 6
  sources for which both indicators are available have a $\Delta(I-K)$
  excess, while 3 of them fall in the region of reddened normal
  photospheres in Fig.~\ref{fig:ircc}. In the whole sample of 1616
  COUP sources for which optical data are available, 472 have a
  $\Delta(I-K)$ determination; of these, 388 have $\Delta(I-K) > 0.02$
  (82\%), compatible with an excess due to circumstellar material,
  while 71 have $\Delta(I-K) < 0.02$. In the present sample,
  $\Delta(I-K) > 0.02$ for 30 out of 32 sources (94\%), a comparable
  number as the complete sample given the low statistics.}

\ffcomm{Further evidence that disks may be present also when no excess
  in $JHKL$ NIR colors is observed comes from the COUP data directly.}
\cite{tfg+2004} have recently searched the COUP databases for sources
showing significant Fe 6.4 keV fluorescent emission, of which they
found 7 cases. In all cases, fluorescent emission was detected during
intense flares, a fact easily explained by the need for a sufficient
number of hard ($E>7.11$ keV) photons to excite the fluorescence. Two
of the \cite{tfg+2004} sources are also in our large flare sample,
namely COUP 649 and 1040.  The remaining 5 fluorescing sources have
too low statistics (either because of the short duration of the flare
or because of the low peak count rate) for our analysis. Flares are
however not a necessary condition for fluorescent emission, as shown
by the detection of 6.4 keV fluorescence in the $\rho$ Oph YSO Elias
29 in quiescence by \cite{fms+2005}.  At the same time, the large
equivalent widths of the 6.4 keV line observed in the COUP sample (as
well as in Elias 29) are compatible only (as discussed by
\citealp{fms+2005} and \citealp{tfg+2004}) with the fluorescence
originating in a centrally illuminated disk observed face on. A disk
illuminated from above, or other topologies (such as fluorescence from
the photosphere) would result in a much lower equivalent width, likely
too low to be detected on COUP sources. Yet most of the
\cite{tfg+2004} sources (and in particular COUP 649 and 1040), for
which a disk is certainly present, given the observed X-ray
fluorescence, show no $JHKL$ NIR excess, although some of them (e.g.\ 
COUP 649) do show a significant excess in $\Delta(I-K)$.  Again,
$JHKL$ NIR excesses appear to be, in the ONC, a sufficient condition
for a disk to be present, but not a necessary one, \ffcomm{with
  $\Delta(I-K)$ providing a more sensitive indicator}.

\ffcomm{Given the biases affecting the samples and the problems
  affecting disk diagnostics, we refrain from formal correlation
  analysis on e.g.\ flaring loop size and the presence of disk,
  limiting the analysis to qualitative considerations. Using
  $\Delta(I-K)$ as an indicator, the only two sources with
  $\Delta(I-K) < 0$ in the present sample (which we take to imply that
  disks are most likely not present) both have flares confined to
  compact magnetic structures (COUP~7 with $L = 0.1\,R_*$ and COUP~960
  with $L = 0.3\,R_*$), while both compact and large magnetic
  structures are present in the sources with $\Delta(I-K) > 0$.}

%It is however suggestive that COUP 597 (the only one in the sample
%with a disk with a NIR excess which, according to the \cite{mch97}
%models, indicates significant accretion and a moderate size central
%hole), shows a flare compatible with a modestly sized loop ($L \le
%7.5\times 10^{11}$ cm, with a most likely value $L \simeq 2.2 \times
%10^{11} {\rm cm} \simeq 1.6\,R_*$), compatible with a coronal loop
%anchored to the star's photosphere at both ends. 

\ffcomm{The 6 sources with significant $K-L$ excess comprise flares
  spanning both large loops and more compact structures (comparable to
  the stellar size): large structures appear present on COUP~332
  (although the determination of loop size suffers from a large
  uncertainty), COUP~454, with $L\simeq 3\times 10^{12}$ cm, COUP~848,
  with $L\simeq 2\times 10^{12}$ cm, and COUP~1343, with $L\simeq
  10^{12}$ cm. On the other hand COUP~597 has $L< 6 \times 10^{11}$ cm
  and COUP~976 has $L< 8 \times 10^{11}$ cm. Large loops are also
  present in sources with no $K-L$ excess, such as COUP 669 ($L\simeq
  10^{12}$) and COUP 1083 ($L\simeq 2\times 10^{12}$ cm), so that no
  clear correlation appears present between the presence of large
  loops and $K-L$ excess. With one single exception (COUP 1608) none
  of the sources in the present sample show significant $H-K$
  excesses. }

Only one strongly accreting source (COUP 141, with Ca\,{\sc ii}
equivalent width $-17.8$ \AA) is present in our sample of large
flares; two sources (COUP 1114 and COUP 1608) show very modest
emission (and presumably accretion rates), and all other sources show
(when available) Ca\,{\sc ii} in absorption or filled in and thus
little or no ongoing accretion. The statistical significance of this
is difficult to assess given the biases of our sample (discussed in
Sect.~\ref{sec:sample}), and the fact \citep{pre+2005} that accreting
sources in the COUP sample are statistically intrinsically fainter
than the non-accreting sample (as well as more absorbed), introducing
a small bias against strong flares from accreting sources being
present in our sample.  Interestingly, COUP 141 is one of the cases of
compact loops, with $L \le R_*$, so that in this case no
disk-photosphere magnetic structures need to be postulated, and the
flare is likely to be a normal coronal event, similar to the ones
observed in more evolved stars. COUP 1114 and COUP 1608 both only
result in upper limits to the size of the flaring structure.

\ffcomm{In the complete COUP sample, out of 1616 sources for which
  optical data are available, Ca\,{\sc ii} data are available for 537
  sources. Of these, 198 have Ca\,{\sc ii} in absorption, 189 have
  Ca\,{\sc ii} in emission and 150 have filled lines (equivalent width
  of the Ca\,{\sc ii} lines reported as ``0.0'' in \citealp{gf+2005}).
  Among the sources with Ca\,{\sc ii} determination the fraction of
  sources in emission is then 35\%, while in the flaring sample
  discussed in the present paper only 3 sources out of 20 for which
  the data are available have Ca\,{\sc ii} in emission, i.e.\ 15\%.
  Given the small numbers, and the biases present in the various
  samples, it's difficult to assess the significance of the
  difference.} If the lack of long flaring structures in accreting
YSOs is however real, given that long magnetic structures connecting
the disk to the star are certainly present, it implies that active
accretion inhibits the formation of strong flares in the extended
magnetic structures connecting the disk to the star.  Long and intense
flares (and thus very hot plasma) appear to be present in
non-accreting structures connecting the disk to the star.

\section{Conclusions}
\label{sec:concl}

In the complete COUP sample of YSOs in the ONC, 32 flares have
sufficient statistics (in terms of flare duration and photon count
rate) and regular evolution to grant a detailed analysis of their
decay. The type of flare decay analysis employed here allow us to
derive the size of the flaring loops in the presence of heating
extending into the decay phase, and, under simple assumptions,
estimates of the plasma density and confining magnetic field.

The most notable result from our analysis is the strong evidence for
very large flaring structures in these stars. The magnetic structures
confining the plasma in a number of cases are much larger than the
stars themselves. Among the events analyzed, a large fraction have
very high peak temperatures ($T > 100$ MK), and some are very
long-lasting, with the longest flares extending to up to a week in
duration. The peak X-ray luminosity for the most intense flare reaches
$8 \times 10^{32}$ erg s$^{-1}$, comparable to other very bright
flares observed in YSOs.

Our sample is limited to the brightest 1\% of COUP flares, and the
results may not be representative of the average magnetic reconnection
event in Orion stars.  Nevertheless, the present results show that
very extended magnetic structures confining hot flaring plasma exist
in YSOs.  Structures of comparable sizes have never been seen in more
evolved stars, and, given the short rotational periods of many of the
flaring COUP stars, would not be stable if anchored onto the
photosphere with both footpoints. These long magnetic structures are
in the present paper interpreted as linking the stellar photosphere
with the inner rim of the circumstellar disk.

\ffcomm{The available indicators of the presence of disks do not allow
for a formal analysis of whether large flaring loops are linked to the
presence of disks; in particular stars without NIR excesses are not
necessarily diskless, as they may e.g.\ have disks with little
dust. However in a limited number of sources the NIR data clearly
point to disks being present, either as $K-L$ excess in color-color
diagrams or as strong excesses above the de-reddened photospheric
colors ($\Delta(I-K) > 0.4$). Both small flaring structures (likely
anchored on the photosphere only) and long flaring structures, with
sizes which would extend well into the disk (assuming the disk to be
truncated at the co-rotation radius) are present in these stars.  Both
types of structures (loops anchored on the star only and loops
connecting the star to the disk) are then perhaps likely to coexist in
YSOs with disks. }

\ffcomm{For the two sources in our sample without any $\Delta(I-K)$
excess, and for the star with a strong Ca\,{\sc ii} accretion
signature, the flaring structures are relatively compact.  This
suggests that inner disks are needed for reconnection in star-disk
magnetic loops and that heating to X-ray temperatures is inhibited by
mass loading associated with YSO accretion. More statistics are
clearly needed to corroborate the present conclusions, and hopefully
future, already approved long X-ray observations of other star-forming
regions will be able to supply the needed additional evidence.}

The data presented in the present paper thus constitute the first
direct evidence for the presence of very long magnetic structures in
YSOs, and, by inference, of magnetic structures linking the stellar
photosphere with the circumstellar disk. Such structures are
postulated by magnetospheric models of YSO accretion, but had not been
directly detected to date.

\begin{acknowledgements}
  
  We would like to thank Frank Shu (National Tsing Hua Univ.) and
  Ronald Taam (Northwestern Univ.)  for the thoughtful and stimulating
  discussions, and an anonymous referee for the careful reading of the
  original manuscript.  COUP is supported by \emph{Chandra} guest
  observer grant SAO GO3-4009A (E.\,D.  Feigelson, PI).  EDF is also
  supported by NASA contract NAS8-38252.  E. F., F.  R., G. M. and
  S. S. acknowledge financial support from the {\em Ministero
  dell'Istruzione dell'Universit\`a e della Ricerca}.

\end{acknowledgements}

\section*{Appendix A: Notes on selected individual events}
\label{sec:individ}
\renewcommand{\thesection}{A}

The present Appendix presents a detailed discussion (together with the
relevant light curves and diagrams from the analysis) for a number of
individually selected COUP sources. Its purpose is both to discuss in
detail some peculiar or particularly representative event and to show
the variety of flare shapes and characteristics present among the ONC
YSOs. 

%\textsc{To fill up and completed for all stars which deserve an
%  individual discussion.}

\subsection{COUP 28}

\begin{figure}
%\epsscale{0.5}
\plotone{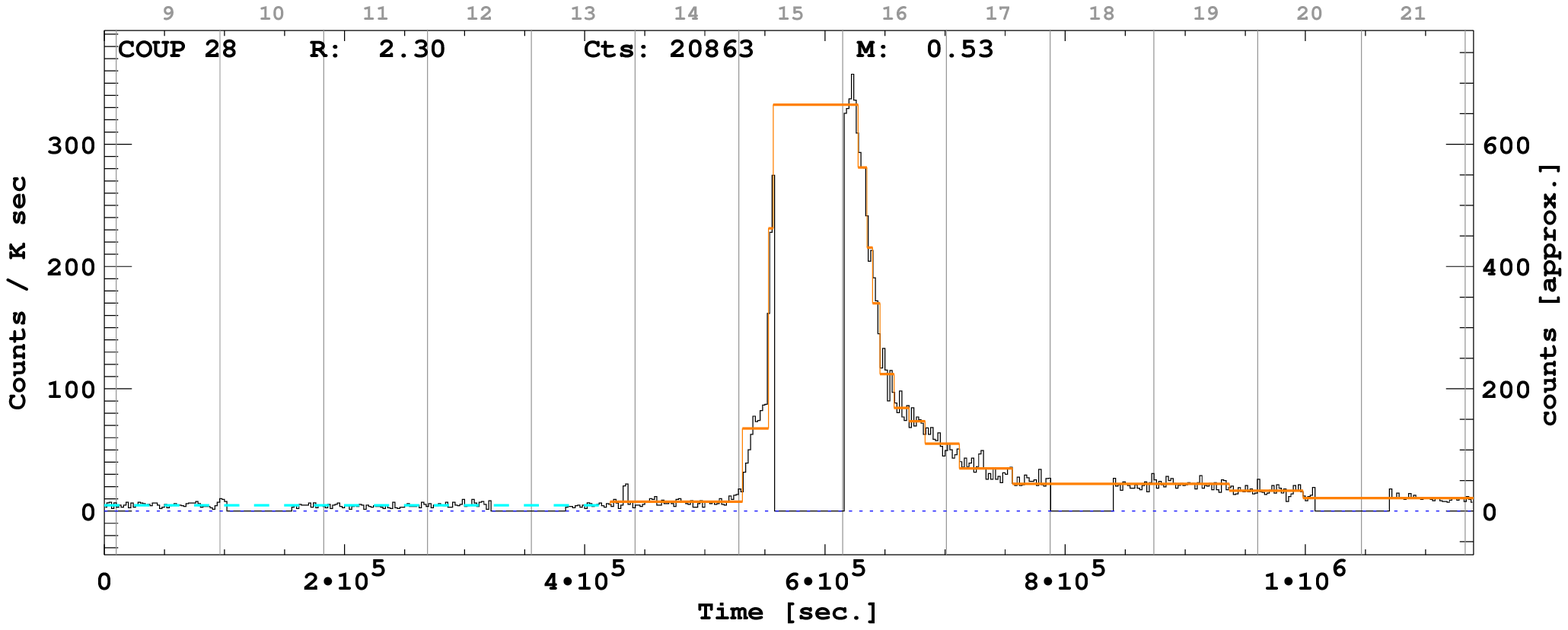}
\plotone{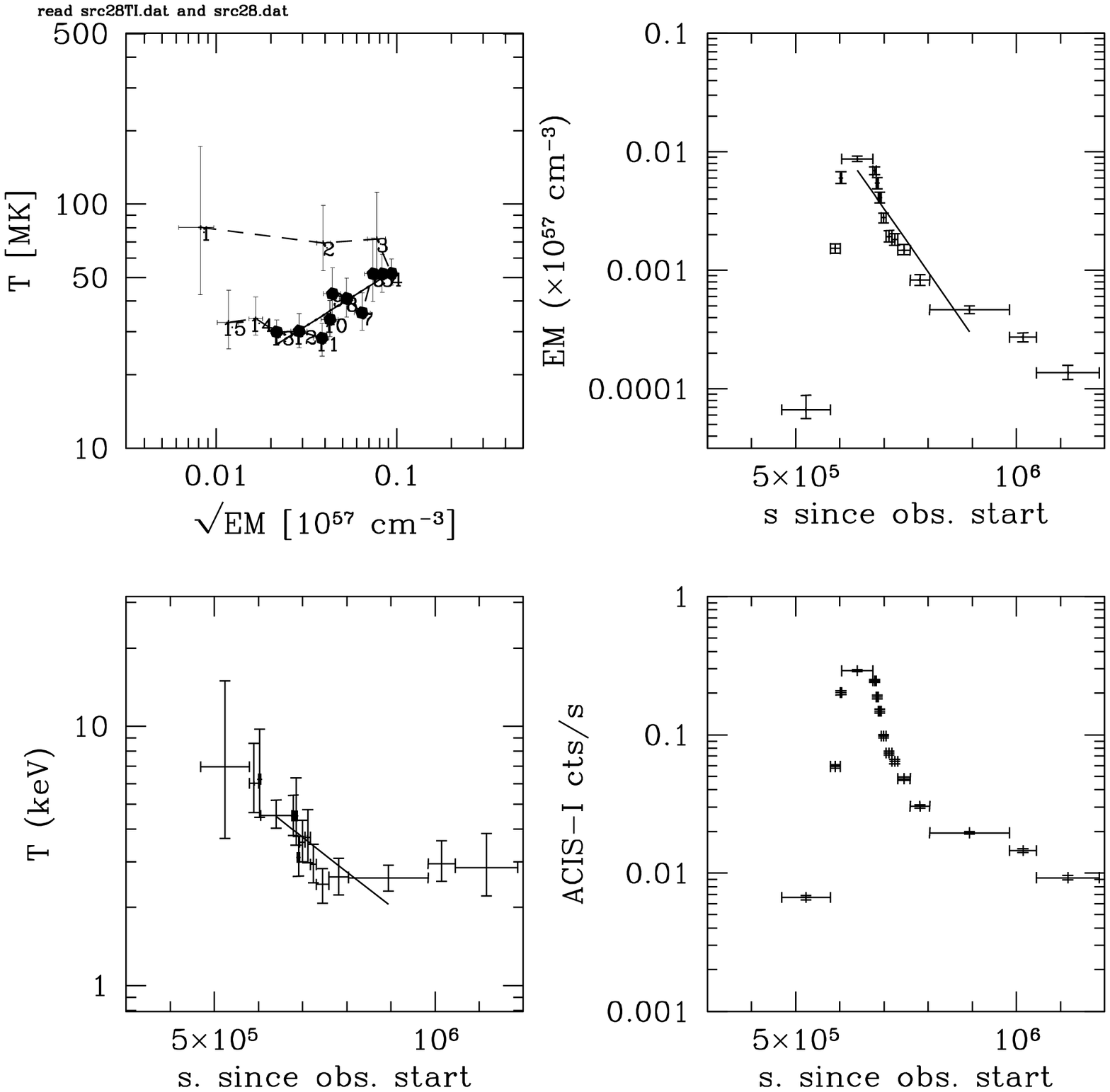}
\caption{Top panel: the light curve of COUP 28. Bottom panels:
  evolution of the flare (see Fig.~\ref{fig:src1343} for
  details).\label{fig:src28}}
\end{figure}

COUP 28 is a $0.53\,M_\odot$, $2.3\,R_\odot$ star of spectral type M0,
showing a NIR excess ($\Delta(I-K) = 0.30$) and no evidence for active
accretion.  Its rotational period is 4.4\,d. It displays one of the
longest flares (Fig.~\ref{fig:src28}) of the COUP sample, with the
total duration of the event extending over more than a
week. Unfortunately an observation gap has blocked much of the flare
rise, which however appears to be slow and extending over more than
one day. The peak seems to be observed, and this allows a reliable
analysis of the decay phase. The sources is at a large off-axis angle, so
that no pile-up is present even if the flare at peak reaches a high
count rate. The flare's peak count rate is almost $100\, \times$ the
source characteristic level. The peak temperature, at $\simeq 80$ MK,
is a moderate one for COUP flaring sources (although it is very high
by the standard of flares observed in older stars), and is still well
determined on ACIS spectra. The $\log T$ vs.\ $\log \sqrt{E\!M}$
diagram shows, as typical in both solar and stellar flares, the
temperature peaking well before the $E\!M$. The excellent stastistics
allows us to follow the flare's evolution in detail, and the temperature
decay shows evidence for reheating (in block 9, Fig.~\ref{fig:src28}),
as also reflected in the change in slope of the light curve decay,
which is not a simple exponential but rather shows evidence for two
different time scales.  In this case, fitting an average decay to both
the light curve and the slope in the $\log T$ vs.\ $\log \sqrt{E\!M}$
diagram still provides a good estimate of the parameters of the
flaring region \citep{rgp+2004}. At $L = 5.5\times 10^{11}~ {\rm cm} =
1.9\,R_*$ the flare is a relatively compact one among the ones
observed in COUP.

\subsection{COUP 43}

\begin{figure}
\plotone{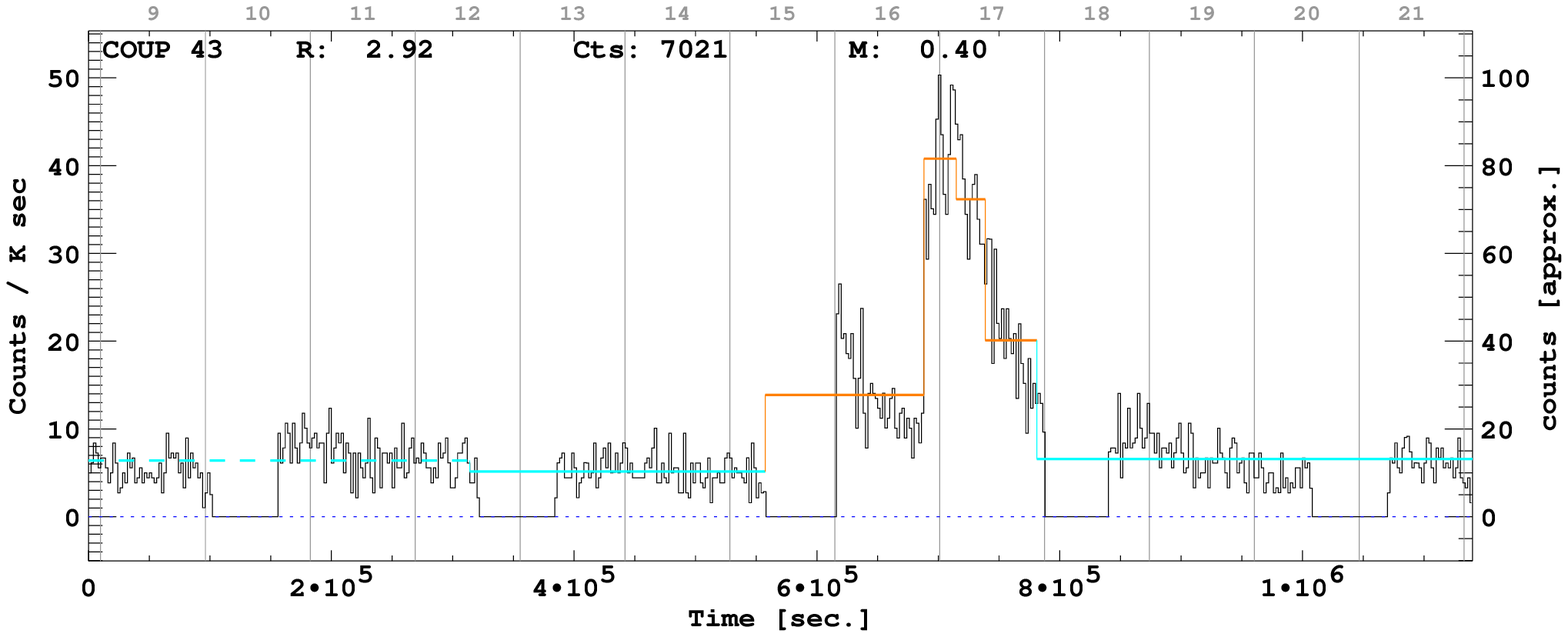}
\plotone{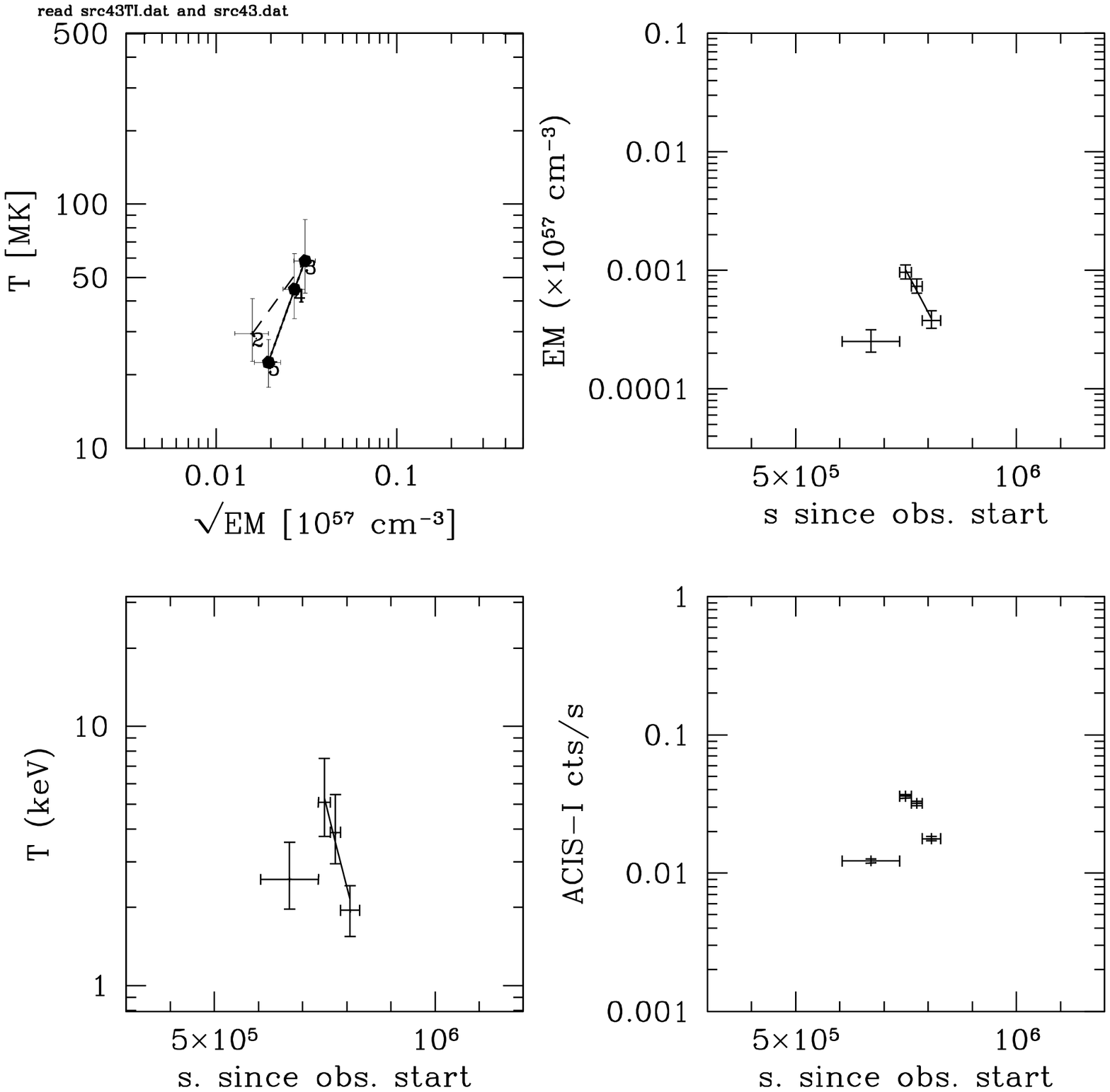}
\caption{Flare evolution of COUP 43.\label{fig:src43}}
\end{figure}

COUP 43 is a known SB2 binary; \refcomm{the mass and radius computed
by \cite{hil97} for the unresolved source are $0.40\,M_\odot$ and
$2.9\,R_\odot$}. Its spectral type is M1, and it shows no evidence of
active accretion; an $I-K$ excess ($\Delta(I-K) = 0.50$) is present.
The flare (Fig.~\ref{fig:src43}) is a double event, with the second
flare beginning while the first one is still in its decay. The rise
phase of the first event is lost in an observation gap, and therefore
only the second event, with a duration of about two days, and which
shows an impulsive rise, has been analyzed. While the limited
statistics only allow us to subdivide the decay in three intervals, the
steep $\zeta$ implies very limited sustained heating, and thus a long
structure ($L = 1.1 \times 10^{12}~ {\rm cm} = 5.5\,R_*$) driven by the
slow decay ($\tau_{\rm lc} = 62$ ks).  The two flares are sufficiently
well separated, so that the first (incomplete) event does not
significantly affect the spectral analysis of the second event.
However, the lack of coverage of the latest phases of the decay imply
caution in the interpretation of the analysis results, as a flattening
of the temperature decay could imply a final shallower $\zeta$ and
thus a smaller loop size.

\subsection{COUP 262}
\label{sec:coup262}

\begin{figure}
\plotone{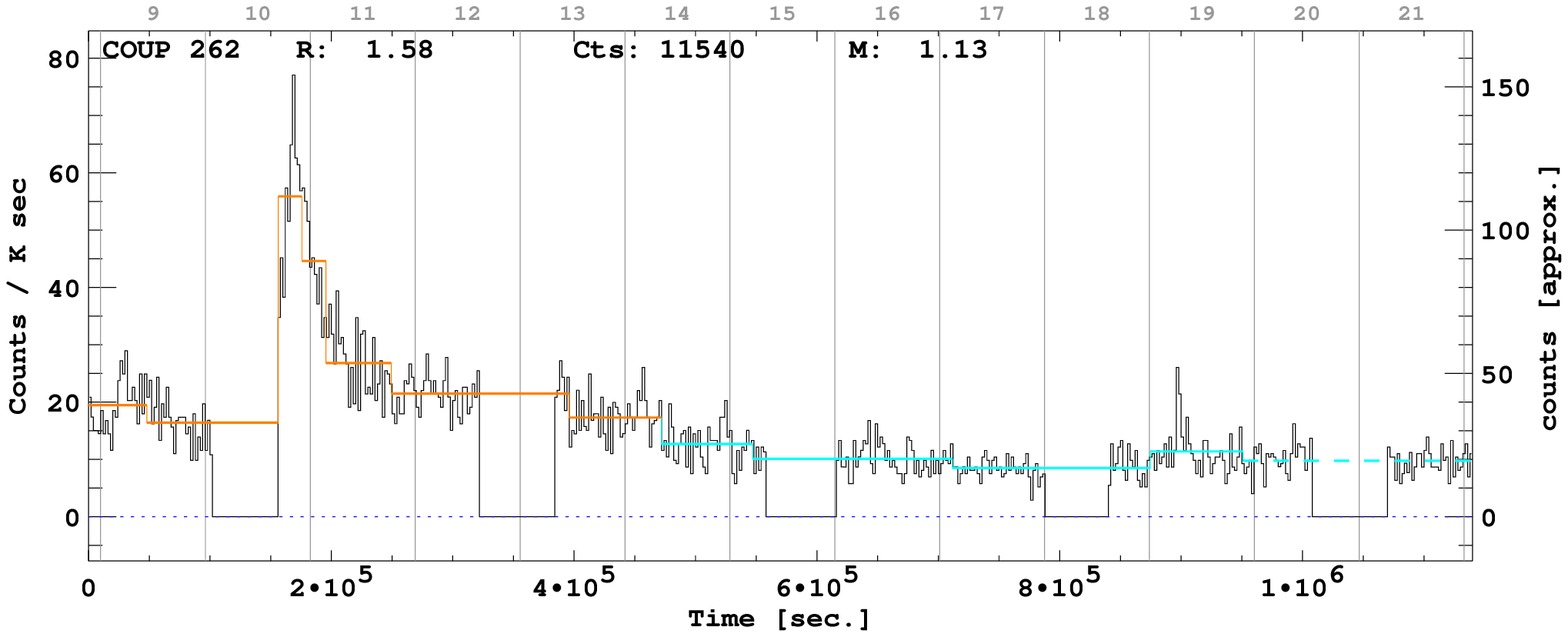}
\plotone{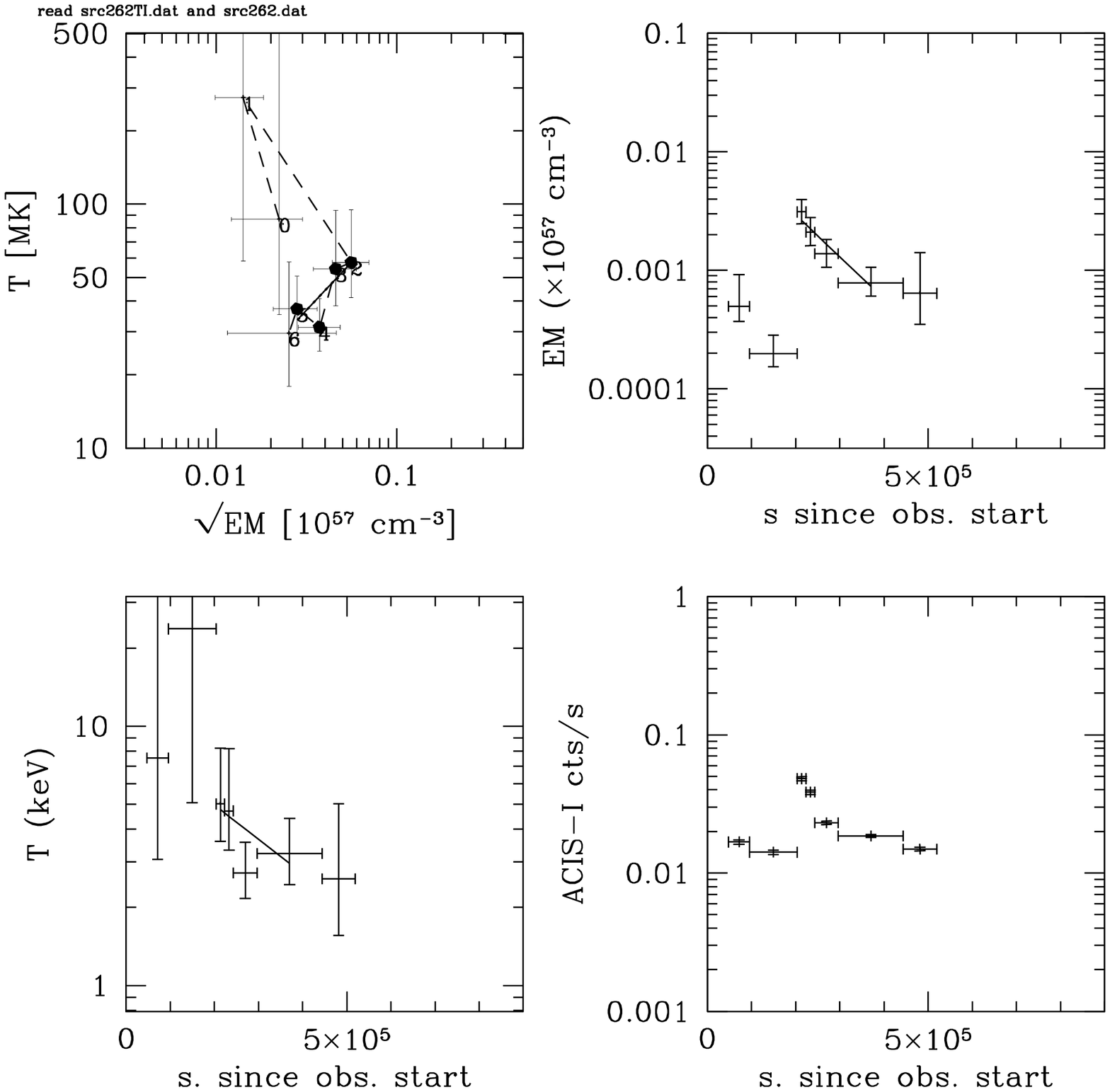}
\caption{Flare evolution of COUP 262.\label{fig:src262}}
\end{figure}

COUP 262, with $1.1\,M_\odot$ and $1.6\,R_\odot$ is a ``young Sun'',
and thus of particular interest for flare studies. In the context of
COUP, its flaring activity is also discussed by \cite{whf+2005}. With
spectral type K5, it shows no evidence for active accretion and has a
rather strong $I-K$ excess ($\Delta(I-K) = 2.2$), pointing to a
relatively massive but inactive disk being present (although no $H-K$
excess is present). Although the flare (Fig.~\ref{fig:src262}) has a
very sharp peak, it decays slowly ($\tau_{\rm lc} = 120$ ks), lasting
for a few days.  The event represents a relatively modest increase
over the characteristic level of a factor of 4, but the shallow
$\zeta$ implies a high level of sustained heating during the decay, so
that (also given the high temperature, $T>100$ MK) the resulting
flaring loop is very large, $L = 2.0 \times 10^{12}~ {\rm cm} =
18\,R_*$, although the significant uncertainty in $\zeta$ implies a
lower confidence range of $L = 3.6\,R_*$.  Nevertheless, even at the
lower end of the confidence range, the flaring region is larger than
normally found in active stars, providing evidence that large loops,
likely connecting the star to the disk, must have existed for the
young Sun.

\subsection{COUP 597}

\begin{figure}
\plotone{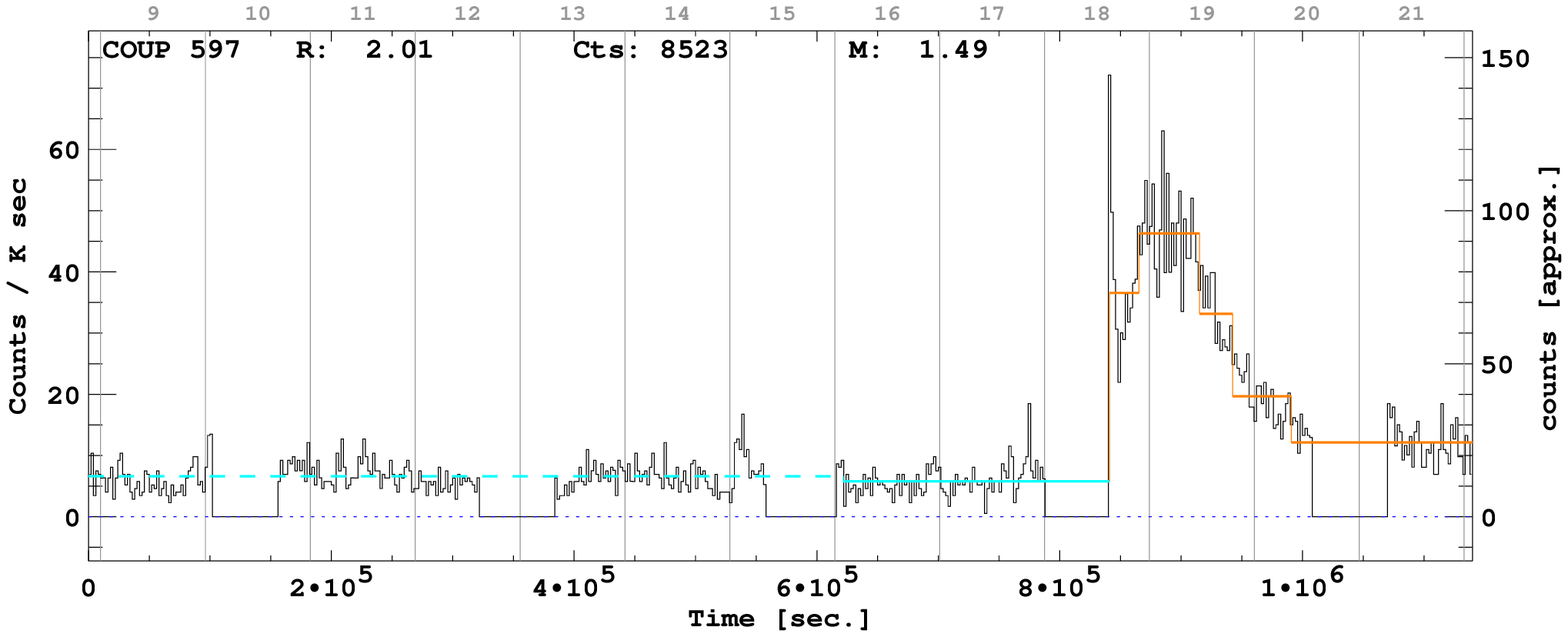}
\plotone{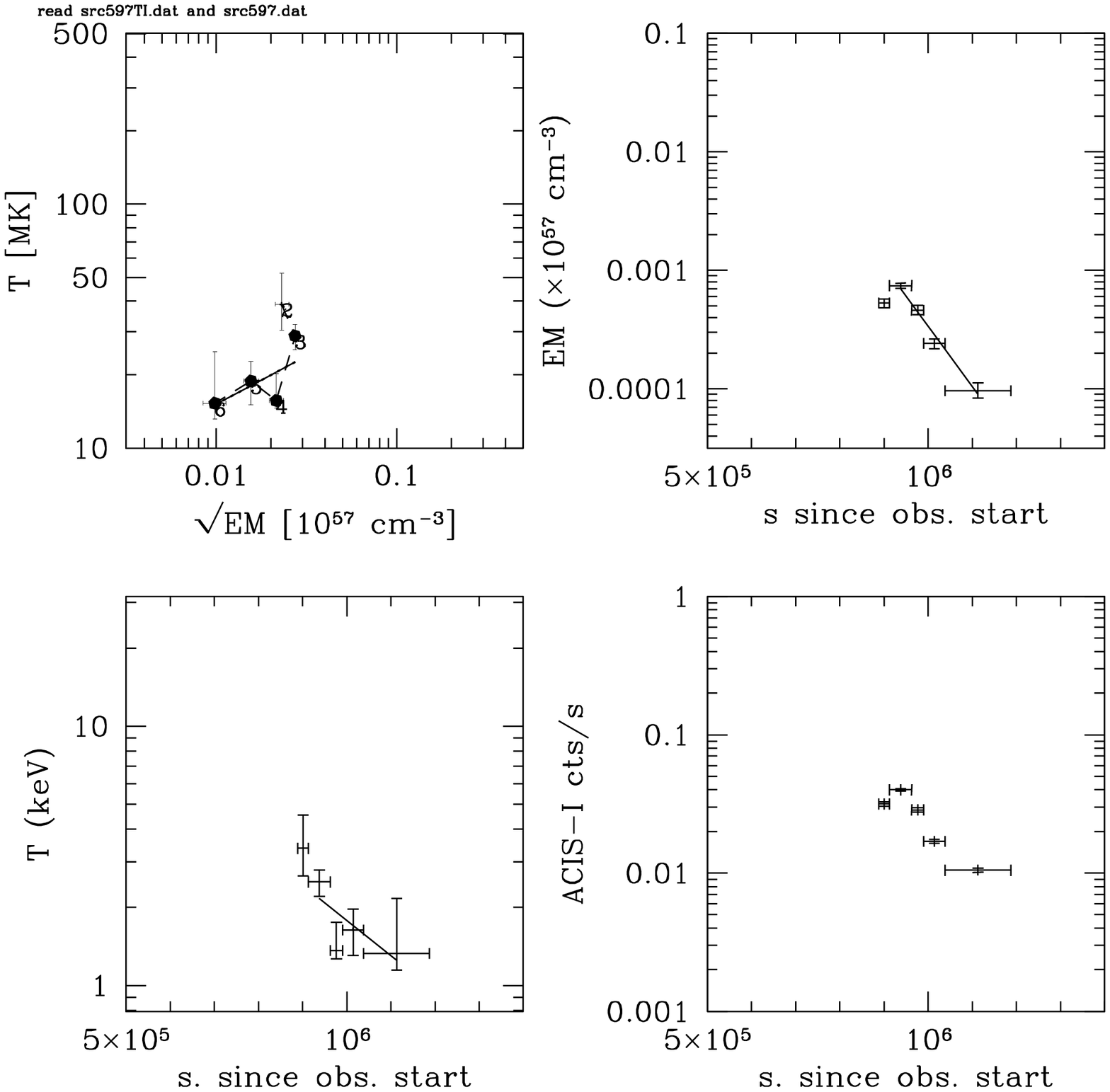}
\caption{Flare evolution of COUP 597.\label{fig:src597}}
\end{figure}

COUP 597 is a somewhat more massive star than COUP 262, with $M =
1.5\,M_\odot$ and $R = 2.0\,R_\odot$. The strong Ca\,{\sc ii}
absorption shows it not to be actively accreting. The flare
(Fig.~\ref{fig:src597}) appears to be the superposition of two events,
a less intense and more long-lasting one and a more intense one of
which only the very fast decay is visible, as the rest falls into an
observation gap. As the analysis relies on the delay of the second,
slower flare, the first, fast event should have no effect on the
results. The peak temperature for the longer event (determined during
the rise phase) may be affected at some level from the presence of the
first flare.  However, given the very moderate peak temperature, and
the weak dependance (as $\sqrt{T}$) of the loop size on peak
temperature, this will not have a strong effect on the results.

Of interest here is the slow rise of the longer flare, which appears
to take place over a day or so. The loop size resulting from the
analysis is $L = 1.6\,R_*$, thus a moderate size loop. The size is
largely driven by the slow temperature decay and resulting shallow
$\zeta$ (with $F(\zeta) = 9.7$, implying strong sustained heating).
The slow rise can therefore imply either a very slowly rising heating
function, or perhaps (using the results of the modeling of COUP 1343,
Sect.~\ref{sec:coup1343}) a heating source distributed along the loop
(or even concentrated near the loop top) rather than localized at the
loop footpoints.

\subsection{COUP 649}
\label{sec:src649}

COUP 649, is a typical low-mass member of the ONC, with $0.4\,M_\odot$
and $2.2\,R_\odot$. It shows no evidence for active accretion and has
an $I-K$ excess ($\Delta(I-K) = 0.4$).

\begin{figure}
\plotone{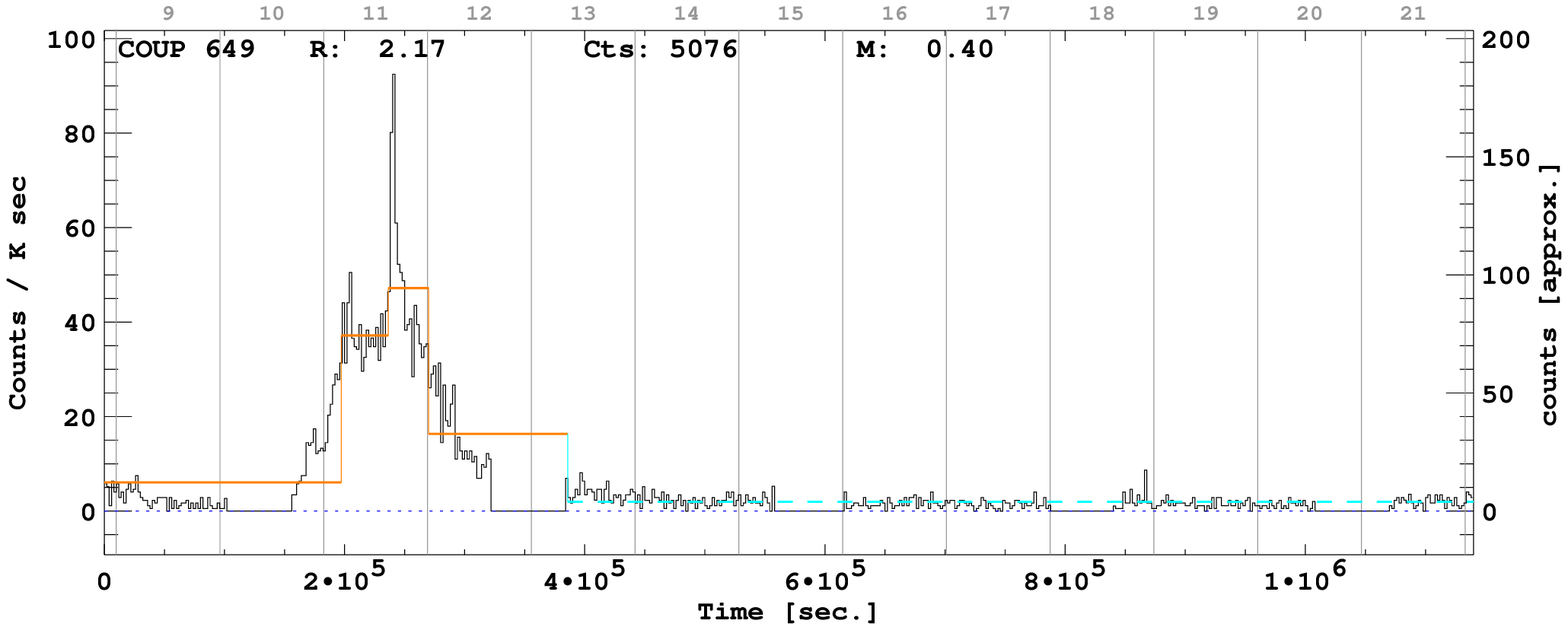}
\plotone{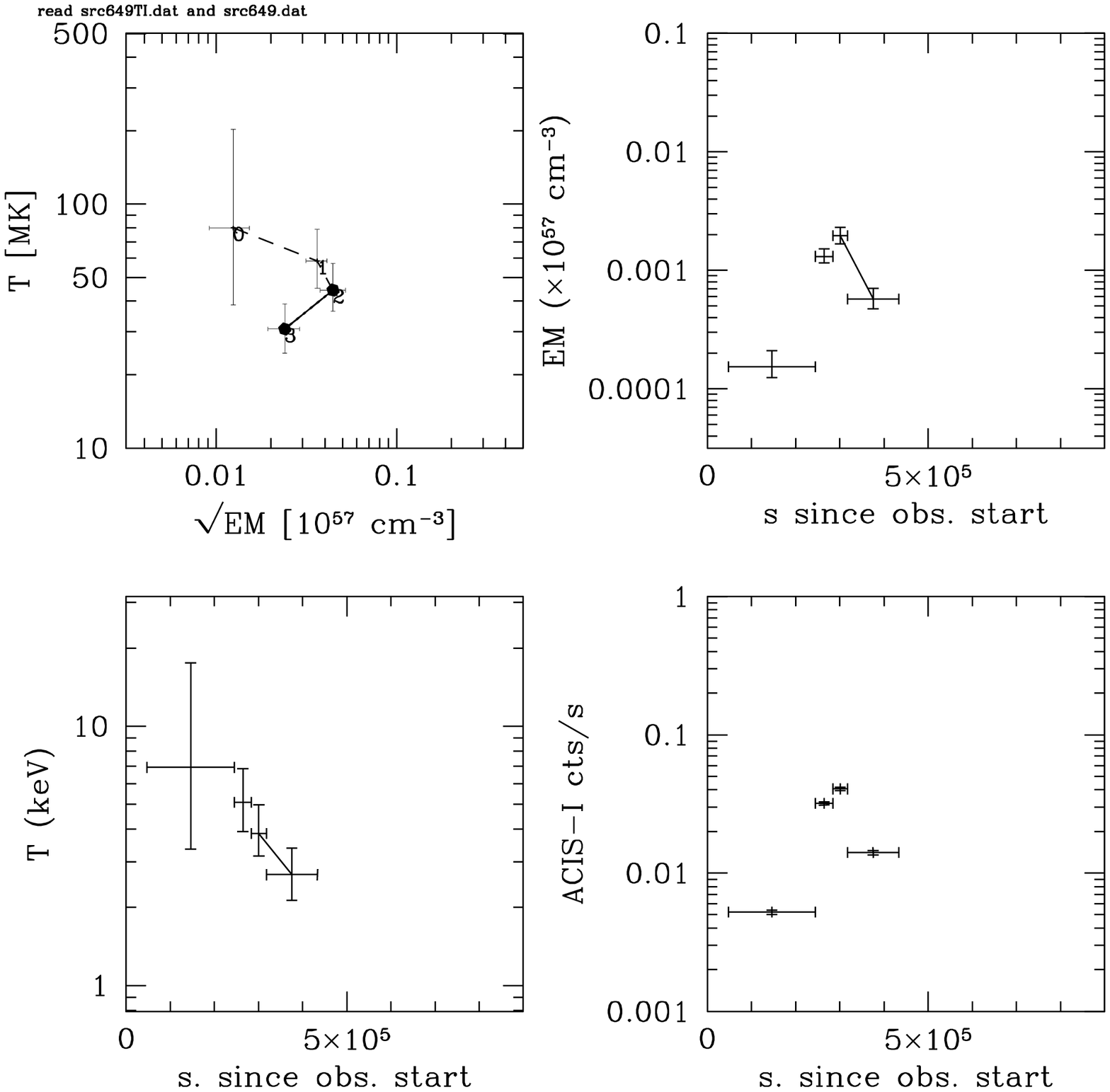}
\caption{Flare evolution of COUP 649.\label{fig:src649}}
\end{figure}

The flaring event has a very peculiar evolution
(Fig.~\ref{fig:src649}), with an initial very slow rise, lasting for
about a day, followed by a plateau at about $10\,\times$ the
characteristic rate; after another half day, an impulsive peak
increases the count rate by another factor of two, and then the decay
begins. Notwithstanding the oddly shaped light curve, the event shows
some of the characteristics of a magnetically confined flaring
structure, in particular the temperature which peaks while the count
rate is still increasing, and has already started to decrease on the
plateau. Somewhat peculiarly, the secondary sharp peak preceding the
decay does not result in an increase in the temperature.

The limited statistics does not allow a detailed analysis, with only
two points on the flare's decay. The peak temperature ($T = 80$ MK) is
moderate by the standard of the other large flares analyzed here, and
the nominal length of the flaring structure is $L = 6.4 \times 10^{11}~
{\rm cm} = 4.2 \, R_*$, although with a large uncertainty driven by
the uncertainty on $\zeta$.

\subsection{COUP 669}

\begin{figure}
\plotone{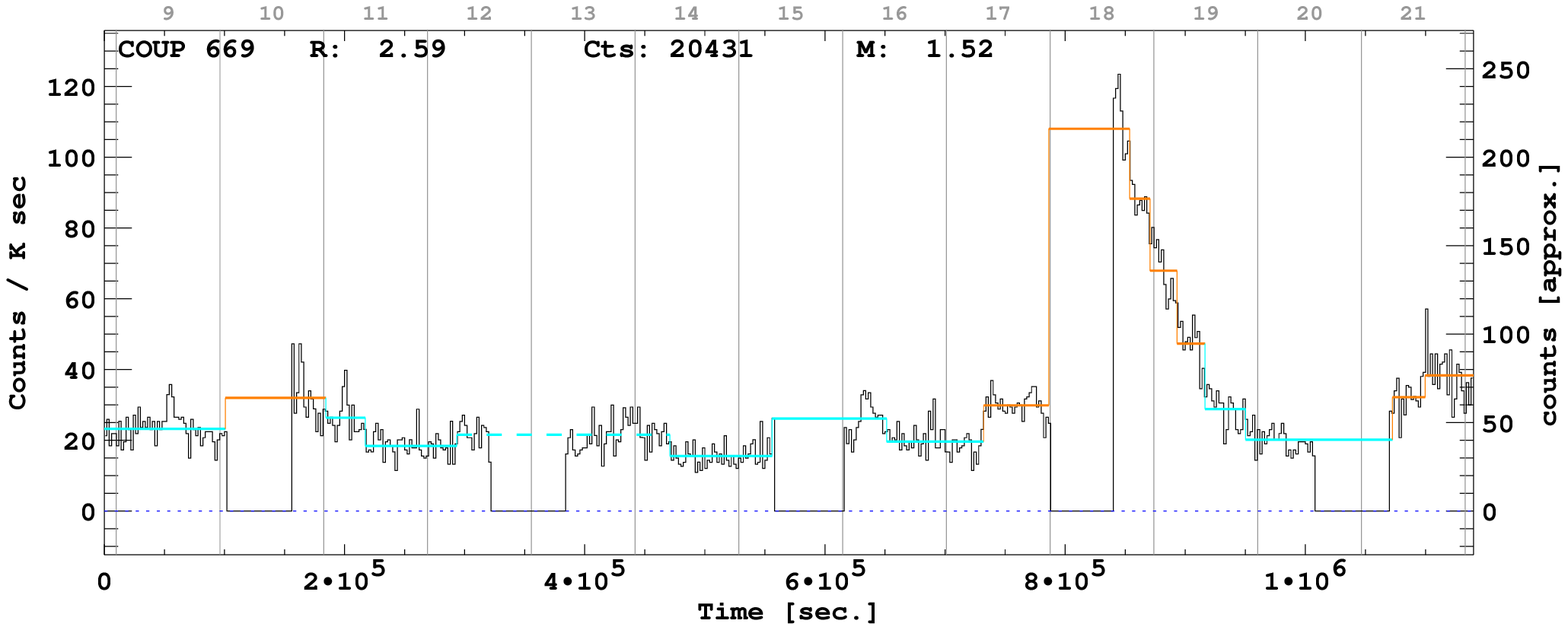}
\plotone{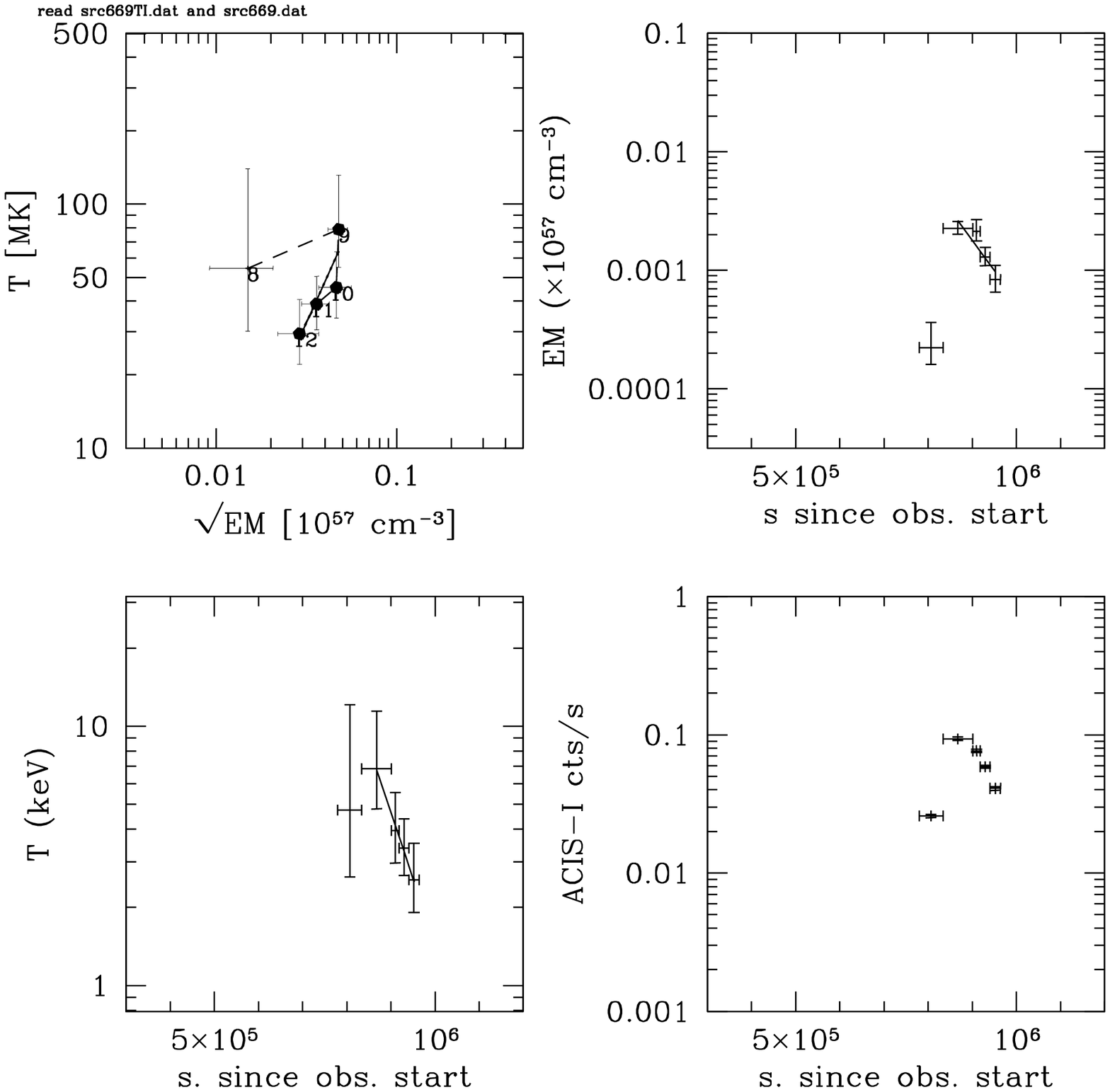}
\caption{Flare evolution of COUP 669.\label{fig:src669}}
\end{figure}

COUP 669 is a relatively massive object, with $1.5\,M_\odot$ and
$2.6\,R_\odot$, and an $I-K$ excess ($\Delta(I-K) = 0.4$). The flare
(Fig.~\ref{fig:src669}) is a well defined impulsive event, for which
the rise phase has however been cut by an observing gap. The peak
temperature, at $T=79$ MK is relatively modest, but the fast decay of
the temperature implies limited sustained heating, resulting in a
flaring loop with a reasonably well determined size, $L = 9.2 \times
10^{11}~ {\rm cm} = 5.1\,R_*$, typical of the large flaring loops in
the COUP sample.

\subsection{COUP 752}

\begin{figure}
\plotone{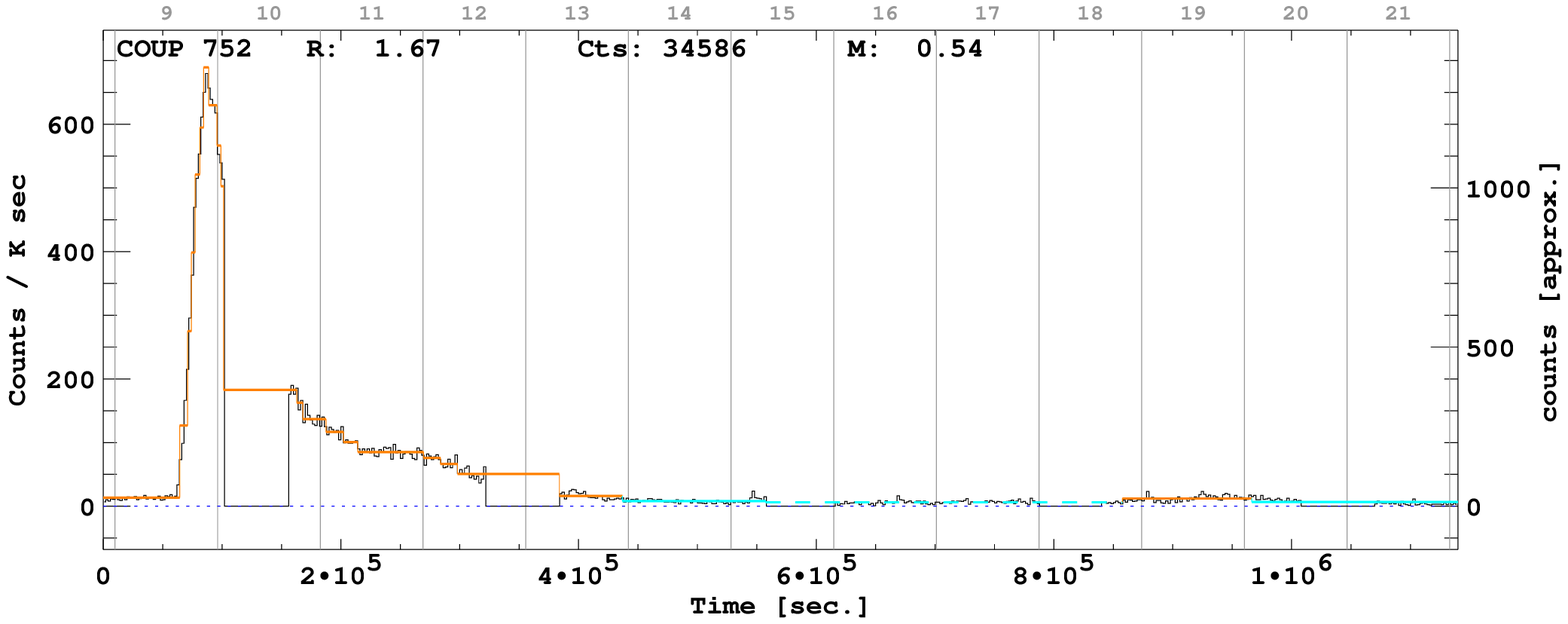}
\plotone{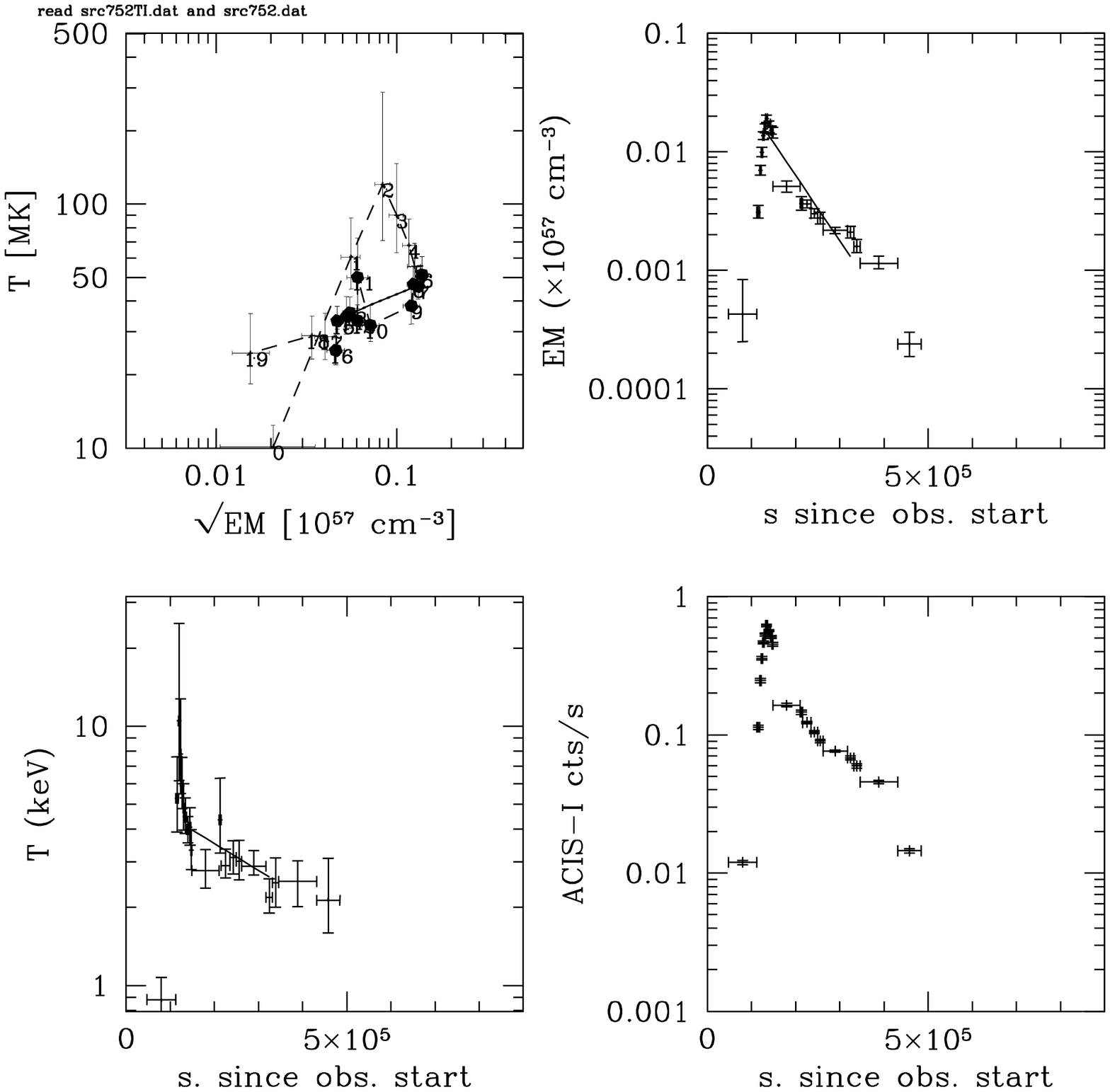}
\caption{Flare evolution of COUP 752.\label{fig:src752}}
\end{figure}

Another typical low-mass object, COUP 752, has $0.5\,M_\odot$ and
$1.7\,R_\odot$, no evidence for ongoing accretion, and an $I-K$ excess
($\Delta(I-K) = 0.6$). COUP 752 undergoes a very intense flare
(Fig.~\ref{fig:src752}), with a peak count rate more than $100\,
\times$ the characteristic one.  The event lasts over 5 days, and the
rise phase is long (half a day) and well observed. The excellent
statistics allow for a detailed analysis of the event, although the
initial part of the decay unfortunately falls in an observing gap. The
flare's peak temperature slightly exceeds 100 MK, still falling in the
domain in which the ACIS detector gives reliable temperature
determinations.

Notwithstanding the very large increase in count rate, the event
follows well the evolution in the $\log T$--$\log \sqrt{E\!M}$ plane
for a flare confined in a single loop.  The temperature peaks (block
1) when the emission measure has just started to rise (and still is
only about one tenth of the peak value), and then decreases to $\simeq
50$ MK when the emission measure has peaked. The decay does not follow
a simple exponential law; rather, there is clear evidence for
sustained heating, with well defined reheating events. In particular,
in block 11 (immediately after the first observing gap) the
temperature increases again briefly, and the decay of both the light
curve and the emission measure changes slope, clear evidence for
reheating of the plasma.

Our analysis shows a very shallow $\zeta$, indicative of the presence
of strong ongoing heating during the flare decay, as it is also
evident from the increase in plasma temperature. As a consequence,
only an upper limit to the flaring structure's size can be derived, at
$L \le 7.4 \times 10^{10}~ {\rm cm}$, with a best-fit value $L = 6.5
\times 10^{10} {\rm cm} = 5.6\,R_*$. The resulting minimum confining
magnetic field, at $B \ge 250$ G is relatively strong.

\subsection{COUP 891}

COUP 891 is the most massive object in the COUP flaring sample, with
$2.4\,M_\odot$, $4.9\,R_\odot$, a strong $I-K$ excess ($\Delta(I-K) =
1.10$), no evidence for accretion and significant obscuration ($A_V =
8.0$ mag). The X-ray light curve suffers from significant pile-up, and
therefore we have used, for our analysis, the list of photons
extracted from an annular region free from pile-up, as described by
\cite{gf+2005}. This strongly reduces the number of photons available
for a spectral analysis, so that only moderate statistics are
available for this event.

\begin{figure}
\plotone{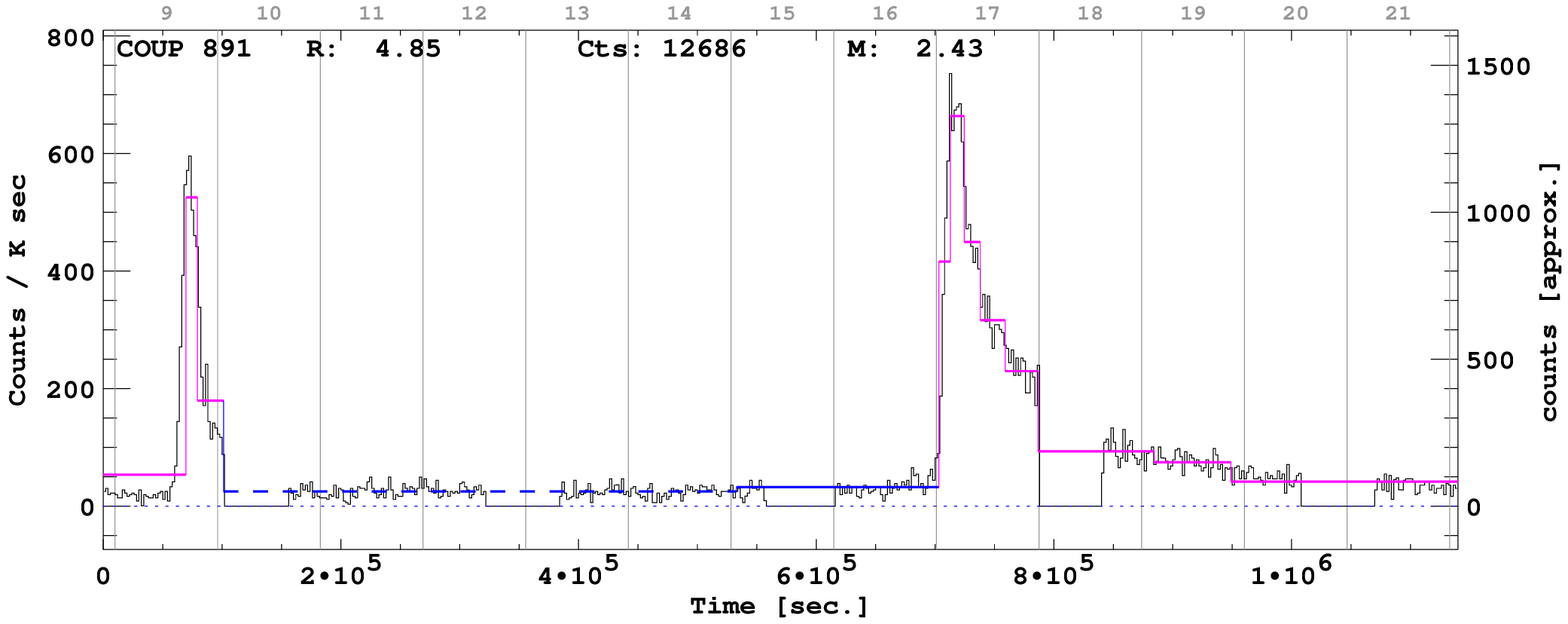}
\plotone{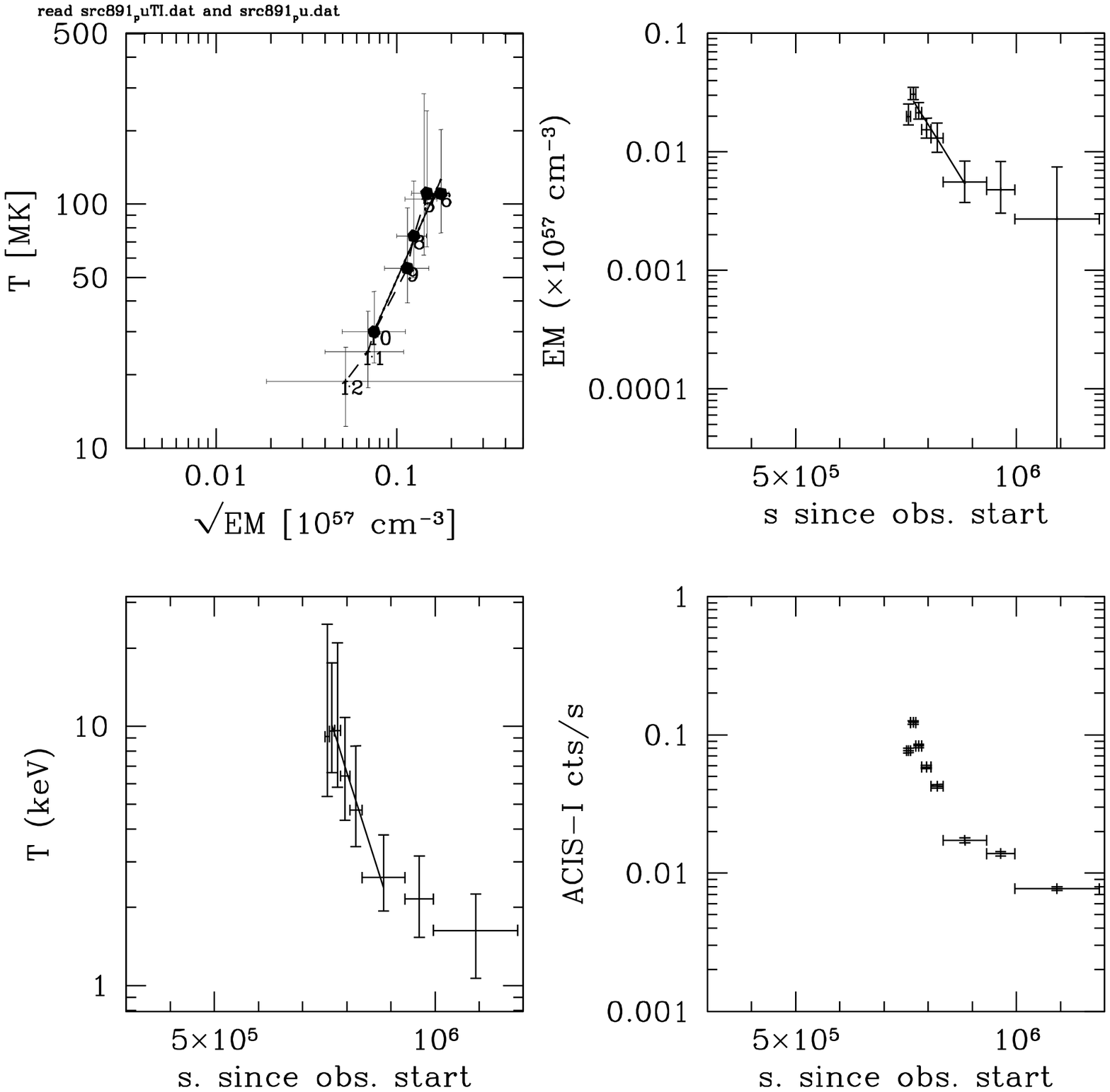}
\caption{Flare evolution of COUP 891.\label{fig:src891}}
\end{figure}

Two flares are present in the COUP light curve
(Fig.~\ref{fig:src891}); however the first one is too short and cannot
be analyzed with the approach adopted here. The second event is a well
defined impulsive event, with a fast rise and a regular, exponential
decay, well traced for more than four days. At the peak, the flare
count rate is $\simeq 50 \times$ larger than in quiescence. The peak
temperature is slighly above 100 MK, but still within the range of the
ACIS instrument. The fast decay of temperature results in a fairly
steep $\zeta$, and thus in little sustained heating. The long decay
time ($\tau_{\rm lc} = 73$ ks) therefore results in a long flaring
structure, $L = 1.7\times 10^{12}~ {\rm cm} = 5.1\,R_*$.

\subsection{COUP 971}

\begin{figure}
\plotone{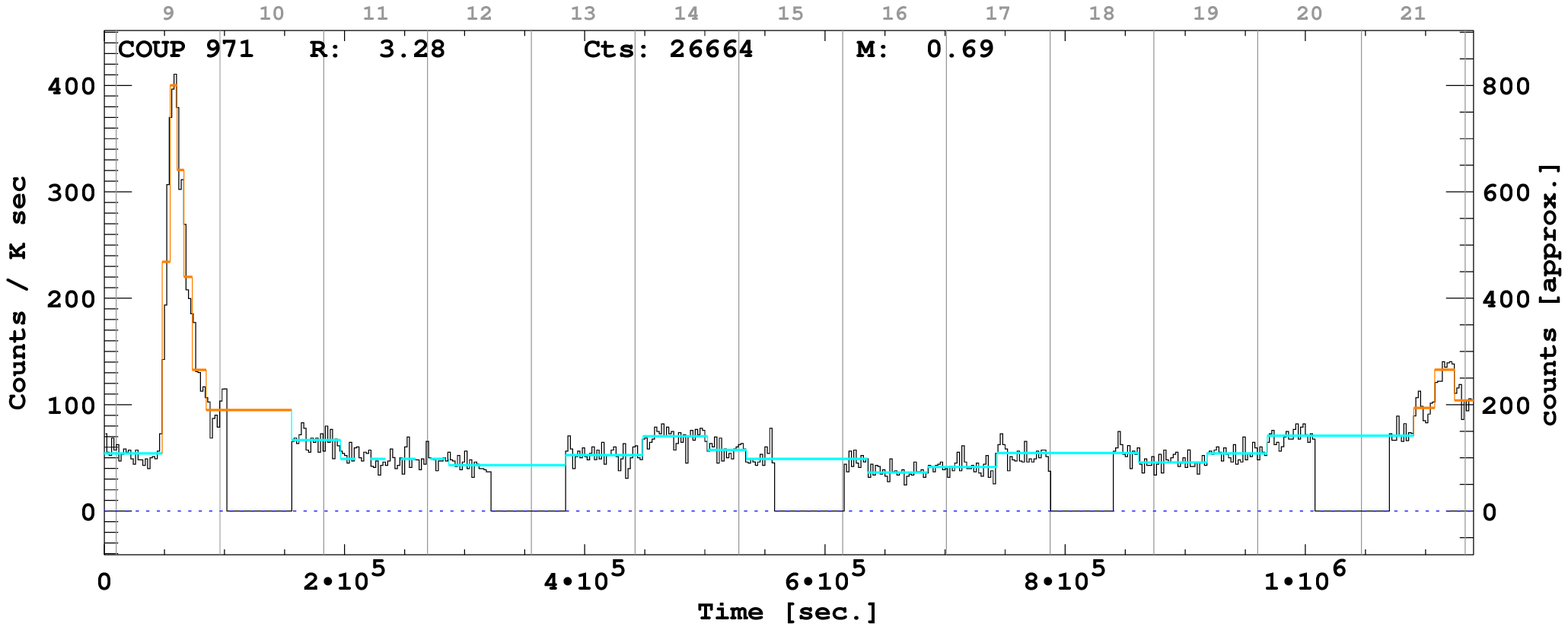}
\plotone{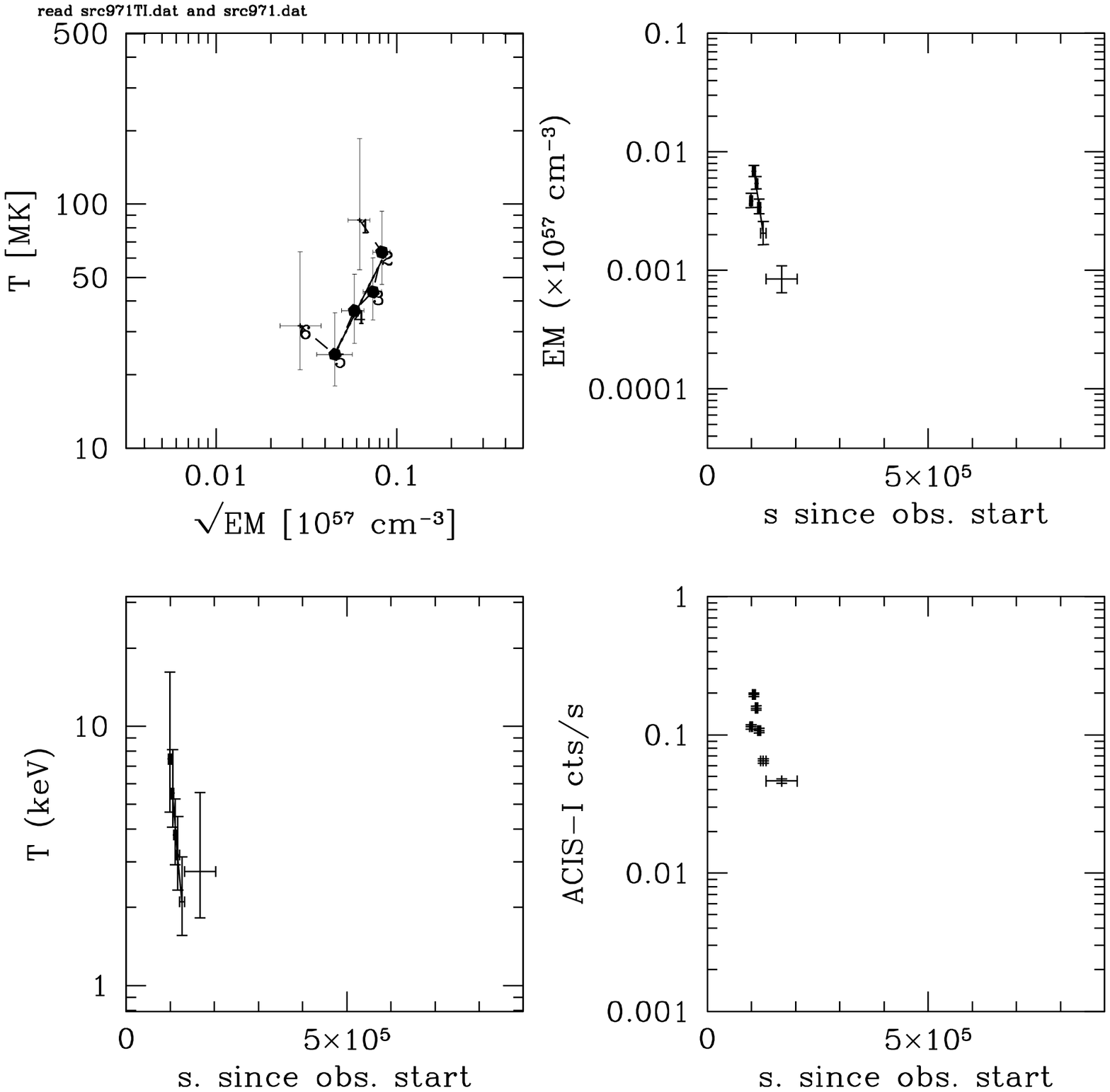}
\caption{Flare evolution of COUP 971.\label{fig:src971}}
\end{figure}

COUP 971 is a low mass ($0.7\,M_\odot$) star with $R = 3.3\,R_\odot$.
The Ca\,{\sc ii} in absorption points to a non-accreting system. The
flaring event (Fig.~\ref{fig:src971}) is a typical example of a short
impulsive flare. The decay in the $\log T$--$\log \sqrt{E\!M}$ plane
is steep and regular, pointing to little if any residual heating. The
result is a moderate size loop ($L = 1.6\,R_*$), similar to the ones
resulting from the analysis of intense flares in more evolved stars;
this is likely an example of an event confined in a structure anchored
to the stellar photosphere only.

\subsection{COUP 1083}

Little optical information is available for COUP 1083, and in
particular no mass and radius estimate is known. A rotational period
has however been derived ($P = 5.9$\,d) for this star. Its spectral
type has been estimated at M0, showing that the object must have a
moderate mass.

\begin{figure}
\plotone{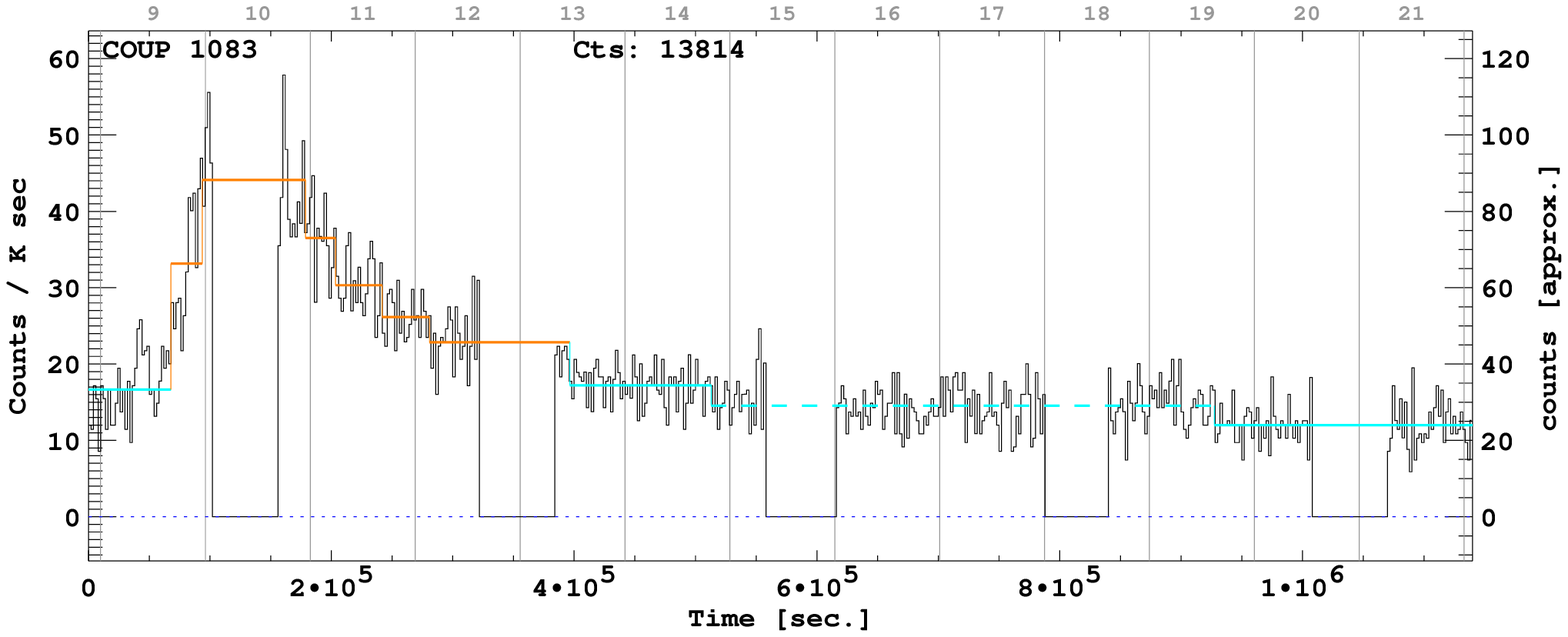}
\plotone{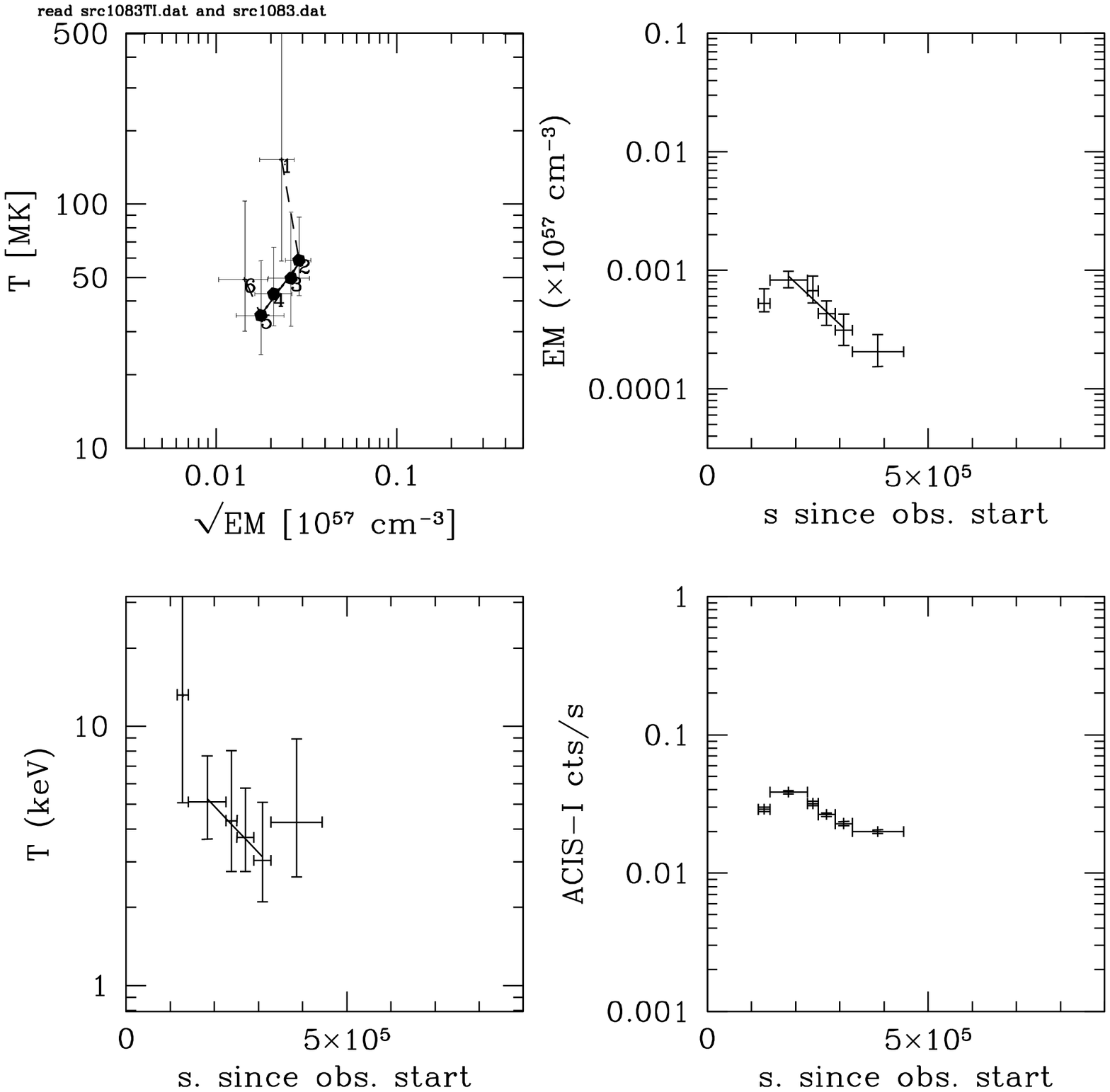}
\caption{Flare evolution of COUP 1083.\label{fig:src1083}}
\end{figure}

The flare present in the COUP 1083 lightcurve has a peculiar shape,
well visible in Fig.~\ref{fig:src1083}, with a very slow rise, lasting
more than one day, and a decay lasting about 3 days. The peak of the
flare unfortunately falls in an observing gap, and it thus not
visible. We have analyzed the event assuming that the observed maximum
is the actual peak, but in practice the peak might be higher, although
this would not substantially affect our conclusions.

Notwithstanding the peculiar light curve, the evolution of the event
follows the one theoretically expected for a confined flare, with the
temperature peaking prior to the ${E\!M}$ and a regular decay in the
$\log T$--$\log \sqrt{E\!M}$ plane. A second event may be starting in
the final phase of the decay of the flare, on day 12, as shown both by
the increasing count rate and by the increasing temperature. The
second flare is cut by the observation gap and thus nothing can be
derived for it. We have not included the last ML block in the
analysis.

Although the statistics are limited (and thus the error bars large),
the derived steep $\zeta$ is compatible with limited sustained
heating, so that a long structure, $L = 2.4 \times 10^{12}$ cm, is
derived from the analysis due to the slow flare decay ($\tau_{\rm lc}
= 125$ ks).  The large size is also compatible with the very slow rise
time, as discussed for the detailed model of COUP 1343
(Sect.~\ref{sec:coup1343}).

\subsection{COUP 1246}

\begin{figure}
\plotone{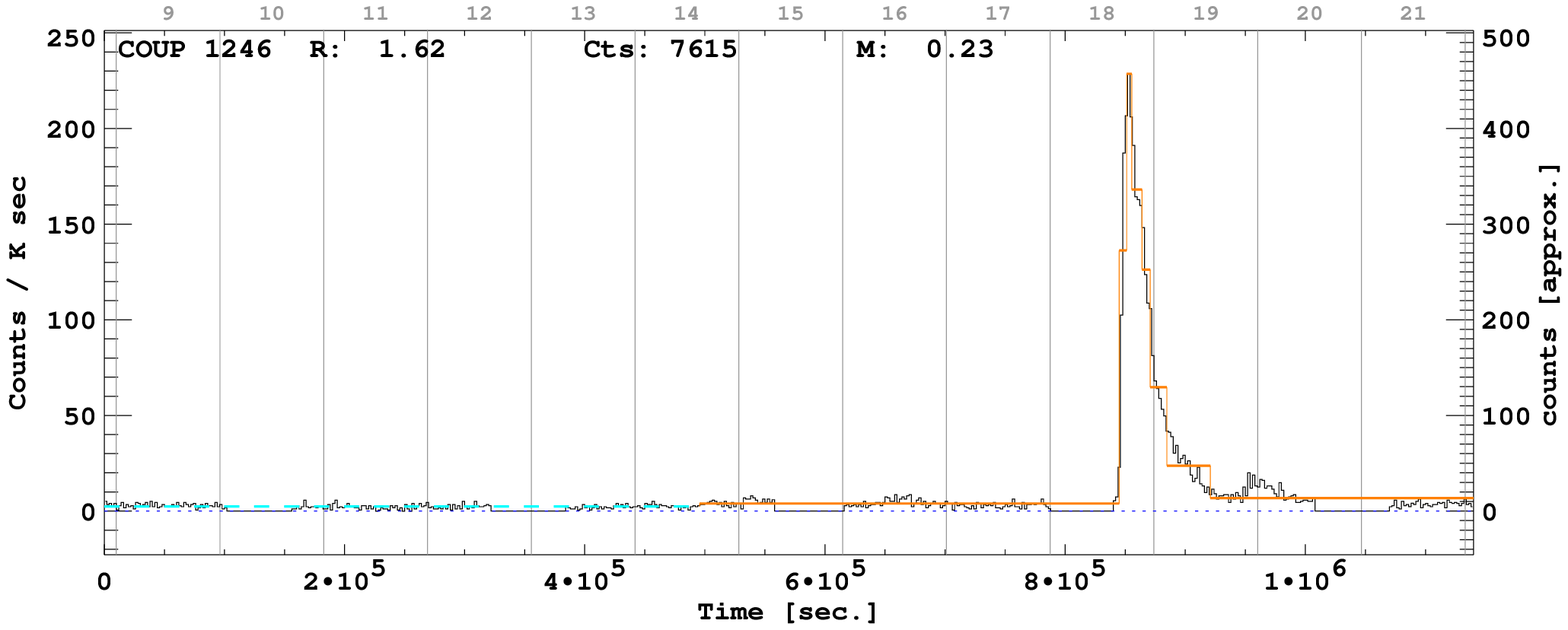}
\plotone{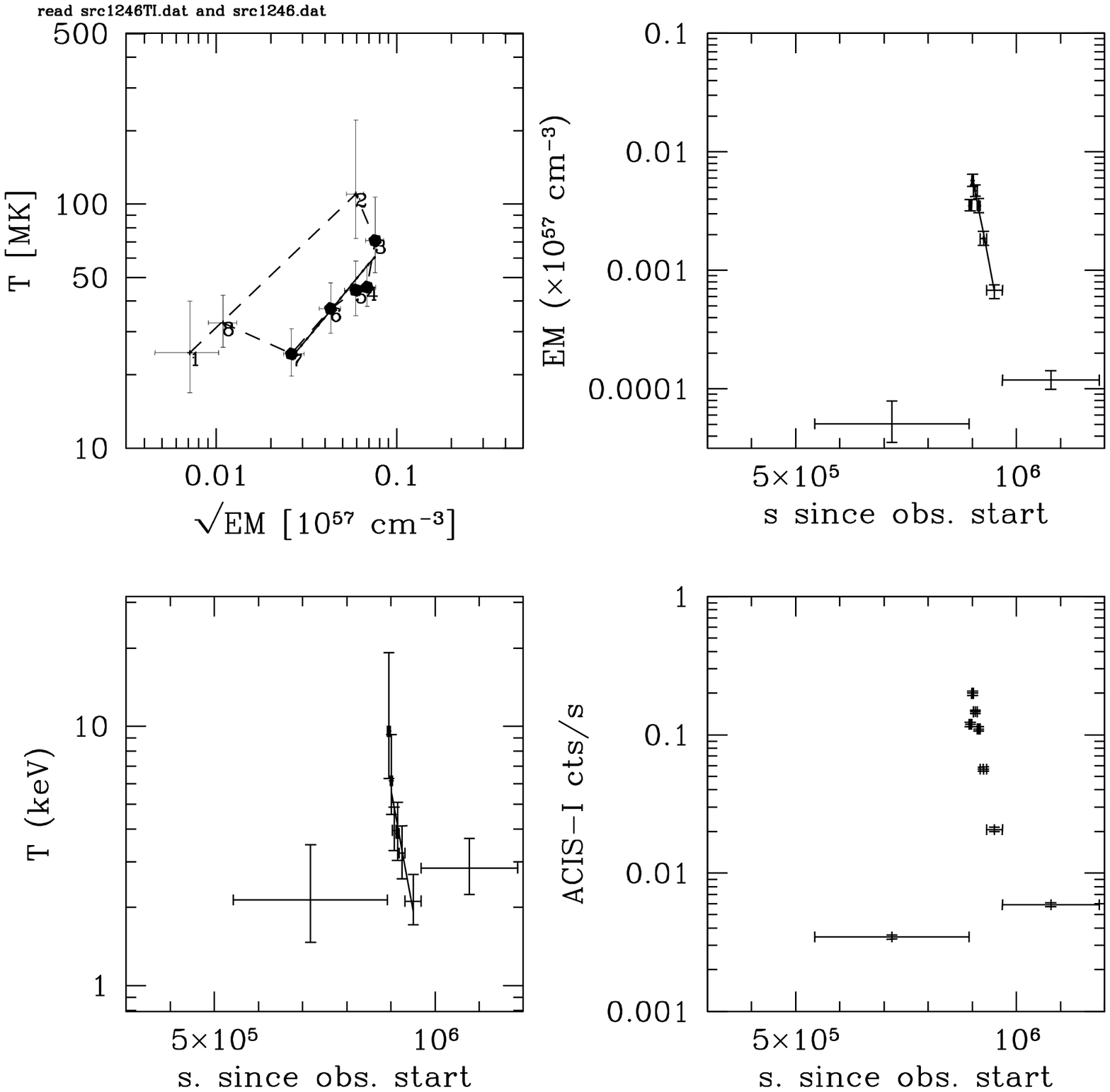}
\caption{Flare evolution of COUP 1246.\label{fig:src1246}}
\end{figure}

COUP 1246 is one of the lowest-mass stars in the present sample, with
$0.2\,M_\odot$ and $1.6\,R_\odot$. It shows no evidence for active
accretion, but it has a sizable $K$-band IR excess ($\Delta(I-K) =
0.8$) and is one of the sources showing an $K-L$ excess but not a
$H-K$ band one; its rotational period has been determined at
$P=5.2$\,d.  The flare (Fig.~\ref{fig:src1246}) is a very well defined
impulsive event, with a fast rise and a peak count rate $\simeq 100
\times$ the characteristic rate. Also in this case the peak observed
temperature is somewhat above 100 MK, but still within the range of
reliability of ACIS. The decay in the $\log T$--$\log \sqrt{E\!M}$
plane is well defined, and closely follows the one expected from the
models of confined flares, allowing to analyze the event in some
detail. The light curve plotted in the top panel of
Fig.~\ref{fig:src1246} shows clear evidence for a second, smaller
flare developing on the tail of the first one, resulting in an
increase in temperature at the flare end. Statistics are insufficient
to analyze the second flare at any level of detail, and therefore the
last block has not been included in the analysis. The resulting size
of the flaring region is $L = 4.0 \times 10^{10}~ {\rm cm} = 3.8\,R_*$,
i.e.\ a typical event for the COUP sample.

\subsection{COUP 1343}

Very little is known about the optical counterpart to COUP 1343 (the
analysis for the flare is shown in Fig.~\ref{fig:src1343}), with no
mass or radius estimates known, although a rotational period
measurement ($P = 8.7$\,d) is available. The flare is a very well
defined event, with a clean impulsive rise and an undisturbed decay.
The peak flare luminosity is $> 10 \times$ larger than the
characteristic source luminosity, allowing a clean analysis of the
flare spectral characteristics. In the analysis we have ignored the
last segment of the flare (the one extending into day 13), for which
the temperature has too large an error (also because of the data gap).
The peak temperature of the flare is very high, with a nominal
best-fit temperature of $\simeq 400$ MK. As discussed above such high
temperatures cannot be reliably determined with ACIS, and therefore
the resulting flare size (based on the assumption $T_{\rm obs} = 100$
MK) is to be considered as a lower limit.

The rapid temperature decrease implies that little sustained heating
is present, so that the observed decay is close to the thermodynamic
decay of the loop. The smooth decay also argues for the lack of
significant reheating. Therefore, with $\tau_{\rm lc} = 39$ ks and
$T_{\rm obs} \ge 100$ MK, the analysis results in $L \simeq 1 \times
10^{12}$ cm, or $L \simeq 14\,R_\odot$. Even if COUP 1343 were a
relatively massive star, with a radius of a few $R_\odot$, the flaring
structure would still be several times the size of the star. The
resulting magnetic field, while moderate in itself (150 G), must be
coherently organized over some 0.1 AU.

As described in Sect.~\ref{sec:coup1343} the evolution of the same
event has also been simulated in detail, confirming that the
temperature and emission measure evolution are the result of a flare
confined in a large loop of the same size of the one found by the
decay analysis discussed here, with impulsive heating.

\begin{figure}
\plotone{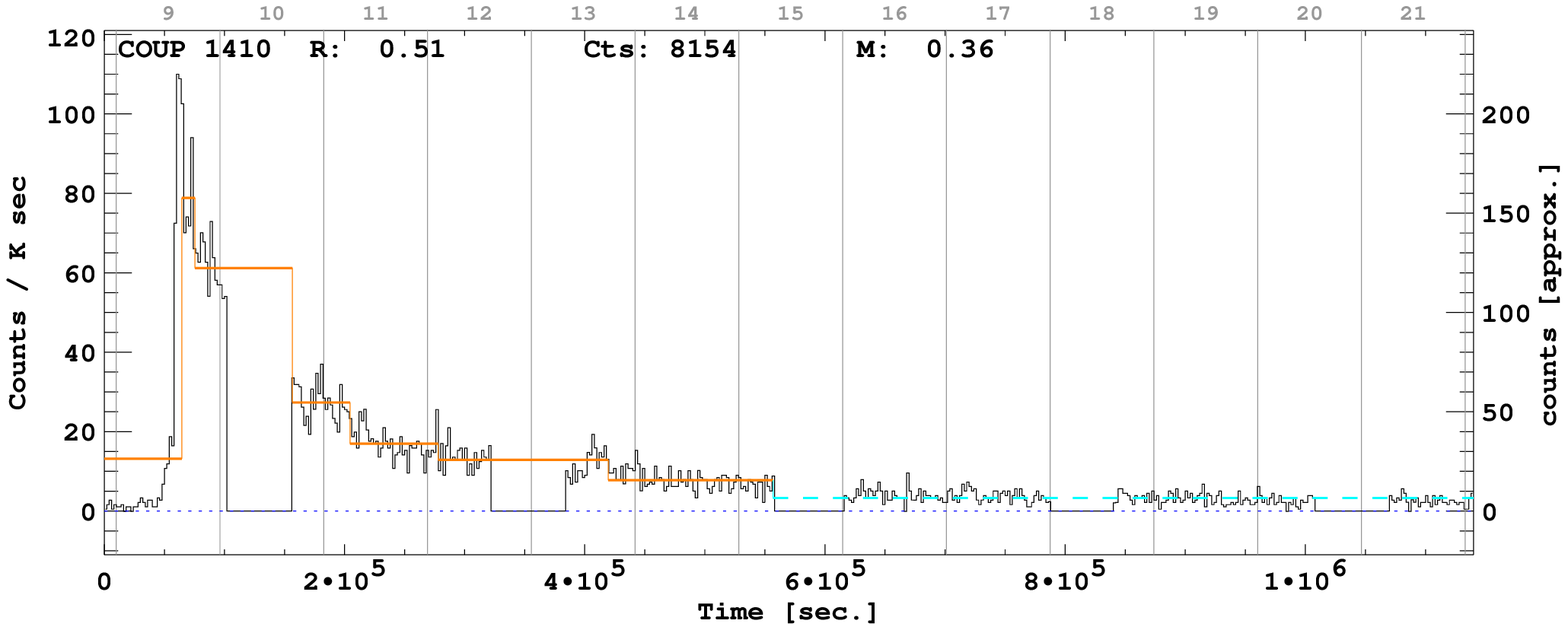}
\plotone{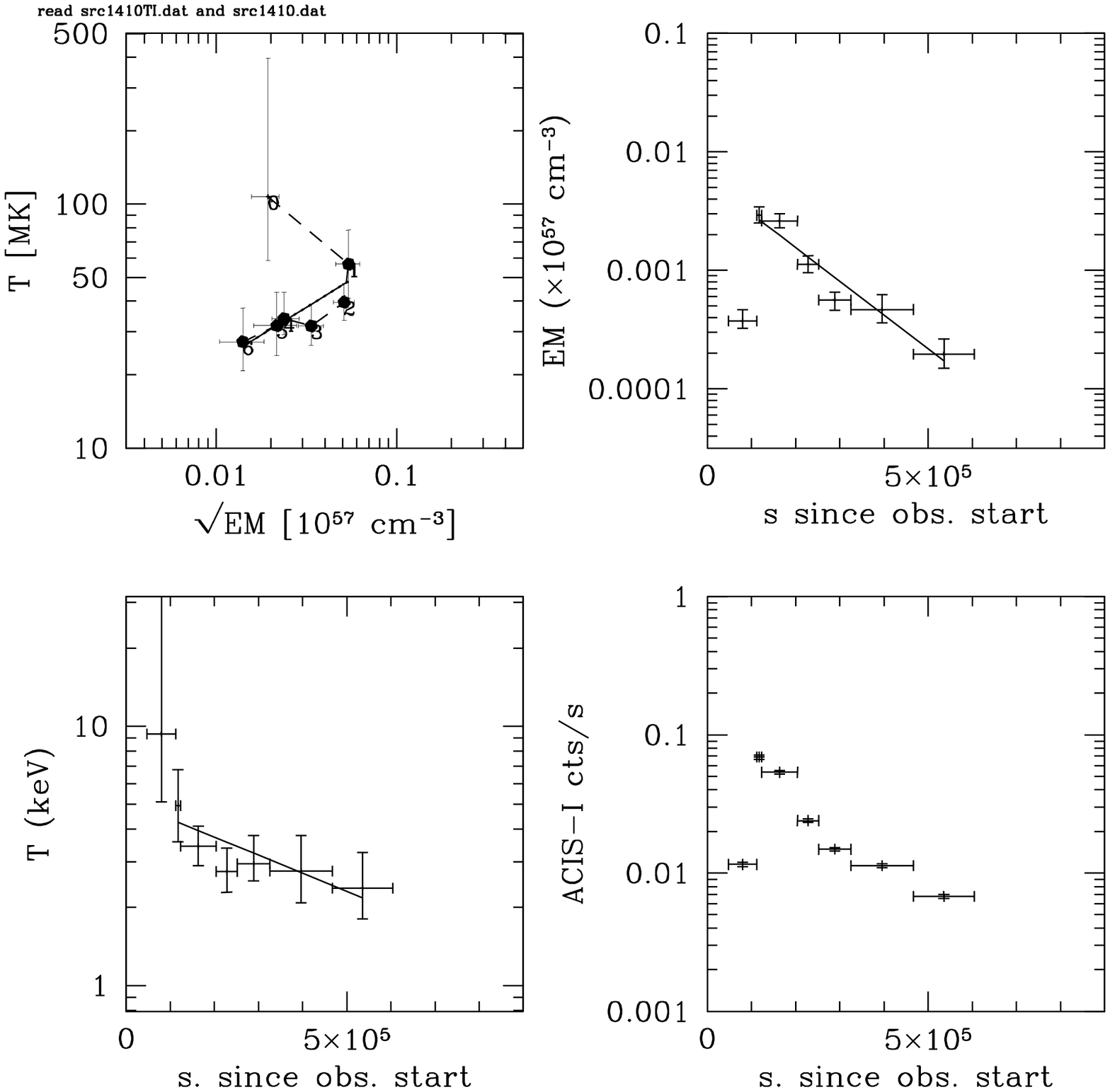}
\caption{Flare evolution of COUP 1410.\label{fig:src1410}}
\end{figure}

\subsection{COUP 1410}

COUP 1410 appears more evolved than the rest of the sample, with an
estimated age $t = 3.6\times 10^7$ yr, and a consequently small radius
($0.51\,R_\odot$) for its mass ($0.36\,M_\odot$). It shows no evidence
for accretion from Ca\,{\sc ii} emission, but it has a significant
$K$-band excess ($\Delta(I-K) = 2.3$). The rotational period has been
measured at $P = 6.76$\,d.

The flare (Fig.~\ref{fig:src1410}) has a well defined impulsive rise,
followed by a very long decay extending for over a week, with
$\tau_{\rm lc} = 150$\,ks. Even though significant heating is present
(as shown by the shallow $\zeta$), the resulting size of the flaring
structure is $L = 1.1 \times 10^{12}~ {\rm cm} = 55\,R_*$, very long
with respect to the (small) stellar radius. However, the uncertainty
on $\zeta$ results in a significant uncertainty on $L$, which has a
lower range of $L = 1 \times 10^{11}~ {\rm cm} = 5\,R_*$, the same
large size as found for the best determined large events in the COUP
sample, making this event a good candidate for a star-disk connecting
structure.

\begin{figure}
\plotone{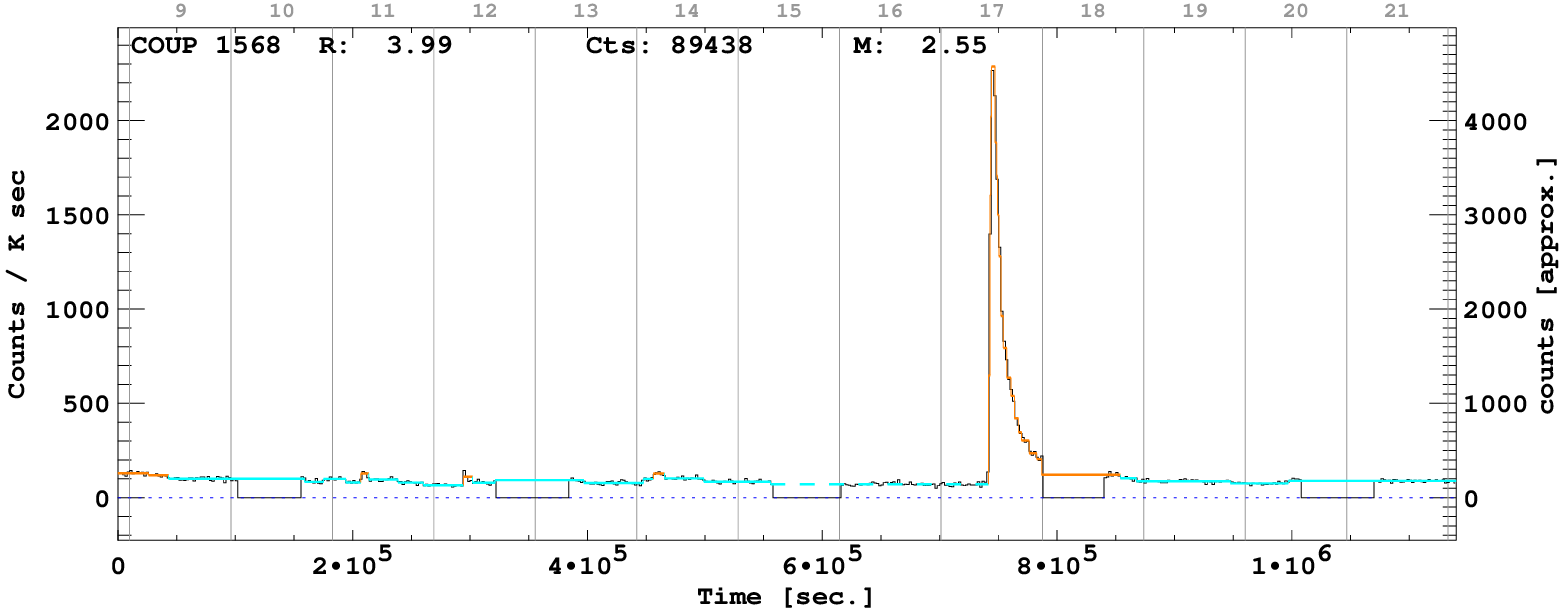}
\plotone{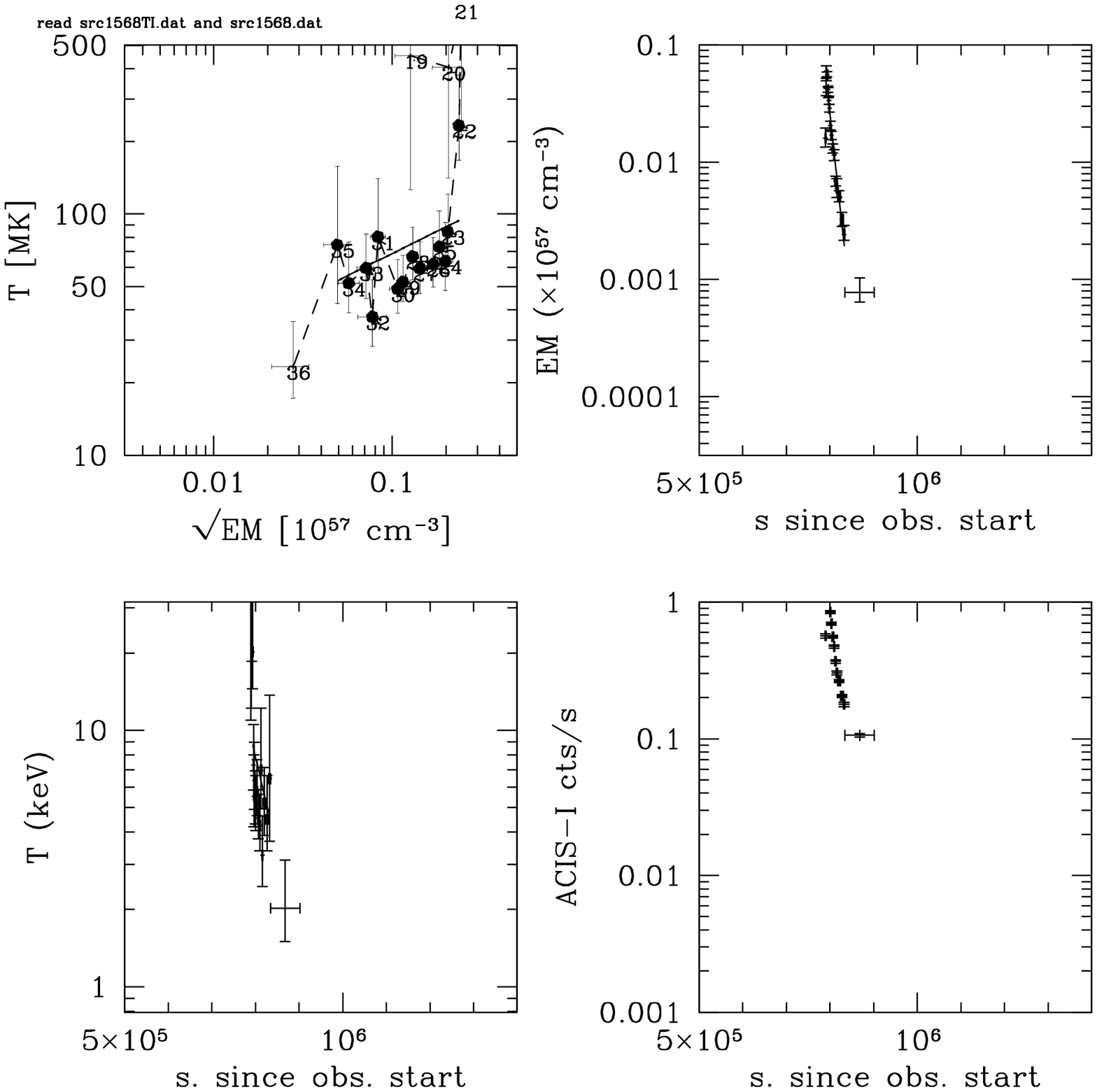}
\caption{Flare evolution of COUP 1568.\label{fig:src1568}}
\end{figure}

\subsection{COUP 1568}

COUP 1568 is another relatively massive object in the flaring sample,
with $2.6\,M_\odot$ and $4.0\,R_\odot$.  It has a small IR excess
($\Delta(I-K) = 0.2$).

The flare (Fig.~\ref{fig:src1568}) is relatively short, lasting just
one day, but it is intense, peaking at over $100 \times$ the
characteristic rate, with excellent statistics due to the source's
off-axis angle which prevents pile-up. The impulsive phase is very
sharp, and the decay almost a perfect exponential. The peak
temperature is very high: the best-fit values for the ACIS spectra at
peak are $T \simeq 500$ MK, however with a very large uncertainty
which always include values as low as 100 MK, due to the ACIS spectral
response.

Once in the decay phase, the temperature decays very slowly, resulting
in a very shallow $\zeta$. Therefore only an upper limit to size of
the flaring structure can be obtained, $L \le 1.5 \times 10^{11}$ cm,
with a best fit value $L = 3.7 \times 10^{10}~ {\rm cm} = 0.4\,R_*$.
Unlikely many of the large COUP flares studied in this paper, this
event is very likely to be compact, with size smaller than the star
itself. A small size is also implied by the very fast rise phase
(which constrains the time scale with which the chromospheric
evaporation will fill the flaring loop). Given the compact size,
together with the large peak emission measure and high temperature,
the peak flare density will likely be high ($n_e = 1.2\times 10^{12}$
cm$^{-3}$), as must the confining magnetic field ($B = 3.5$ kG)

%\bibliographystyle{natbib}
%\bibliography{references}

%\appendix
%\twocolumn

%\clearpage

\end{document}